\documentclass[twocolumn]{autart}

\pdfoutput=1

\usepackage{amsmath} 
\usepackage{amssymb}  
\usepackage{amsfonts}  
\usepackage{graphicx}
\usepackage{pdflscape}
\usepackage{rotating}
\usepackage{color}
\usepackage[labelfont=bf]{caption}
\usepackage{subcaption}
\usepackage{hyperref}

\newcommand{\col}{\mbox{col}}

\def\cale{\mathcal{E}}
\def\cali{\mathcal{I}}

\def\L2e{{\cal L}_{2e}}

\def\rea{\mathbb{R}}

\def\adj{\mbox{adj}}
\def\liminf{\lim_{t \to \infty}}

\def\begequarr{\begin{eqnarray}}
\def\endequarr{\end{eqnarray}}
\def\begequarrs{\begin{eqnarray*}}
\def\endequarrs{\end{eqnarray*}}
\def\begarr{\begin{array}}
\def\endarr{\end{array}}
\def\begequ{\begin{equation}}
\def\endequ{\end{equation}}
\def\label{\label}
\def\begdes{\begin{description}}
\def\enddes{\end{description}}
\def\begenu{\begin{enumerate}}
\def\begite{\begin{itemize}}
\def\endite{\end{itemize}}
\def\endenu{\end{enumerate}}
\def\lef[{\left[\begin{array}}
\def\rig]{\end{array}\right]}

\def\begcen{\begin{center}}
\def\endcen{\end{center}}

\def\caly{\mathcal{Y}}

\def\call{\mathcal{L}}
\def\calz{\mathcal{Z}}

\def\epst{\epsilon_t}

\def\rea{\mathbb{R}}

\def\IJACSP{{\it Int. J. on Adaptive Control and Signal Processing}}

\def\TAC{{\it IEEE Trans. Automatic Control}}

\def\SCL{{\it Systems \& Control Letters}}

\def\CST{{\it IEEE Trans. Control Systems Technology}}

\usepackage{color}

\def\begmat#1{\begin{bmatrix}#1\end{bmatrix}}

\usepackage{color}

\usepackage[prependcaption,colorinlistoftodos]{todonotes}

\begin{document}

\begin{frontmatter}

\title{State Observers for Sensorless Control of Magnetic Levitation Systems\thanksref{footnoteinfo}}

\thanks[footnoteinfo]{This paper was not presented at any IFAC 
meeting. Corresponding author A.~A.~Vediakov.}

\author[ITMO]{Alexey Bobtsov}\ead{bobtsov@mail.ru},
\author[ITMO]{Anton Pyrkin}\ead{pyrkin@corp.ifmo.ru},
\author[SPLC]{Romeo Ortega}\ead{romeo.ortega@lss.supelec.fr},
\author[ITMO]{Alexey Vedyakov}\ead{vedyakov@corp.ifmo.ru}

\address[ITMO]{Department of Control Systems and Informatics, ITMO University, Kronverksky av., 49, 197101, Saint Petersburg, Russia}
\address[SPLC]{Laboratoire des Signaux et Systèmes, CNRS-SUPELEC, Plateau du Moulon, 91192, Gif-sur-Yvette, France}

										
\begin{abstract}
In this paper we address the problem of state observation for sensorless control of nonlinear magnetic levitation systems, that is, the regulation of the position of a levitated object measuring only the voltage and current of the electrical supply. Instrumental for the development of the theory is the use of parameter estimation-based observers, which combined with the dynamic regressor extension and mixing parameter estimation technique, allow the reconstruction of  the magnetic flux. With the knowledge of the latter it is shown that the mechanical coordinates can be estimated with suitably tailored nonlinear observers. Replacing the observed states, in a certainty equivalent manner, with a full information globally stabilising law completes the sensorless controller design. We consider one and two-degrees-of-freedom systems that, interestingly, demand totally different mathematical approaches for their solutions. Simulation results are used to illustrate the performance of the proposed schemes.
\end{abstract}

\end{frontmatter}

%
\section{Introduction}
\label{sec1}
%
The use of magnetically levitated (MagLev) technology eliminates mechanical contact between moving and stationary parts in the system, attenuating the cumbersome friction problem. An additional benefit is the possibility of actively changing the position of the levitated object and to change the stiffness of the levitation system. Therefore, it finds many application areas such as magnetic bearings \cite{2}, vibration isolation \cite{3}, bearingless motors \cite{4}, bearingless pumps \cite{RAGetal}, microelectromechanical systems \cite{5}, and high speed rail transportation \cite{6}. In addition, MagLev can also control a floating object which is performing linear or rotary motion \cite{7}. See \cite{HANKIM,SCHMAS} for recent overview of MagLev systems.

Since MagLev systems are inherently unstable, position control of the levitated object is of paramount importance. Clearly, the knowledge of the position is necessary to accomplish this task, making MagLev systems highly expensive because of the cost (and low reliability) of existing position sensors. To overcome this limitation a lot of research has been devoted to the development of sensorless (also called self-sensing) MagLev systems. In these schemes the position sensor is replaced by some kind of estimation algorithm that reconstructs the position from the measurement of voltages and currents. These estimation algorithms  may be classified in two groups: (i) technologically-based techniques that exploit the functional relationship between the systems inductance and the position of the levitated object; (ii) theoretically-based designs of state observers proceeding from the mathematical model of the system. The interested reader is referred to \cite{GLUetal,MASetalifac,MIZetal,VAN} for a review of the existing literature on sensorless control of MagLev systems reported in the control community and to \cite{RANetal,SCHMAS} for results found in the application journals.  

The present contribution belongs to the second category mentioned above. Namely,  proceeding from the full nonlinear mathematical model derived from physical laws, we design a state observer for the flux, position and velocity of the MagLev system measuring only voltages and currents. We consider one and two-degrees-of-freedom  ($1$ and $2$-dof) systems that, interestingly, demand totally different mathematical approaches for their solutions. As is well-known, the dynamic behavior of these systems is highly nonlinear. Therefore, to ensure good performance in a wide operating range it is necessary to avoid the use of linearized models that, to the best of the authors' knowledge, is the prevailing approach reported in the literature \cite{GLUetal,MIZetal}. See \cite{MASetal,MON} for a detailed analysis of the deleterious implications of linearization in sensorless Maglev models. The first step in our design is the reconstruction of the flux, which is done by combining the parameter estimation-based observers (PEBO) recently reported in \cite{ORTetalscl} with the dynamic regressor extension and mixing (DREM) parameter estimation technique of \cite{ARAetaltac}. The combination of these two new techniques has been proven highly successful in the solution of several complex practical problems \cite{ARAetalalcosp,BOBetalaut,PYRetal}---see also \cite{ORTetalaut} for the reformulation of DREM as a functional Luenberger observer.  With the knowledge of the flux we propose suitably tailored nonlinear observers for the mechanical coordinates, obtaining in this way a globally convergent solution to the posed observation problem. To complete the sensorless controller design the observed state is then replaced in the globally asymptotically stabilizing interconnection and damping assignment passivity-based controller (IDA-PBC) reported in \cite{RODetal}, see also \cite{RODetalacc}. 

Since there are several full-state controllers that achieve the stabilization objective, see {\em e.g.}, \cite{MASetalifac,ORTbook,TORORT}, our main contribution is the solution of the---until now open---problem of state observation that, as shown below, turns out to be significantly involved. In \cite{YIetal} injection of high-frequency sinusoidal probing signals in the voltage is used to generate a virtual output \cite{COMetal} and be able to design a PEBO for a $1$-dof MagLev system. The invasive injection of probing signals is avoided in the present contribution. On the other hand, as always for observer based controller designs for nonlinear systems, some excitation condition needs to be imposed on the signals of the system \cite{ARAetaltac,ENG}. It should be pointed out that the proposed observer can be combined with other controllers, for instance, the well-known ``complementarity control" \cite{LEVetal,BONetal} in which the two magnetic forces are never simultaneously activated yielding a more efficient  energy consumption.

The remaining of the paper is organized as follows.  The model of a $2$-dof MagLev system is presented in Section \ref{sec2}---from which the more widely known $1$-dof system is obtained as a particular case. Interestingly, but not surprisingly, state observation of the former system is significantly simpler than the latter one. Therefore, we present first  in Section \ref{sec3} the design for the $2$-dof MagLev system. The design for the $1$-dof case is given in Section \ref{sec4}. The performance of the proposed observer and sensorless controller is validated in Section \ref{sec5} via simulations. The paper is wrapped--up with some conclusions and future work in Section \ref{sec6}. The design of the PEBO for the 1-dof system, being notationally involved, is deferred to an appendix.
%
\section{Model of the MagLev Systems and Problem Formulation}
\label{sec2}
\begin{figure}[htp]
\centering
	\subcaptionbox{\label{fig1} $2$-dof}{\includegraphics[width=0.26\textwidth]{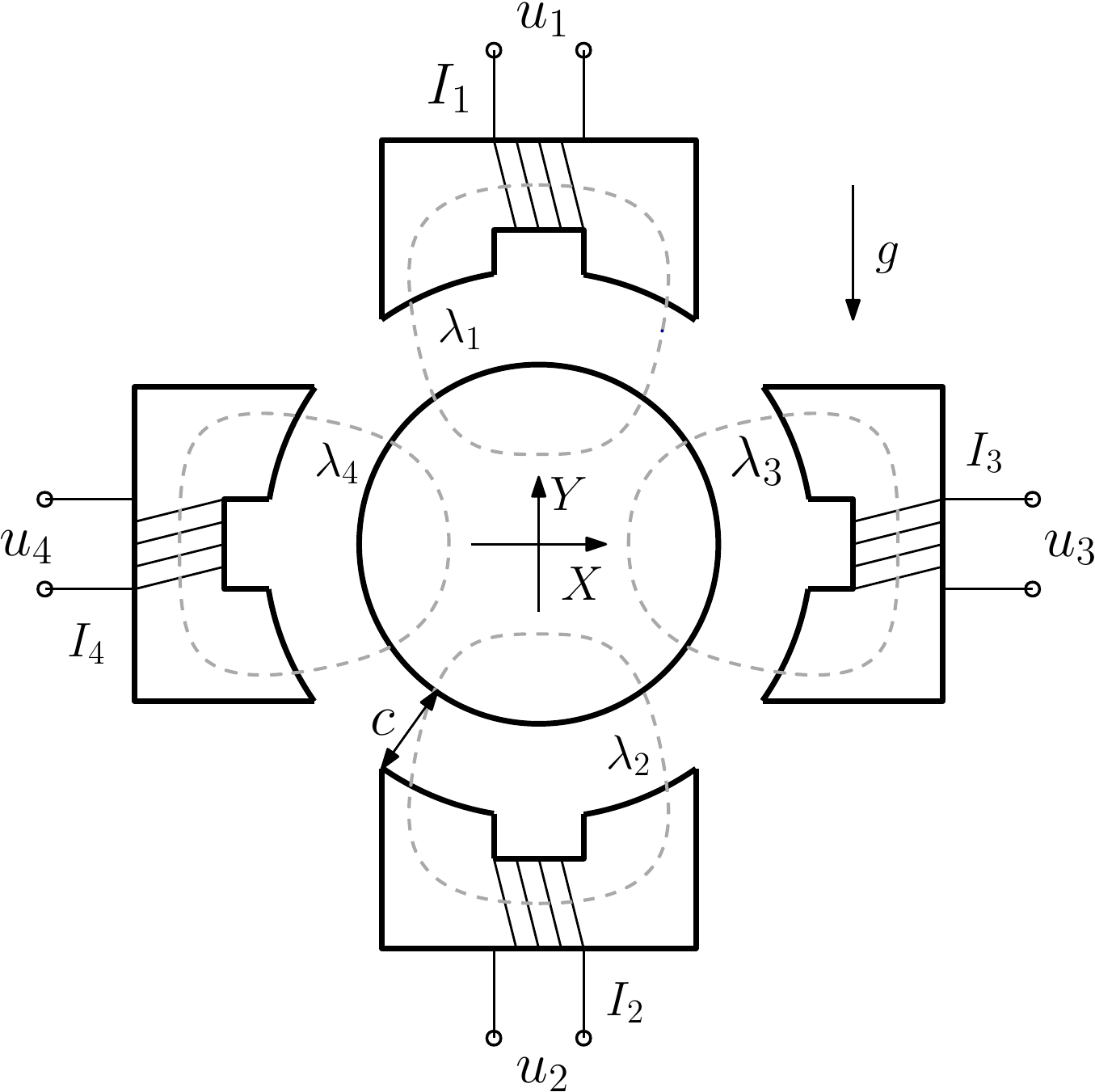}}
	\subcaptionbox{\label{fig2} $1$-dof}{\includegraphics[width=0.21\textwidth]{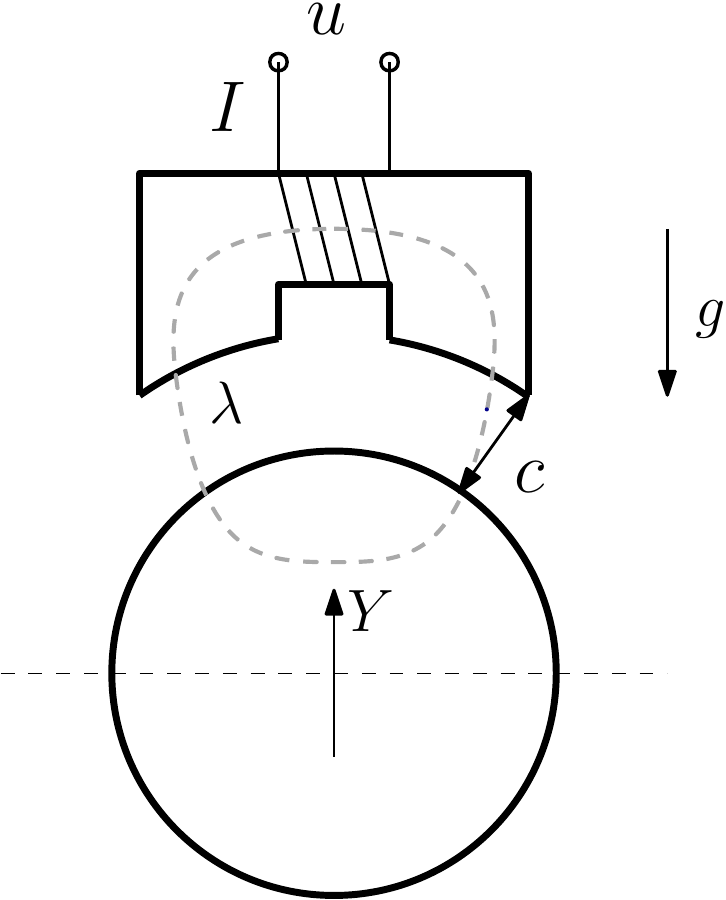}}
	\caption{\label{fig0} MagLev systems.}
\end{figure}

%
The model of the $2$-dof MagLev system depicted in Fig. \ref{fig1} is obtained from Faraday's and Newton's laws as
\begin{align}
\label{lamdot}
\dot\lambda_i &=-RI_i+u_i,\quad i=1,...,4\\
\label{Y}
m\ddot Y&=f_1-f_2-mg,\\
\label{X}
m\ddot X&=f_3-f_4,
\end{align}
where $X$, $Y$ are the rotor positions in the horizontal and vertical directions, respectively, $R$ are the coils resistances, $m$ is the mass of the rotor, $g$ is the acceleration of gravity, and $\lambda_i$, $I_i$, $f_i$, $u_i$, $i=1, ..., 4$ denote the total magnetic flux, the current in the coil, the force and the control voltage associated with the $i$-th actuator, respectively. 
The following assumptions on the magnetic device are made:
\begite
\item[(A1)] The magnetic forces of the vertical and horizontal motions are decoupled (see, e.g., \cite{DIXetal}).
\item[(A2)] The total flux, rotor position and coil current are related as
\begin{align}
\nonumber
I_j&={1\over k}(c+(-1)^j Y)\lambda_j,\quad j=1, 2\\
\label{Ij}
I_j&={1\over k}(c+(-1)^j X)\lambda_j,\quad j=3, 4
\end{align}
for some positive constants $c$ and $k$.
\item[(A3)] The forces produced by the actuators satisfy
\begin{align}
\label{f}
f_i={1\over 2k} \lambda_i^2,\quad i=1,...,4.
\end{align}
\endite

From the equations above it is clear that, due to Assumption (A1), the dynamics of the horizontal and vertical motions are decoupled, with independent control signals. This fact will be reflected in the observer and sensorless controller design that, as will become clear below, can be carried out independently.  

The model of the 1-dof system depicted in Fig. \ref{fig2} is obtained as a particular case of the one above and is summarized in the following equations. 
\begin{align}
\nonumber
\dot\lambda &=-R i +u,\\
\nonumber
m\ddot Y&=f-mg,\\
\nonumber
f&={1\over 2k} \lambda^2,\\
\label{I1dof}
i&={1\over k}(c-Y)\lambda.
\end{align}
It is assumed that all the systems parameters are known and that the only signals available for measurement are the currents $I_i, i=1,\dots,4$, that we arrange in a vector $I:=\col(I_1,\dots,I_4)$. The control objective is to design a dynamic output feedback controller 
\begin{align*}
\dot \zeta & = F(\zeta,I)\\
U & = H(\zeta,I),
\end{align*}
where $\zeta \in \rea^{n_\zeta}$ is the controller state and $U:=\col(u_1,\dots,u_4)$, such that 
\begequ
\label{tracla}
\sup \liminf |Y(t) - Y_*(t)| \leq \varepsilon, \sup \liminf |X(t) - X_*(t)| \leq \varepsilon
\endequ
where $|\cdot|$ is the Euclidean norm, $Y_*$ and $X_*$ define a desired trajectory for the levitated ball and $\varepsilon$ is some small constant, which equals zero when the position reference is constant.  A similar control objective is posed for the $1$-dof system.
\vspace{-2mm}
\begin{rem}
\label{rem1}
Full-state feedback stabilizing controllers for Maglev systems, applying various nonlinear control techniques, are available in the literature, {\em cf.} \cite{MASetalifac,ORTbook,TORORT}. Therefore, the main task to be solved is the observation of the systems state from the measurement of the currents $I$. To the best of the authors' knowledge this nonlinear observation problem is totally open. It should be mentioned that in Chapter 15 of \cite{SCHMAS}, it is assumed that the position of 1-dof levitated object is constant,\footnote{An ad-hoc technological modification is proposed in \cite{MON} to (partially) overcome this restriction.} and is treated as a parameter that can be estimated with classical adaptation techniques.
\end{rem}
\vspace{-2mm}
\begin{rem}
\label{rem2}

Equations \eqref{lamdot}-\eqref{f} is the standard model for radial magnetic bearings. Assumptions (A1) and (A3) are practically reasonable in most applications. However, Assumption (A2), which neglects saturation effects in the coils, may be restrictive. Notice, however, that magnetic coupling between the two orthogonal subsystems, which is not necessarily more negligible than magnetic saturation is not considered. Also, in some $1$-dof systems leakage inductance cannot be neglected as done here.
\end{rem}
\vspace{-2mm}
\begin{rem}
\label{rem3}
Implicit to Assumption (A2) is the fact that the state space where the system lives is restricted to
$$
|Y| < c,\quad |X| < c,
$$
constraint that is also imposed to $Y_*$ and $X_*$. However, as is often the case in control theoretical developments, these constraints are not taken into account in the analysis. 
\end{rem}
\vspace{-2mm}
\begin{rem}
\label{rem4}
It is clear that the MagLev benchmark example considered in \cite{RODetalacc} is a particular case of the 2-dof considered above. Therefore, the observer technique developed to solve the latter will be directly applicable to the former example.
\end{rem}
%
\section{Sensorless Control of the $2$-dof Maglev System}
\label{sec3}
%
To enhance the readability of the paper we split the presentation of the controller in five parts. 
\begite
\item[(i)] Application of PEBO to translate the problem of observation of the flux into a parameter estimation problem and derivation of the required regression model for their estimation. 
\item[(ii)] Application of the DREM parameter estimator to the aforementioned regressor. 
\item[(iii)] Derivation of a nonlinear observer for the speed.
\item[(iv)] Derivation of a nonlinear observer for the position.
\item[(v)] Presentation of the certainty equivalent sensorless IDA-PBC. 
\endite
As indicated in the previous section, since the dynamics of the horizontal and vertical motions are decoupled, the observer and controller designs for each one of them can be carried out independently. However, in the interest of brevity we present it all together.

\subsection{Regression model for the PEBO of the flux}
\label{subsec31}
%
\vspace{-2mm}
\begin{prop}
\label{pro1}
Consider the model of the $2$-dof Maglev system \eqref{lamdot}--\eqref{f}. Define the dynamic extension
\begequ
\label{dotpsi}
\dot \psi  =  -RI+U,
\endequ 
and the measurable signals
\begin{align}
\nonumber
z_1 & := -I_1\psi_2-I_2\psi_1+{2c\over k}\psi_1\psi_2,\\
\nonumber
z_2& := -I_3\psi_4-I_4\psi_3+{2c\over k}\psi_3\psi_4,\\
\label{zxi}
\xi & := I -{2c\over k}\psi.
\end{align}
The following (nonlinearly parameterised) regression model holds
\begequ
\label{regmod}
z = \Phi_0(\theta) \xi - \Phi_1(\theta) {2c\over k},
\endequ
where 
\begequ
\label{phi}
 \Phi_0(\theta):=  \begmat{ \theta_2 &  \theta_1 & 0 & 0  \\ 0 & 0 & \theta_4 & \theta_3},\quad  \Phi_1(\theta):=\lef[{c} \theta_1\theta_2\\ \theta_3\theta_4 \rig],
\endequ
and $\theta:=\col(\theta_1,\dots,\theta_4)$ is a constant vector that satisfies 
\begequ
\label{lampsi}
\lambda =\psi + \theta +\epst
\endequ
with $\epst$ an exponentially decaying signal stemming for the initial conditions of \eqref{dotpsi}.\footnote{Without loss of generality, all additive exponentially decaying terms are neglected in the sequel---see Remark 3 in \cite{ARAetaltac} where the effect of these terms is rigorously analysed.}
 
\end{prop}
\begin{pf}
Equation \eqref{lampsi} follows via integration of \eqref{lamdot} and \eqref{dotpsi} and defining
$$
\theta :=\lambda(0)-\psi(0).
$$
From \eqref{Ij}  it follows that
\begin{align}
I_1\lambda_2+I_2\lambda_1 & = {2c\over k}\lambda_1\lambda_2,\\
I_4\lambda_3+I_3\lambda_4 & = {2c\over k}\lambda_3\lambda_4.
\end{align}
The proof is completed replacing \eqref{lampsi} in the equations above and grouping terms. 
\end{pf}
\vspace{-2mm}
It is clear from \eqref{lampsi} that the flux observer design is completed generating an estimate for the parameters $\theta$, called $\hat \theta$, and defining the flux estimate as
\begin{align}
\label{fluest}
\hat\lambda &=\psi+\hat\theta,
\end{align}
In the next subsection we generate $\hat\theta$ using the DREM procedure. Clearly, if the parameter estimator is consistent, that is $\liminf \hat \theta(t)=\theta$,  the flux estimate \eqref{fluest} will satisfy
\begequ
\label{estflucon}
\liminf |\hat \lambda(t)-\lambda(t)|=0.
\endequ
\vspace{-2mm}
\begin{rem}
\label{rem5}
Besides the additional difficulty of needing to estimate $\theta$, the main drawback of PEBO is that it relies on the open-loop integration \eqref{dotpsi}, which might be a problematic operation in practice.\footnote{For a discussion on this matter see \cite{MASIWA} where the open-loop integration \eqref{dotpsi} is proposed---but without the essential parameter estimation step.} In spite of that, PEBO  has proven instrumental in the solution of numerous physical systems problems \cite{BOBetalaut,BOBetalijc,CHOetal}---some of them being unsolvable with other observer design techniques. 
\end{rem}
\vspace{-2mm}
\subsection{Flux observer}
\label{subsec32}
%
In Proposition \ref{pro1} we derived the regression model \eqref{regmod} for the $2$-dof Maglev system \eqref{lamdot}--\eqref{f} that is, alas, nonlinearly parameterised. One of the motivations to use DREM to estimate the parameters from this regression is that it allows us to deal with these cases. A second motivation to use DREM is that it ensures that, element by element,  the parameter errors decrease monotonically---see Remark \ref{rem7}. The reader is referred to  \cite{ARAetaltac,ORTetalaut} for further details on DREM.
\begin{prop}
\label{pro2}
Consider the model of the $2$-dof Maglev system \eqref{lamdot}--\eqref{f} with the PEBO \eqref{dotpsi}, \eqref{zxi} and \eqref{fluest}. Fix four stable filters ${\kappa_j\over p+\nu_j}$, $j=1,\dots,4$, with $p:={d \over dt}$ and $\kappa_j,\nu_j$ some positive tuning gains. Define the filtered signals 
\begequ
\label{deffil}
(\cdot)^{f_j}:={\kappa_j\over p+\nu_j}\left(\cdot\right),\;j=1,\dots,4.
\endequ
Generate the  DREM parameter estimates as
\begin{align}
\label{decesti}
\dot{\hat{\theta}}_i & = \gamma_i\Delta_1 (\caly_{1,i} - \Delta_1 \hat\theta_i),\;i=1,2,\\
\dot{\hat{\theta}}_i & = \gamma_i\Delta_2 (\caly_{2,i} - \Delta_2 \hat\theta_i),\;i=3,4,
\end{align}
with adaptation gains $\gamma_i>0,\;i=1,\dots,4$, where we defined the signals
\begin{align}
\label{blozomei}
\calz_1 &:= \lef[{c} z_1 \\ z_1^{f_1} \\ z_1^{f_2} \rig]\!,\quad \calz_2  :=  \lef[{c} z_2 \\ z_2^{f_3} \\ z_2^{f_4} \rig]\!,\\
\label{omei}
\Omega_1 &:= \lef[{ccc} \xi_1 & \xi_2 & -{2c\over k}\\ \xi_1^{f_1} & \xi_2^{f_1} & -({2c\over k})^{f_1}\\ \xi_1^{f_2} & \xi_2^{f_2} & -({2c\over k})^{f_2} \rig]\!,\; \Omega_2 := \lef[{ccc} \xi_3 & \xi_4 & -{2c\over k}\\ \xi_3^{f_3} & \xi_4^{f_3} & -({2c\over k})^{f_3}\\ \xi_3^{f_4} & \xi_4^{f_4} & -({2c\over k})^{f_4} \rig]\\
\label{deli}
\caly_i & = \lef[{c} \caly_{1,i} \\ \caly_{2,i} \\ \caly_{3,i}\rig] := \adj\{\Omega_i\} \calz_i,\; \Delta_i  :=  \det \{\Omega_i\},\;i=1,2,
\end{align}
where $\adj\{\Omega_i\}$  is the adjunct of $\Omega_i$. The following implication is true
$$
\Delta_i(t) \notin \call_2,\;i=1,2\; \Rightarrow \; \liminf |\hat \lambda(t)-\lambda(t)|=0,
$$
with $\call_2$ the space of square integrable functions. 
\end{prop}
\vspace{-2mm}
\begin{pf}
Consider the first element  of the regressor model \eqref{regmod}, that is,
\begin{align}
\label{regmod1}
z_1=\theta_1\xi_1+\theta_2\xi_2-\theta_1\theta_2{2c\over k}.
\end{align}
Operating with the filter ${\kappa_j\over p+\nu_j},j=1,2$, on \eqref{regmod1} we obtain two additional regression models that we pile on a vector as
\begequ
\label{calz}
\calz_1=\Omega_1  \lef[{c}   \theta_1 \\ \theta_2 \\ \theta_1\theta_2 \rig] 
\endequ
where $\calz_1$ and $\Omega_1$ are defined in \eqref{blozomei} and \eqref{omei}, respectively.  Premultiplying  \eqref{calz} by the {adjunct} of $\Omega_1$  and using the fact that
 $$
 \adj\{\Omega_1\} \Omega_1=\Delta_1 \cali_3 
 $$
where $\cali_3$ is the $3 \times 3$ identity matrix and $\Delta_1$ is defined in \eqref{deli}, we get the $3$ {scalar} regressors 
\begin{align}
\nonumber
\caly_{1,i}&=\theta_i\Delta_1,\;i=1,2,\\
\label{scareg}
\caly_{1,3}&=\theta_1\theta_2\Delta_1,
\end{align}
where $\caly_1$ is defined in \eqref{deli}. 

The estimation of the parameters $\theta_1,\theta_2$, can be easily carried out using the first two scalar regressions in \eqref{scareg} via the gradient descent \eqref{decesti}. Replacing \eqref{scareg} in \eqref{decesti}, and defining the parameter errors $\tilde \theta_i:=\hat \theta_i - \theta_i$, we get the error equations
\begequ
\label{errequ}
\dot{\tilde{\theta}}_i = -\gamma_i \Delta_1^2 \tilde\theta_i,\;i=1,2.
\endequ
Solving the simple scalar differential equation \eqref{errequ} we conclude that
$$
 \lim_{t\to \infty} \tilde \theta_i(t)=0,\;i=1,2 \quad \Longleftrightarrow \quad \Delta_1(t) \notin \call_2.
$$

Proceeding as done above with the second element  of the regression model \eqref{regmod} 
$$
z_2 =\theta_3\,\xi_3+\theta_4\,\xi_4-{2c\over k}\theta_3\theta_4
$$
we obtain the DREM parameter estimator for the parameters $\theta_3$ and $\theta_4$ given in the proposition. The proof is completed invoking \eqref{fluest}.
\end{pf}

\begin{rem}
\label{rem7}
As always in observer designs for systems with inputs, some kind of excitation on the signals must be imposed to guarantee convergence. In our case it is the condition of non-square integrability of the determinants $\Delta_1$ and $\Delta_2$ of the extended regressor matrices $\Omega_1$ and $\Omega_2$, respectively. A thorough discussion on the implications of this condition may be found in \cite{ARAetaltac,ORTetalaut} 
\end{rem}

\begin{rem}
\label{rem8}
An important advantage of DREM is that the {individual} parameter errors satisfy
$$
|\tilde \theta_i(0)| \geq |\tilde \theta_i(t)|,\;\forall\; t \geq 0,\;i=1,\dots,4,
$$
which is significantly stronger than the well-known property---of the norm of the parameter errors---of standard gradient and least squares methods, that is,
$$
|\tilde \theta(0)| \geq |\tilde \theta(t)|,\;\forall\; t \geq 0.
$$
This property was used in \cite{GERetal} to tackle a classical open problem in model reference adaptive control.
\end{rem}

\subsection{Speed observer }
\label{subsec33}
%
In Proposition \ref{pro2} it was shown that it is possible to reconstruct the flux---up to an additive exponentially decaying term. As indicated in Remark \ref{rem5} the presence of these terms does not affect our analysis, hence in the sequel we assume that $\lambda$ is known. Notice also, from \eqref{lamdot}, that  $\dot\lambda$ is also known.

\begin{prop}
\label{pro3}
Consider the model of the $2$-dof Maglev system \eqref{lamdot}--\eqref{f} with known $\lambda$ and $\dot \lambda$. Define the speeds observers
\begin{align}
\nonumber
\dot\chi_1 & = -\gamma_Y\left[(\lambda_1^2+\lambda_2^2)\hat v_Y - 2 k(I_1\dot\lambda_1-I_2\dot\lambda_2)\right] \nonumber \\
	& \quad + {1\over 2k m}\left(\lambda_1^2-\lambda_2^2-2kmg\right), \nonumber \\
\dot\chi_2 & = -\gamma_X \left[(\lambda_3^2+\lambda_4^2)\hat v_X-2k(I_3\dot\lambda_3-I_4\dot\lambda_4)\right] \nonumber \\
	& \quad + {1\over 2km}\left(\lambda_3^2-\lambda_4^2\right) \nonumber \\
\hat v_Y & = \chi_1-\gamma_Y k(I_1\lambda_1-I_2\lambda_2), \nonumber \\
\label{speobs}
\hat v_X & = \chi_2-\gamma_X k(I_3\lambda_3-I_4\lambda_4),
\end{align}
where $\gamma_Y,\gamma_X>0$ are tuning gains. The following equivalences are true
\begin{align}
\nonumber
\col(\lambda_1,\lambda_2)\not\in \mathcal{L}_2 & \Leftrightarrow \lim_{t\rightarrow\infty}|\hat v_Y(t)-\dot Y(t)|=0\\
\col(\lambda_3,\lambda_4)\not\in \mathcal{L}_2  & \Leftrightarrow  \lim_{t\rightarrow\infty}|\hat v_X(t)-\dot X(t)|=0.
\label{equspe}
\end{align}
\end{prop}

\begin{pf}
We will present first the proof for the observer of $\dot Y$---the one for $\dot X$ will follow {\em verbatim}. Differentiating the $j=1$ equation in \eqref{Ij} and multiplying by $\lambda_1$ we get
$$
- k\dot I_1\lambda_1+kI_1\dot\lambda_1=\dot Y\lambda_1^2,
$$
Differentiating now the $j=2$ equation in \eqref{Ij} and multiplying by $\lambda_2$ we get
$$
k\dot I_2\lambda_2-kI_2\dot\lambda_2=\dot Y\lambda_2^2.
$$
Adding these two equations we get
\begin{align}
\label{ydotm}
\dot Y(\lambda_1^2+\lambda_2^2)=kI_1\dot\lambda_1-k\dot I_1\lambda_1+k\dot I_2\lambda_2-kI_2\dot\lambda_2.
\end{align}

Defining the observation error 
\begequ
\label{tilvy}
\tilde v_Y:=\dot Y-\hat v_Y,
\endequ
using the system dynamics equation \eqref{Y} and the $\dot Y$ observer equations in \eqref{speobs}  and \eqref{ydotm} we get, after some lengthy but straightforward calculations, the error dynamics
$$
\dot {\tilde v}_Y=-\gamma_Y(\lambda_1^2+\lambda_2^2)\tilde v_Y.
$$
Proceeding, {\em verbatim}, for the $\dot X$ observer we get
$$
\dot {\tilde v}_X=-\gamma_X(\lambda_3^2+\lambda_4^2) \tilde v_X,
$$
where we defined $\tilde v_X:=\dot X - \hat v_X$. The proof is completed integrating the two latter scalar equations.
\end{pf}

\begin{rem}
\label{rem9}
To avoid cumbersome notation we have presented Proposition \ref{pro3} using the actual values of $\lambda$ and $\dot \lambda$. Obviously, in the sensorless controller implementation these signals are replaced by $\hat \lambda$ and $-RI+U$, respectively. 
\end{rem}
\subsection{Position observer}
\label{subsec34}
%
To complete the state observation task we present in this subsection the observers for the positions $Y$ and $X$ of the levitated object.
\vspace{-2mm}
\begin{prop}\em
\label{pro4}
Consider the model of the $2$-dof Maglev system \eqref{lamdot}--\eqref{f} with known $\lambda$. Define the positions observers
\begin{align}
\nonumber
\dot{\hat Y}  =- \mu_Y \Big[ (\lambda_1^2+\lambda_2^2)\hat Y & + (kI_2-c\lambda_2)\lambda_2 \Big. \nonumber \\
	& \quad \Big. - (kI_1-c\lambda_1)\lambda_1 \Big] + \hat v_Y \\
\label{obsx}
\dot{\hat X} =- \mu_X \Big[  (\lambda_3^2+\lambda_4^2)\hat X & + (kI_2-c\lambda_4)\lambda_4 \Big. \nonumber \\
	& \quad \Big. - (kI_1-c\lambda_3)\lambda_3 \Big] + \hat v_X,
\end{align}
where $\mu_Y,\mu_X>0$ are tuning gains and $\hat v_Y,\hat v_X$ are generated as in Proposition \ref{pro3}.

The following implications are true
\begin{align}
\col(\lambda_1,\lambda_2)\not\in\mathcal{L}_2 & \Rightarrow \lim_{t\rightarrow\infty}|\hat Y(t)-Y(t)|=0\\
\col(\lambda_3,\lambda_4)\not\in\mathcal{L}_2  & \Rightarrow  \lim_{t\rightarrow\infty}|\hat X(t)-X(t)|=0.
\end{align}
\end{prop}
\vspace{-2mm}
\begin{pf}
We will present first the proof for the observer of $Y$---the one for $X$ follows {\em verbatim}. From the first two equations of \eqref{Ij} we get
\begin{align}
(\lambda_1^2+\lambda_2^2) Y =(kI_2-c\lambda_2)\lambda_2-(kI_1-c\lambda_1)\lambda_1.
\end{align}
Replacing the latter in the first equation of \eqref{obsx} and defining  the observation error $e_Y:=Y-\hat Y$ we get the first error equation
$$
\dot e_Y=-\mu_Y(\lambda_1^2+\lambda_2^2)e_Y + (\dot Y -\hat v_Y),
$$
Proceeding in the same way with the $X$ dynamics we get
$$
\dot e_X=-\mu_X(\lambda_3^2+\lambda_4^2)e_X + (\dot X -\hat v_X),
$$
where  $e_X:=X-\hat X$. The proof is completed invoking the equivalence \eqref{equspe} and integrating the scalar differential equations of the error dynamics. 
\end{pf}
\subsection{Sensorless controller}
\label{subsec35}
%
In this subsection we implement the sensorless controller replacing the estimated fluxes, positions and velocities generated via the observers of Propositions \ref{pro1}-\ref{pro4} in the full-state feedback IDA-PBC given in \cite{RODetal}, see also \cite{RODetalacc}. Exploiting the fact that the horizontal and vertical motions dynamics are decoupled, the corresponding controllers are designed in \cite{RODetal} in an independent way. However, as indicated in the introduction, it is possible to use any other---possibly coupled---stabilizing controller, for instance the practically attractive ``complementarity control" of \cite{LEVetal}.

As indicated in \cite{RODetal}, for a given constant desired position $(Y_*,X_*)$, the assignable equilibrium points of \eqref{lamdot}--\eqref{f} can be parameterised in terms of the total flux induced by one of the actuators. Taking, without loss of generality, $\lambda_2$ and $\lambda_4$ yields the following parameterisation of the assignable equilibrium set   
\begin{align}
\nonumber
\cale:= & \{ (\lambda,Y,\dot Y,X,\dot X) \in \rea^8\;|\; \lambda_1=\sqrt{2kmg+\lambda^2_{2*}},
\\
& \lambda_{3}=\lambda_{4*},\dot Y=0, \dot X=0\}.
\label{cale}
\end{align}
Notice that, because of the absence of gravity forces in the horizontal dynamics, the third and fourth fluxes should be equal to fix the equilibrium. 

The sensorless controller is then given by 
\begin{align}
u_1 & = RI_1 - {R\over 2k\alpha}(\hat \lambda_1^2-\lambda^2_{1*}) \nonumber \\
	& \quad - \left({R\over\alpha}+\alpha R_a\right)\Gamma\left[{1\over \alpha}\tilde\lambda_1+\tilde Y+R_am\hat v_Y\right] - \alpha\hat v_Y,
\nonumber \\
u_2 & = RI_2
	+ {R\over 2k\beta}(\hat \lambda_2^2-\lambda^2_{2*}) 
	\nonumber \\
	& \quad - \beta R_a\Gamma\left[{1\over \alpha}\tilde\lambda_1+\tilde Y+R_am\hat v_Y\right] - \beta\hat v_Y,
\nonumber \\
u_3 & = RI_3 - {R\over 2k\alpha}(\hat \lambda_3^2-\lambda^2_{3*})
	- \left({R\over2\alpha}+\alpha R_a\right) \bar{D}
	-\alpha\hat v_X,
\nonumber \\
u_4 & = RI_4 + {R\over 2k\beta}(\hat \lambda^2_4-\lambda^2_{4*})
	- \left({R\over2\beta}+\beta R_a\right) \bar{D}
	- \beta\hat v_X,
\label{idapbc}
\end{align}
where $\alpha,\Gamma,R_a>0$ and $\beta<0$ are tuning parameters, $\tilde {(\cdot)}:=\hat {(\cdot)} - (\cdot)_*$ with the desired equilibrium point selected from the set $\cale$  and $\hat \lambda,\hat v_Y,\hat v_X,\hat Y, \hat X$ generated via the observers of Propositions \ref{pro1}-\ref{pro4},
\begin{align}
	\bar{D} = \Gamma\left[{1\over 2\alpha}\tilde\lambda_3+{1\over 2\beta}\tilde\lambda_4+\tilde X+R_am\hat v_X\right].
\end{align}

See  \cite{RODetal,RODetalacc} for further details on the IDA-PBC.

In the light of the convergence results of Propositions \ref{pro2}-\ref{pro4} it is expected that if 
\begequ
\label{exccon}
\lambda,\Delta_1,\Delta_2 \not\in \mathcal{L}_2
\endequ
then for all initial conditions of the overall system starting sufficiently close to equilibrium point we have that 
\begequ
\label{regcla}
\liminf  (\lambda(t),Y(t),\dot Y(t),X(t),\dot X(t))= (\lambda_*,Y_*,0,X_*,0).
\endequ

The following remarks are in order
\begite
\item[(R1)] Given the complexity of the overall system, establishing such a result is beyond the scope of this paper. 
\item[(R2)] As indicated before, the main contribution of the paper is the development of a state observer, whose global convergence under the excitation conditions \eqref{exccon} is rigorously established.
\item[(R3)]  The excitation conditions \eqref{exccon} will hardly be verified in regulation tasks---a fact that has been corroborated by the simulations presented in Section \ref{sec5}. However, good performance was achieved with simple step changes in the desired position of the levitated object. 
\item[(R4)]  The stability claim \eqref{regcla} pertains to regulation to a constant equilibrium. However, it is expected that the tracking objective \eqref{tracla} will also be attained---at least for sufficiently slow desired trajectories. This conjecture is substantiated by the robustness property inherited from the exponential convergence results proven in the propositions. 
\endite 
%
\section{Sensorless Control of the $1$-dof MagLev System}
\label{sec4}
%
Similarly to the 2-dof case, the sensorless control is derived in five steps, which are treated in separate sections.  

\subsection{Regression model for the PEBO of the flux}
\label{subsec41}
%
As will become clear below, in contrast with the 2-dof case, here the computations are pretty cumbersome. Therefore, the proof of the proposition is given in the Appendix. 
 
\begin{prop}
\label{pro5}
Consider the model of the $1$-dof Maglev system \eqref{I1dof}. Define the dynamic extension
\begequ
\label{dotpsi0}
\dot \psi  =  -Ri+u,
\endequ 
The following (nonlinearly parameterised) regression model holds
\begequ
\label{reg9}
z = \phi^\top \Phi(\eta),
\endequ
where $z$ and $\phi$ are measurable signals, 
\begequ
\label{regmodcom}
\Phi(\eta):=\col(\eta,\eta^2,\eta^3,\eta^4,\eta^5)
\endequ
and $\eta$ is a constant parameter that satisfies 
\begequ
\label{lampsi2}
\lambda =\psi + \eta +\epst
\endequ
with $\epst$ an exponentially decaying signal.
\end{prop}
\vspace{-2mm}
\begin{rem}
\label{rem10}
Similarly to the regression model  for the $2$-dof Maglev system \eqref{regmod}, the one for the 1-dof given in \eqref{reg9} is also nonlinearly parameterised. Although it is possible to obtain a linear regression introducing an overparameterisation, we avoid this low performance approach here. Instead,  we use DREM to estimate directly the parameter $\eta$ with just one gradient search.
\end{rem}
\vspace{-2mm}
\subsection{Parameter estimation via DREM}
\label{subsec42}
%
\vspace{-2mm}
\begin{prop}\em
\label{pro6}
Consider the model of the $1$-dof Maglev system \eqref{I1dof} with the regression model \eqref{reg9}. Fix four stable filters ${\kappa_j\over p+\nu_j}$, $j=1,\dots,4$, with $p:={d \over dt}$ and $\kappa_j,\nu_j$ some positive tuning gains. Define the filtered signals \eqref{deffil} and generate the  DREM parameter estimates as
\begequ
\label{decesti0}
\dot{\hat{\eta}}  =  \gamma\Delta (\caly - \Delta \hat\eta),
\endequ
with gain $\gamma>0$, where we introduced the signals
\begequarr
\label{calz2}
\calz &:= & \col( z, z^{f_1}, \dots,  z^{f_4}),\quad \Omega :=  \lef[{cccc} \phi &  \phi^{f_1} & \cdots & \phi^{f_4}\rig]^\top\!\!, \\
\label{caly}
\caly & = & e_1^\top\adj\{\Omega\} \calz,\; \Delta  :=  \det \{\Omega\},
\endequarr
where $e_1:=\col(1,0,0,0,0)$.  Generate the flux estimate as
\begequ
\label{fluest0}
\hat \lambda :=\psi + \hat \eta.
\endequ
The following implication is true
$$
\Delta(t) \notin \call_2 \; \Rightarrow \; \liminf |\hat \lambda(t)-\lambda(t)|=0.
$$
\end{prop}
\vspace{-2mm}
\begin{pf}
The proof follows {\em verbatim} the one of Proposition \ref{pro1}. That is, applying the filters to the regressor model \eqref{reg9}, \eqref{regmodcom} and arranging terms we get $
\calz=\Omega  \Phi(\eta)$. Premultiplying this by the {adjunct} of $\Omega$  and retaining the first {scalar} regressor we get $\caly=\eta\Delta$. Replacing the latter in \eqref{decesti0} we get the error equation
$$
\dot{\tilde{\eta}} = -\gamma \Delta^2 \tilde\eta.
$$
The proof is completed solving this simple scalar differential equation and invoking \eqref{fluest0}.
\end{pf}
\vspace{-2mm}
\subsection{Speed observer }
\label{subsec43}
%
Similarly to the $2$-dof case, in the light of Proposition \ref{pro6} and the first equation in \eqref{I1dof}, we will assume in the sequel that $\lambda$ and $\dot\lambda$ are known.

\begin{prop}
\label{pro7}
Consider the model of the $1$-dof Maglev system \eqref{I1dof} with known $\lambda$ and $\dot \lambda$. Define the speed observer
\begin{align}
	\nonumber
	\dot\chi &= {1\over m} \left(
		{1\over 2k}\lambda^2 -mg
	\right) - \gamma_Y \lambda^2 \hat v_Y + 2 \gamma_Y k i \dot\lambda,\\
	\label{vY_hat}
	\hat v_Y & = \chi - \gamma_Y k i \lambda,
\end{align}
where $\gamma_Y>0$. The following equivalence is true
\begin{align}
\lambda\not\in \mathcal{L}_2 \; \Leftrightarrow \; \lim_{t\rightarrow\infty}|\hat v_Y(t)-\dot Y(t)|=0.
\label{equspe0}
\end{align}
\end{prop}

\begin{pf}
Differentiating the last equation in \eqref{I1dof} and multiplying by $\lambda$ we get
$$
	k {di \over dt} \lambda - k i \dot\lambda =  - \dot Y \lambda^2,
$$
Using this and  the speed observer \eqref{vY_hat} we get, after some simple manipulations, the error model
$$
	\dot {\tilde v}_Y = \gamma_Y \lambda^2 \tilde v_Y,
$$
where $\tilde v_Y$ is defined in \eqref{tilvy}. The proof is completed integrating this scalar equation.
\end{pf}
\vspace{-2mm}
\subsection{Position observer}
\label{subsec44}
%
The final step is to reconstruct the position $Y$.
\vspace{-2mm}
\begin{prop}
\label{pro8}
Consider the model of the $1$-dof Maglev system \eqref{I1dof} with known $\lambda$. Define the positions observer
\begequ
\label{obsy0}
\dot{\hat Y}= - \mu_Y \lambda^2 \hat Y + \mu_Y  (c\lambda - ki)\lambda + \hat v_Y,
\endequ
where $\mu_Y>0$ is a tuning gain and $\hat v_Y$ is generated as in Proposition \ref{pro7}. The following implication is true
$$
\lambda \not\in\mathcal{L}_2  \Rightarrow  \lim_{t\rightarrow\infty}|\hat Y(t)-Y(t)|=0.
$$
\end{prop}
\vspace{-3mm}
\begin{pf}
Multiplying by $\lambda$ the last equation in \eqref{I1dof} we get 
$$
 (c\lambda - ki)\lambda=\lambda^2 Y
$$
which replaced in \eqref{obsy0}  yields 
$$
\dot e_Y=-\mu_Y \lambda^2 e_Y + (\dot Y -\hat v_Y),
$$
where  $e_Y:=Y-\hat Y$. The proof is completed invoking the equivalence \eqref{equspe0} and the same arguments used in the Proof of Proposition \ref{pro4}. 
\end{pf}
\subsection{Sensorless controller}
\label{subsec45}
%
In this subsection we implement the sensorless controller replacing the estimated fluxes, positions and velocities generated via the observers of Propositions \ref{pro5}-\ref{pro8} in the following full-state feedback feedback-linearizing controller (FLC):
\begin{align}
\nonumber
u & = \sqrt{\frac{k}{2F}}mv_{FL} + R(c-Y)\sqrt{\frac{2F}{k}}, \\
\nonumber
v_{FL} & = Y_*^{(3)} - k_2((\frac{F}{m} - g) - \ddot{Y}_*)  \\
	& \quad - k_1(\dot{Y} - \dot{Y}_*) - k_0(Y - Y_*),
\label{feelin}
\end{align}
which is given in Chapter 8, Section 5.1 of \cite{ORTbook}---see also \cite{LINKNO,TORORT}. Replacing this control law in the $1$-dof Maglev system \eqref{lamdot}--\eqref{f} yields the linear dynamics
$$
{\tilde Y}^{(3)} + k_2 \ddot {\tilde Y}  + k_1 \dot {\tilde Y}   + k_0 {\tilde Y}=0,
$$
where the coefficients $k_i,\;i=0,1,2$, are chosen to ensure that the corresponding characteristic polynomial is stable.  

Similarly to the case of 2-dof system, in the light of the convergence results of Propositions \ref{pro6}-\ref{pro8} it is expected that if 
\begequ
\label{exccon0}
\lambda,\Delta \not\in\mathcal{L}_2,
\endequ
 then for all initial conditions of the overall system starting sufficiently close to equilibrium point we have that 
$$
\liminf  (\lambda(t),Y(t),\dot Y(t))= (\sqrt{2kmg},Y_*,0).
$$
The remarks (R1)-(R4) of Subsection \ref{subsec35} apply as well to the $1$-dof case.

\begin{figure*}[htp]
\centering
\vspace{-3mm}
	\subcaptionbox{\label{fig31} Transients for $X(t)$ and desired trajectory \eqref{steps}}{\includegraphics[width=0.31\textwidth]{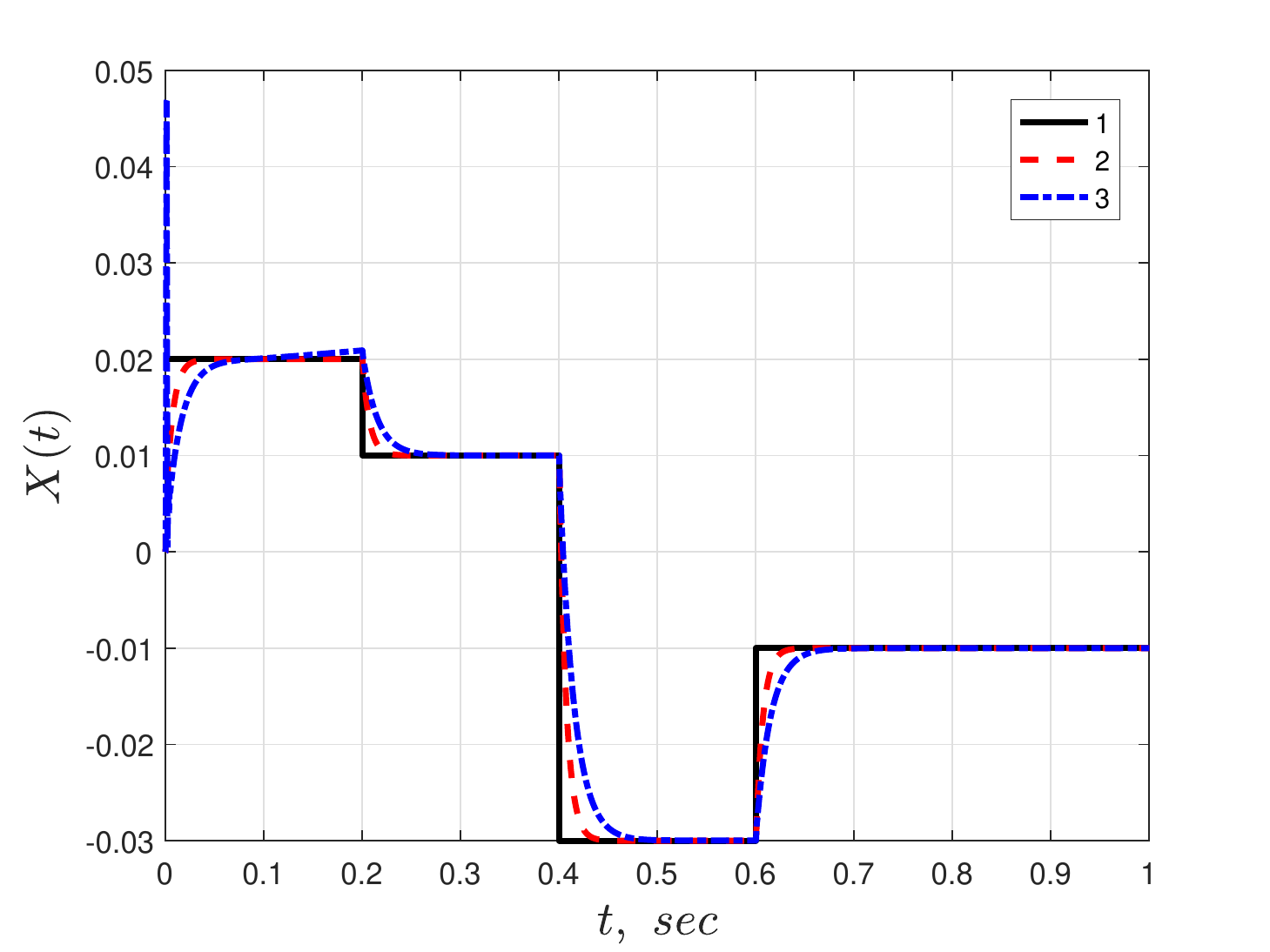}}
    \
	\subcaptionbox{\label{fig32} Transients for $Y(t)$ and desired trajectory \eqref{steps}}{\includegraphics[width=0.31\textwidth]{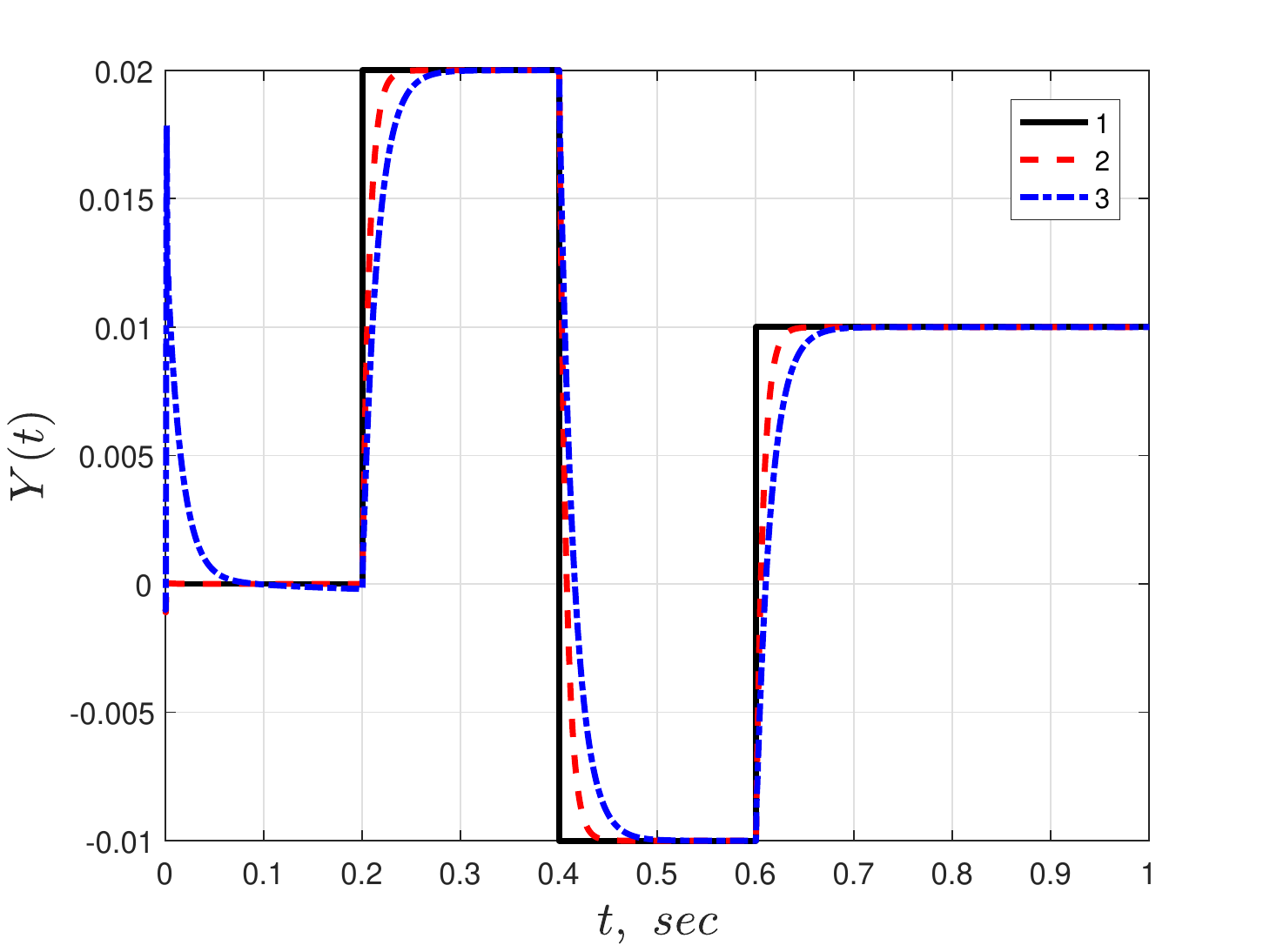}}
    \
	\subcaptionbox{\label{fig4} Transients for $X$ and $Y$ and desired trajectory \eqref{crc}}{\includegraphics[width=0.31\textwidth]{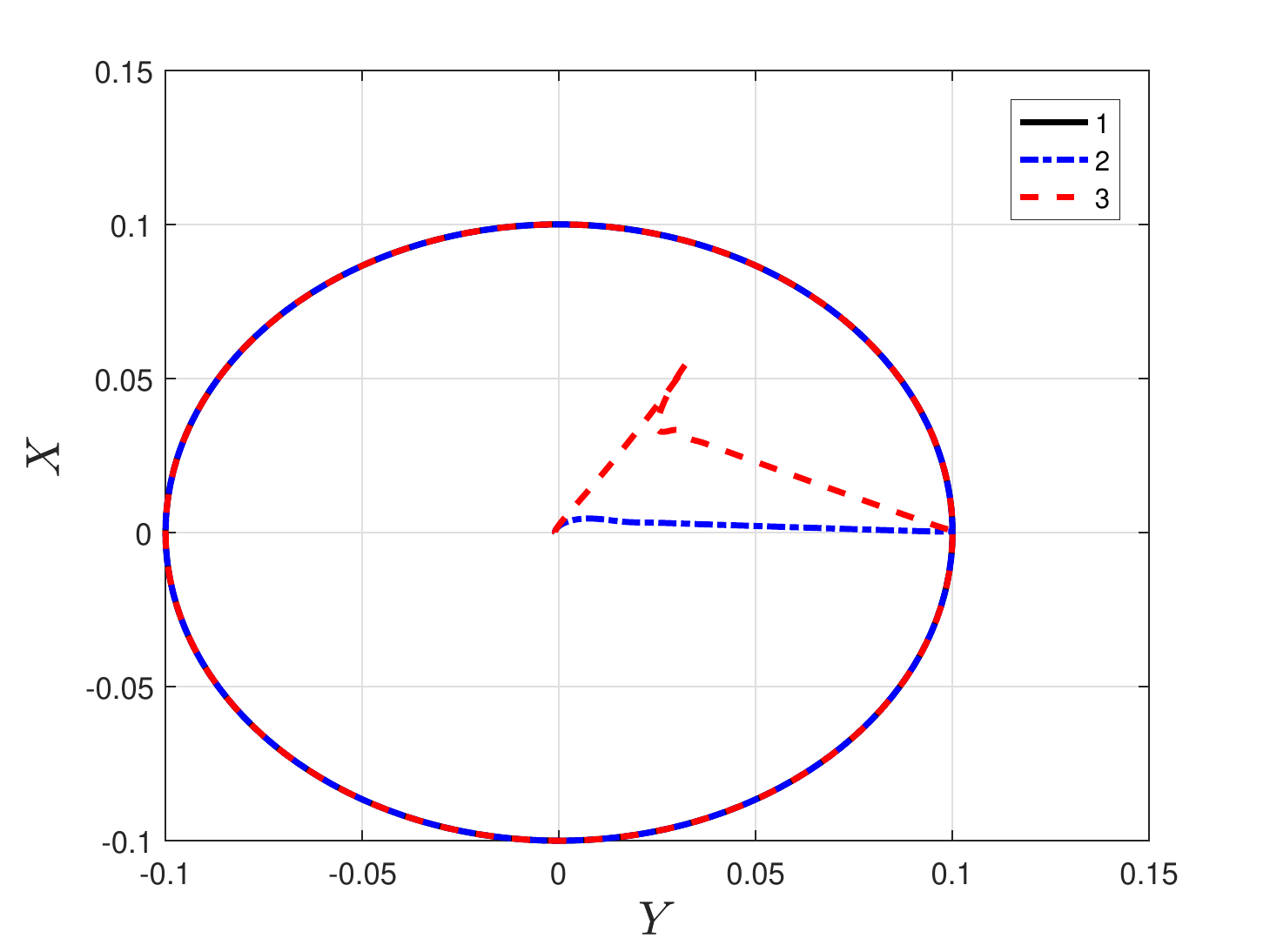}}
\vspace{-2mm}	
	\caption{\label{fig3} Behaviour of the closed loop system for the desired trajectory (black line 1) with the full-state feedback IDA-PBC (red line 2) and its sensorless version (blue line 3).}
\end{figure*}
\begin{figure*}[htp]
\centering
\vspace{-3mm}
	\subcaptionbox{\label{fig:1dof_Y_1} Filtered sum of sinusoids with  $\gamma=1$ and $\eta = 0.01$}{\includegraphics[width=0.3\textwidth]{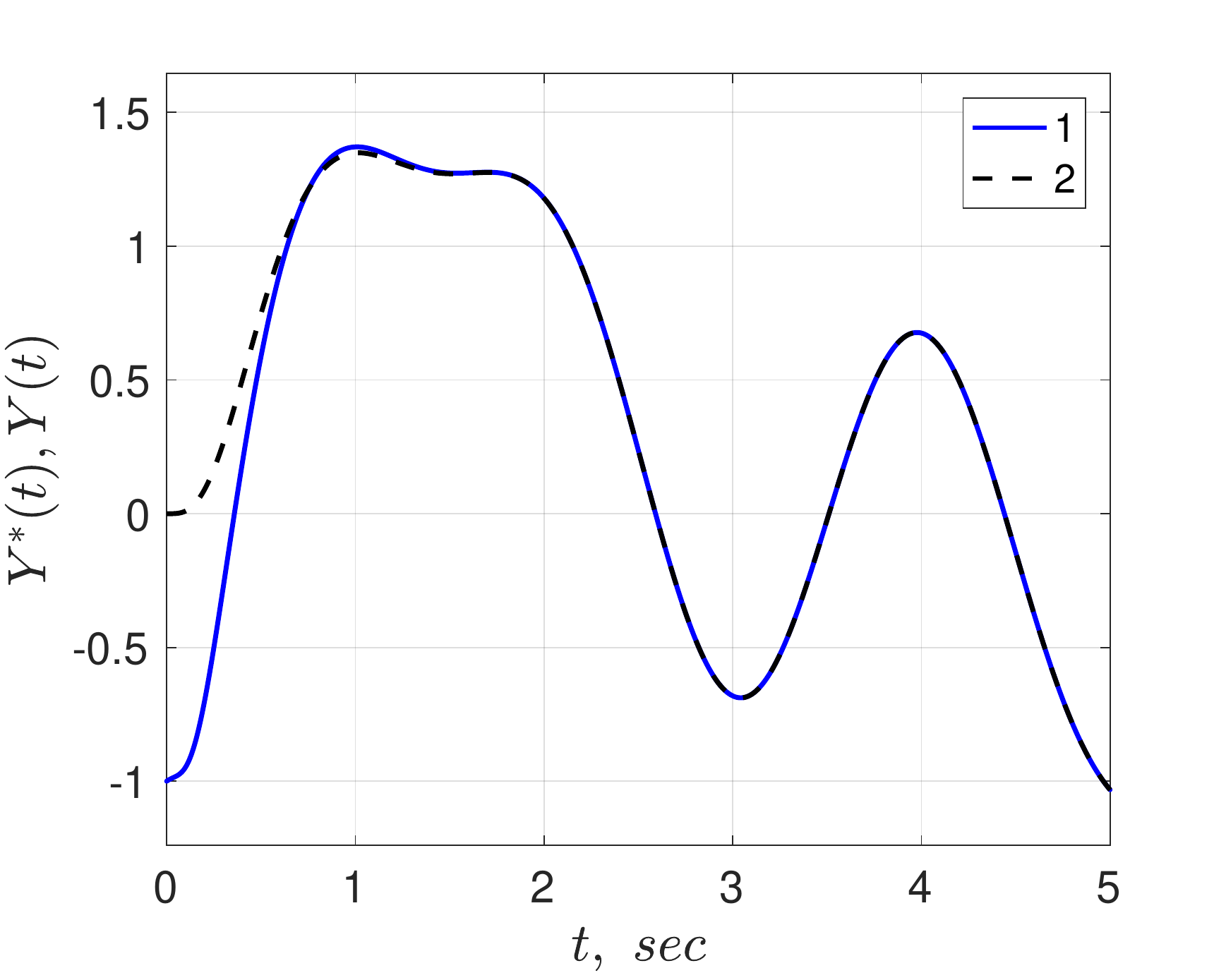}}
	\	
	\subcaptionbox{\label{fig:1dof_Y_2} Filtered steps with $\gamma=10^3$ and $\eta = 0.01$}{\includegraphics[width=0.3\textwidth]{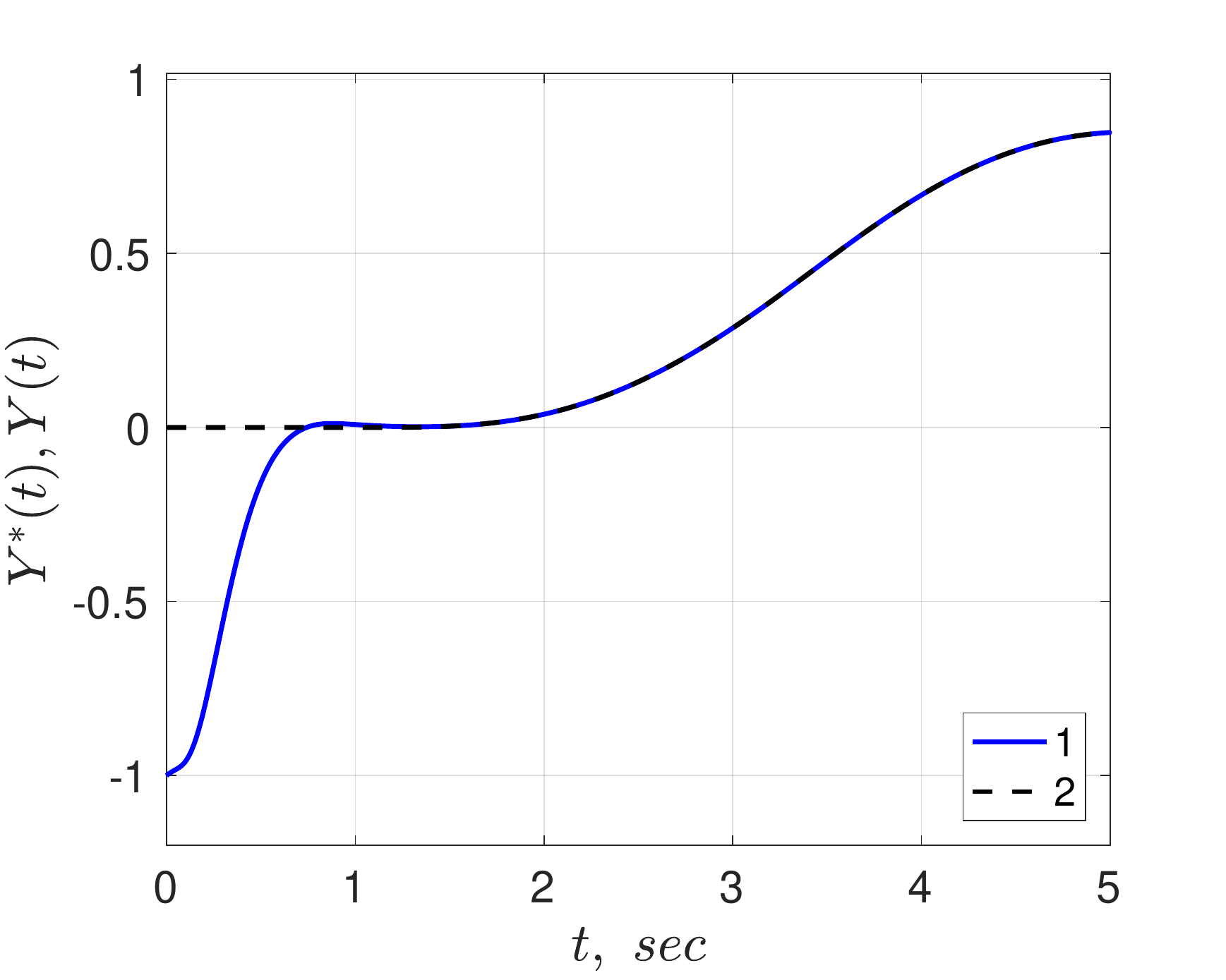}}
	\vspace{-2mm}
	\caption{\label{fig:1dof_Y} The reference signal $Y^*(t)$ (black dashed line 1) and transients for $Y(t)$ with the sensorless version of FLC (blue line 2)}
\end{figure*}

%
\section{Simulations}
\label{sec5}

In this section we present simulations of the proposed sensorless controllers for the $1$- and $2$-dof Maglev systems.
\subsection{ $2$-dof Maglev system}
\label{subsec51}
%
The $2$-dof Maglev system \eqref{lamdot}--\eqref{f} in closed-loop with the IDA-PBC \eqref{idapbc} was simulated with the plant parameters taken from \cite{RODetal}. Namely, $m = 0.0844$, $k = 6.4042e-5$, $R = 2.52$, $c = 0.005$. The controller parameters were fixed at $\alpha=10$, $\beta=-10$, $\Gamma=800$, $R_a=1$, which were tuned to reduce the overshot. For all experiments we set the desired fluxes taking $\lambda_{2*}=2$ and $\lambda_{4*}=1$.

In Figure \ref{fig3} we compare the behaviour of the full-state controller and the sensorless one for the following step changes in the desired position: 
\begin{align}
\nonumber
X_*(t)=\left\{\begin{array}{l}
0.02, \text{ for } 0\le t\le 0.2 \text{ sec, }\\
0.01, \text{ for } 0.2\le t\le 0.4 \text{ sec, }\\
-0.03, \text{ for } 0.4\le t\le 0.6 \text{ sec, }\\
-0.01, \text{ for } t\ge 0.6 \text{ sec, }
\end{array}
\right. \\
Y_*(t)=\left\{\begin{array}{l}
0, \text{ for } 0\le t\le 0.2 \text{ sec, }\\
0.02, \text{ for } 0.2\le t\le 0.4 \text{ sec, }\\
-0.01, \text{ for } 0.4\le t\le 0.6 \text{ sec, }\\
0.01, \text{ for } t\ge 0.6 \text{ sec, }
\end{array}
\right.
\label{steps}
\end{align}
with the following values of the controller parameters,  $\kappa_i=200$, $\nu_i=30$, $\gamma_i=500$, for $i=1,\dots,4$, and $\mu_Y=\mu_X=2000$. The initial conditions are given in Table \ref{tab_1}.

\begin{figure*}[htp]
\centering
	\subcaptionbox{\label{fig51} Transients for $\lambda_1(t)-\hat\lambda_1(t)$}{\includegraphics[width=0.37\textwidth]{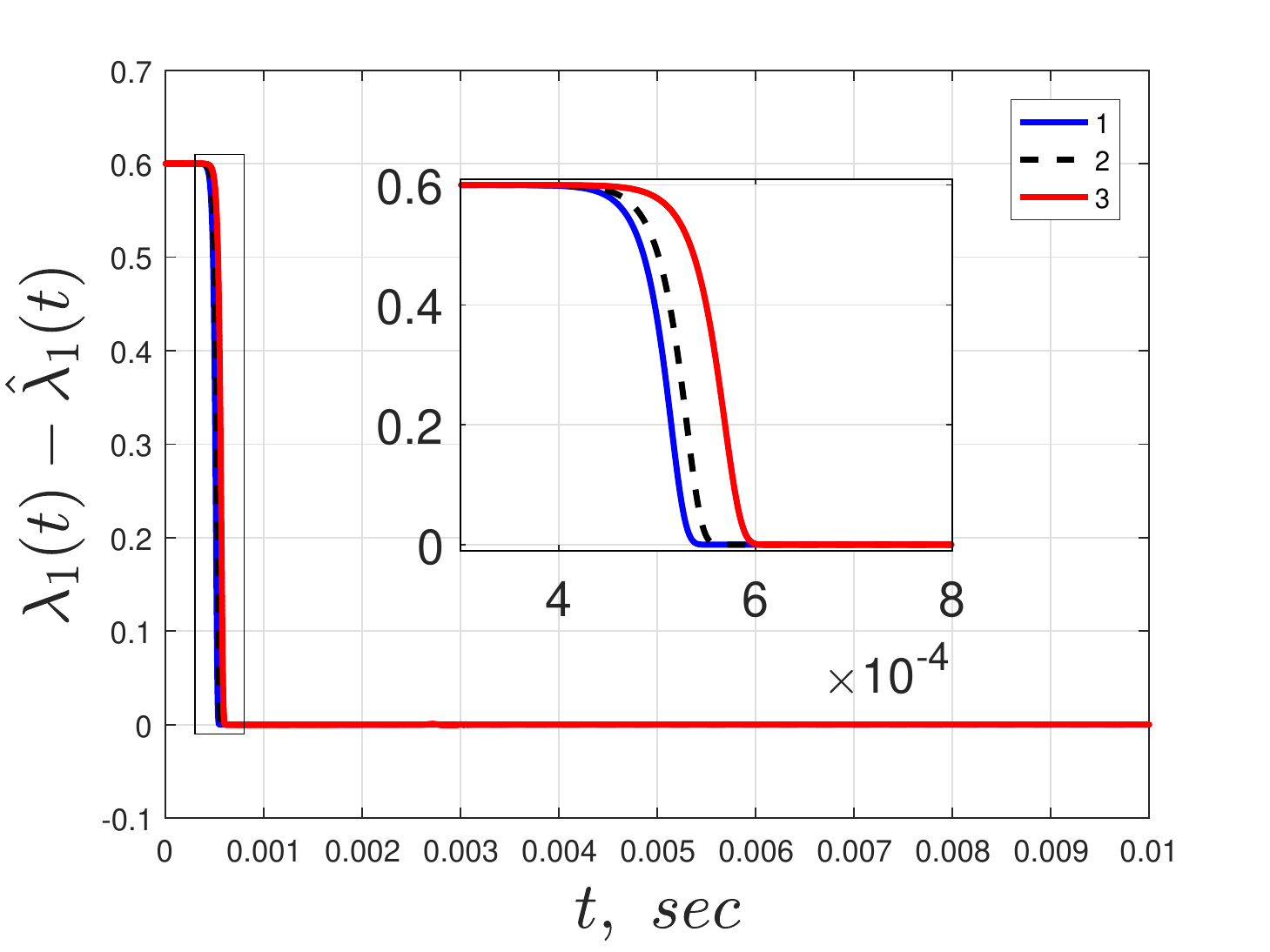}}
	\qquad
	\subcaptionbox{\label{fig52} Transients for $\lambda_2(t)-\hat\lambda_2(t)$}{\includegraphics[width=0.37\textwidth]{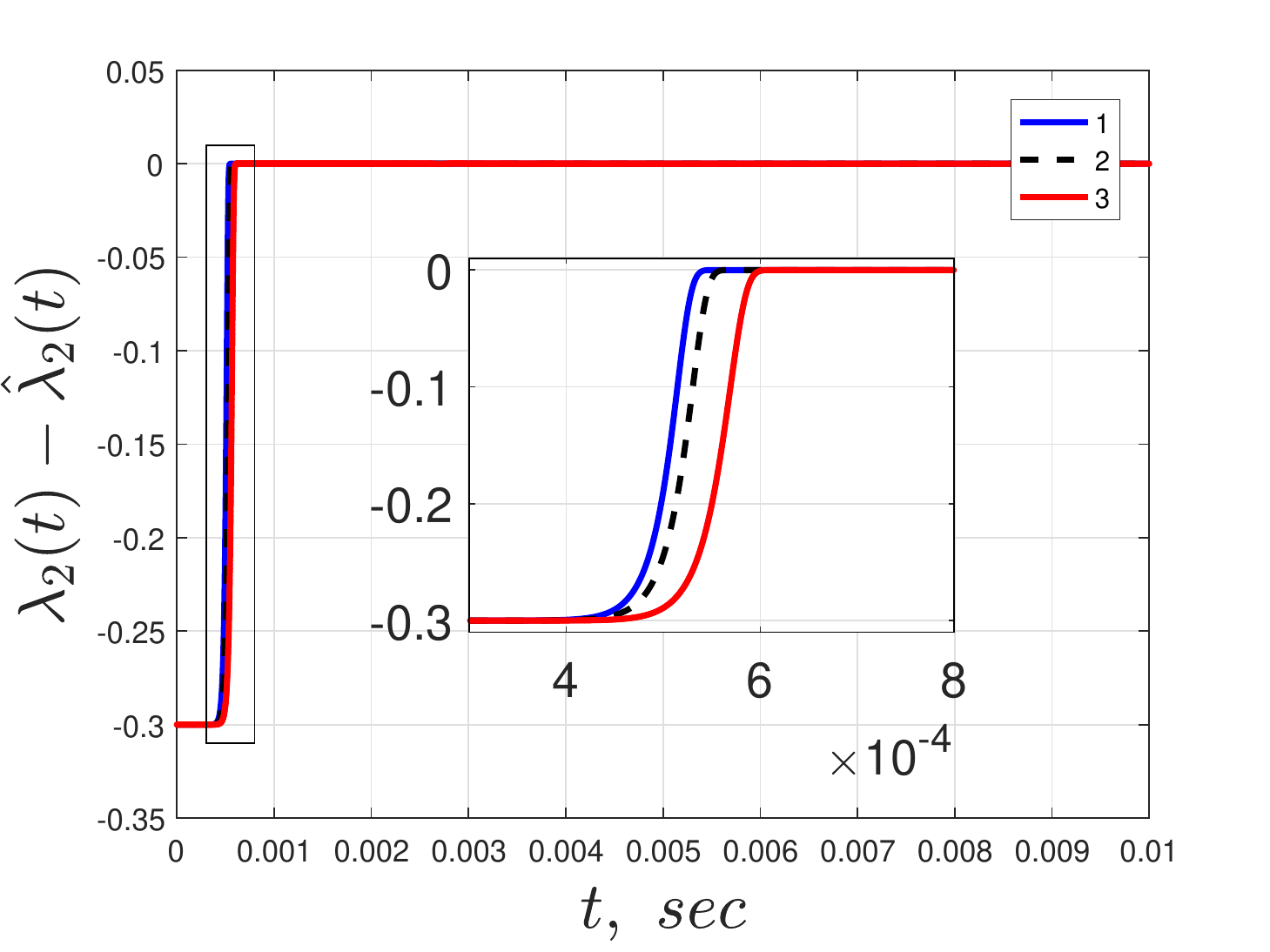}}
	\\
	\subcaptionbox{\label{fig53} Transients for $Y(t)-\hat Y(t)$}{\includegraphics[width=0.37\textwidth]{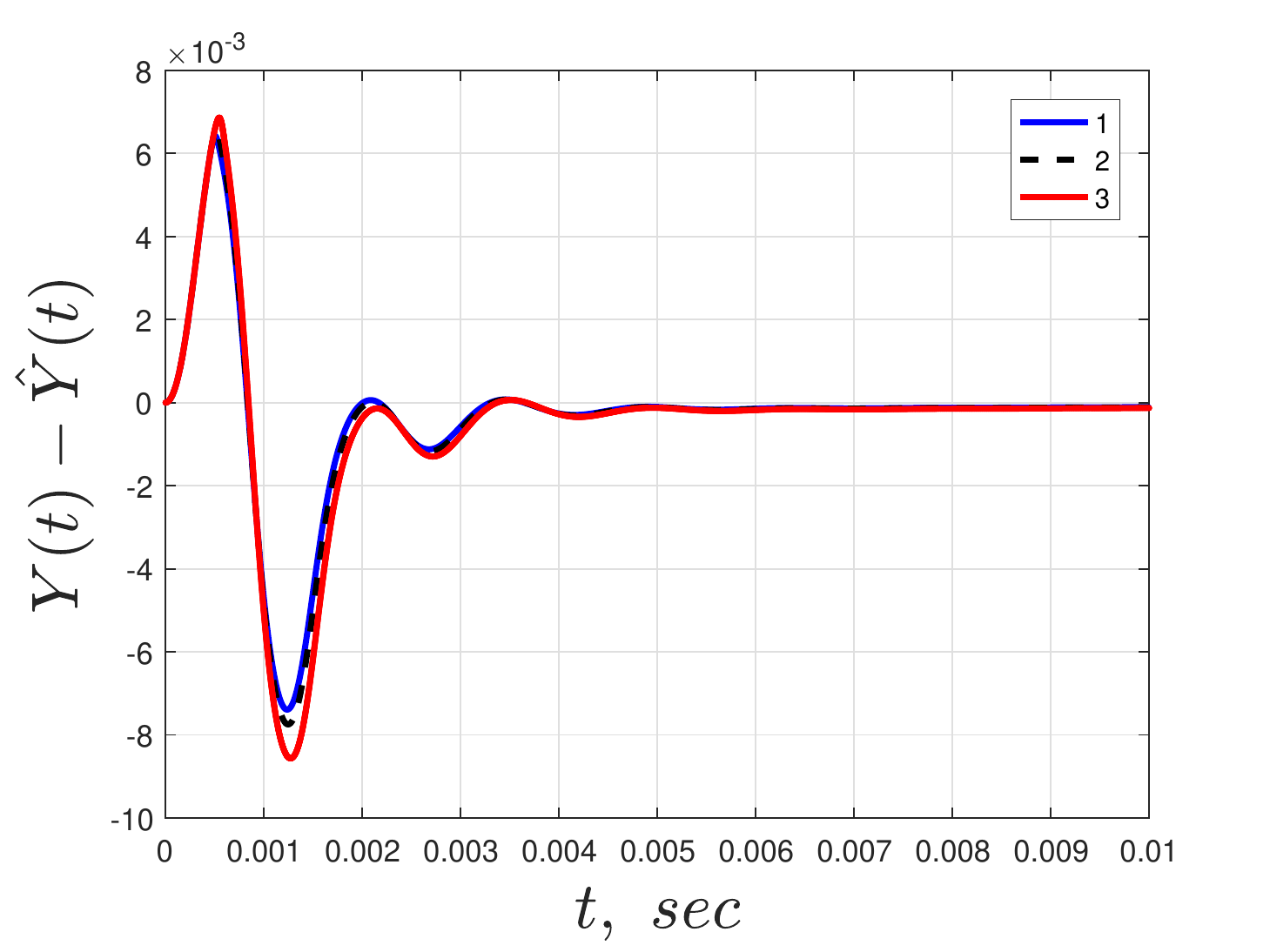}}
	\qquad
	\subcaptionbox{\label{fig54} Transients for $\dot Y(t)-\hat v_Y(t)$}{\includegraphics[width=0.37\textwidth]{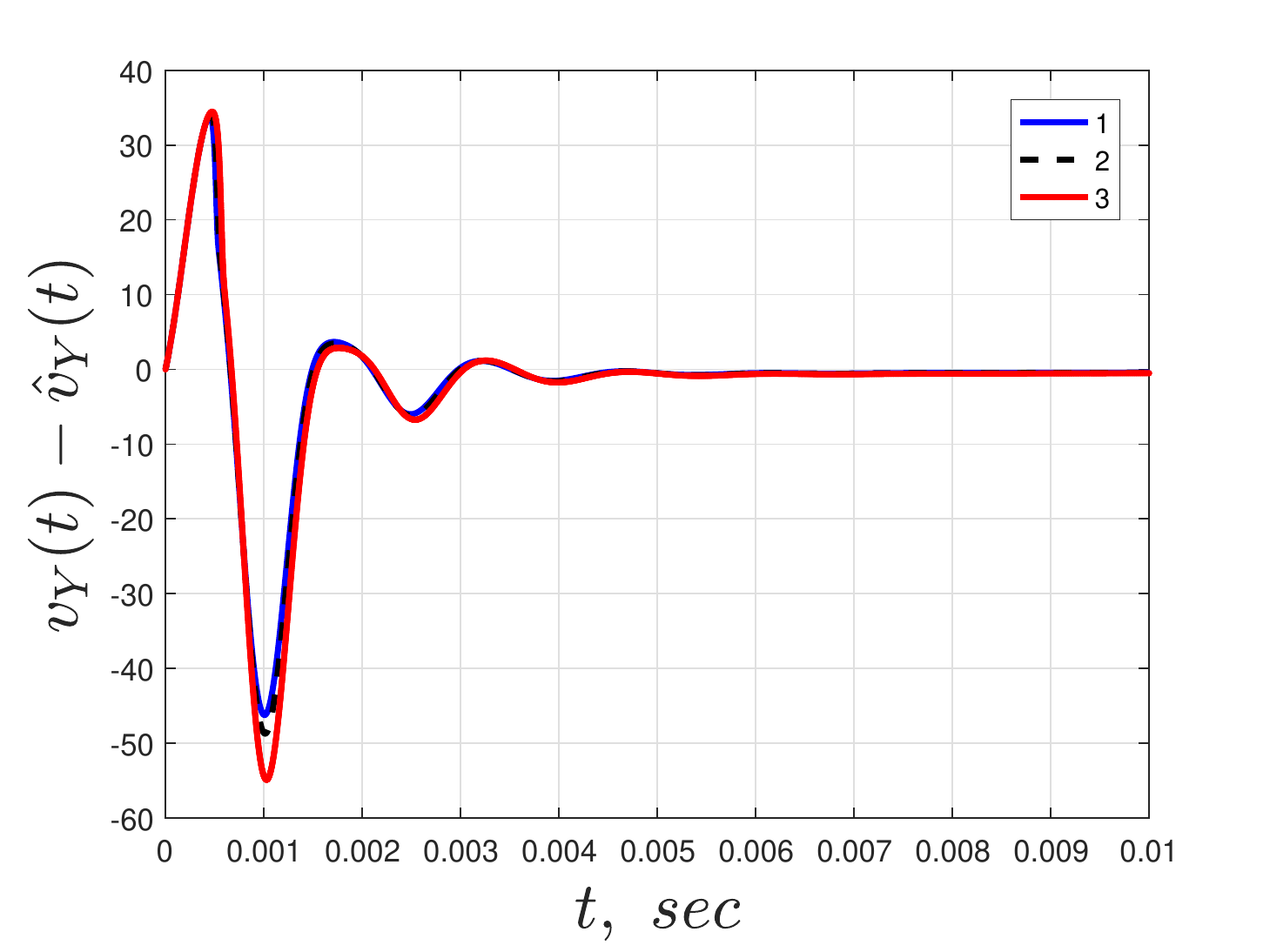}}
	\\
	\subcaptionbox{\label{fig55} Transients for $\lambda_3(t)-\hat\lambda_3(t)$}{\includegraphics[width=0.37\textwidth]{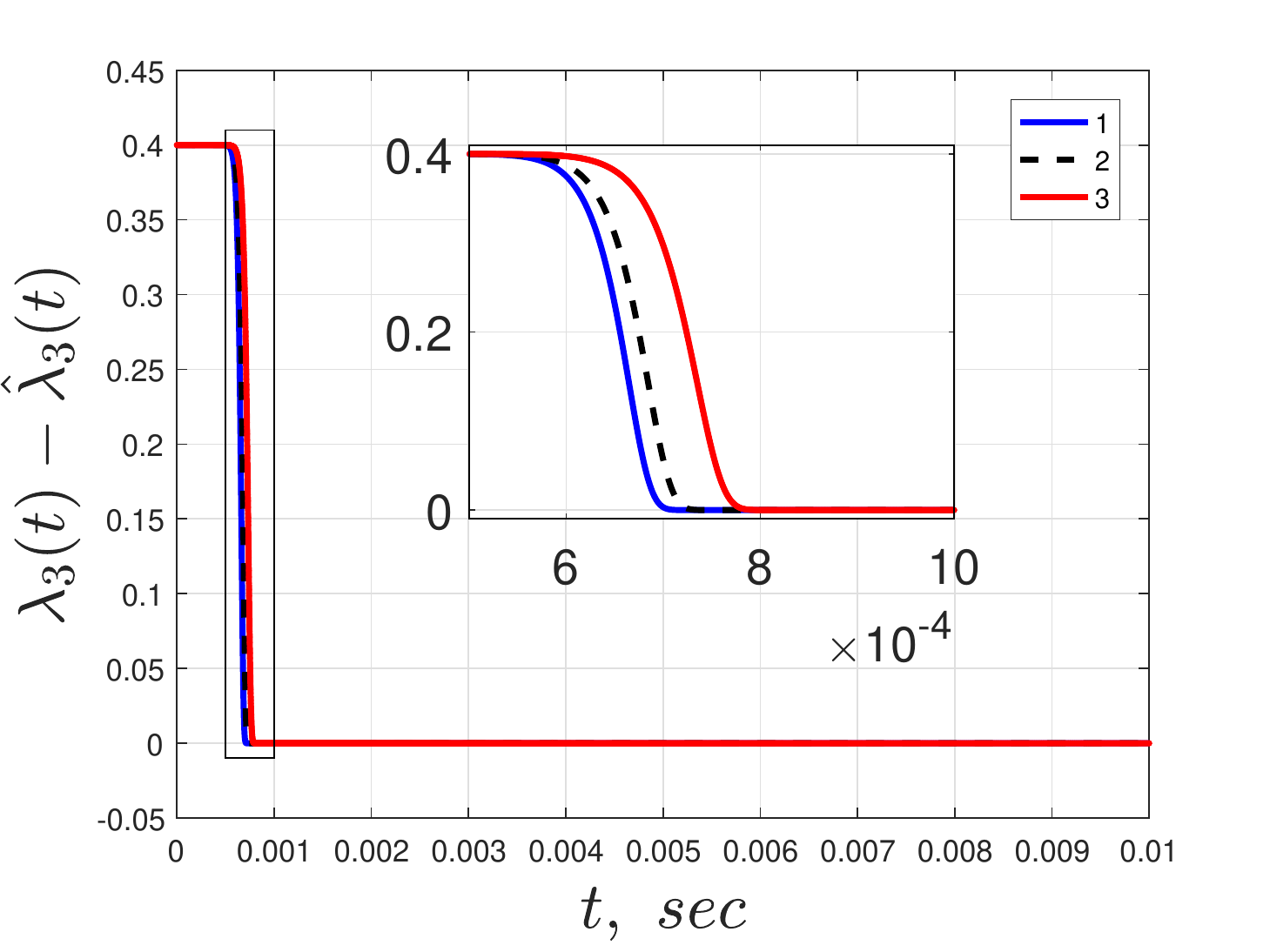}}
	\qquad
	\subcaptionbox{\label{fig56} Transients for $\lambda_4(t)-\hat\lambda_4(t)$}{\includegraphics[width=0.37\textwidth]{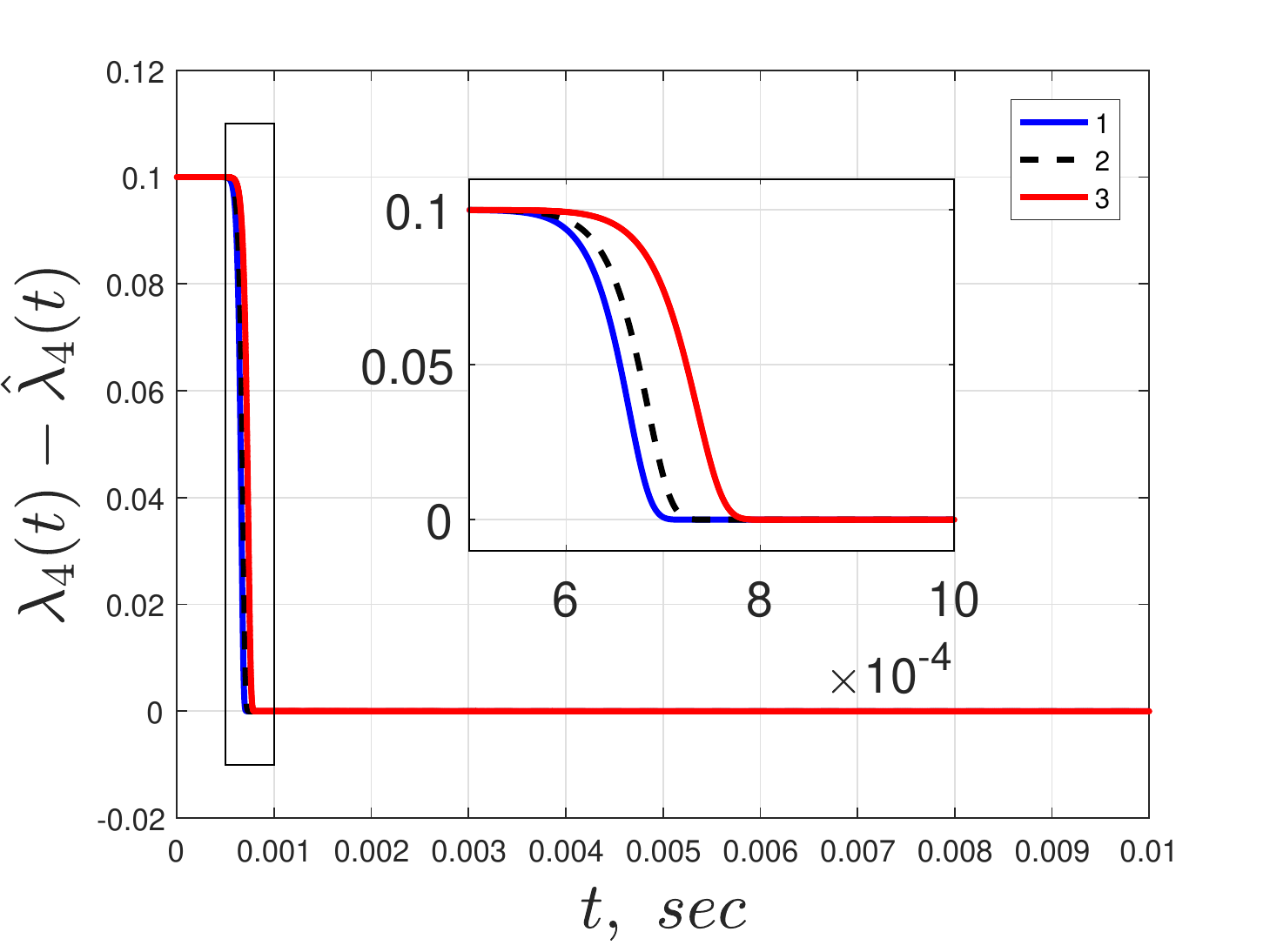}}
	\\
	\subcaptionbox{\label{fig57} Transients for $X(t)-\hat X(t)$}{\includegraphics[width=0.37\textwidth]{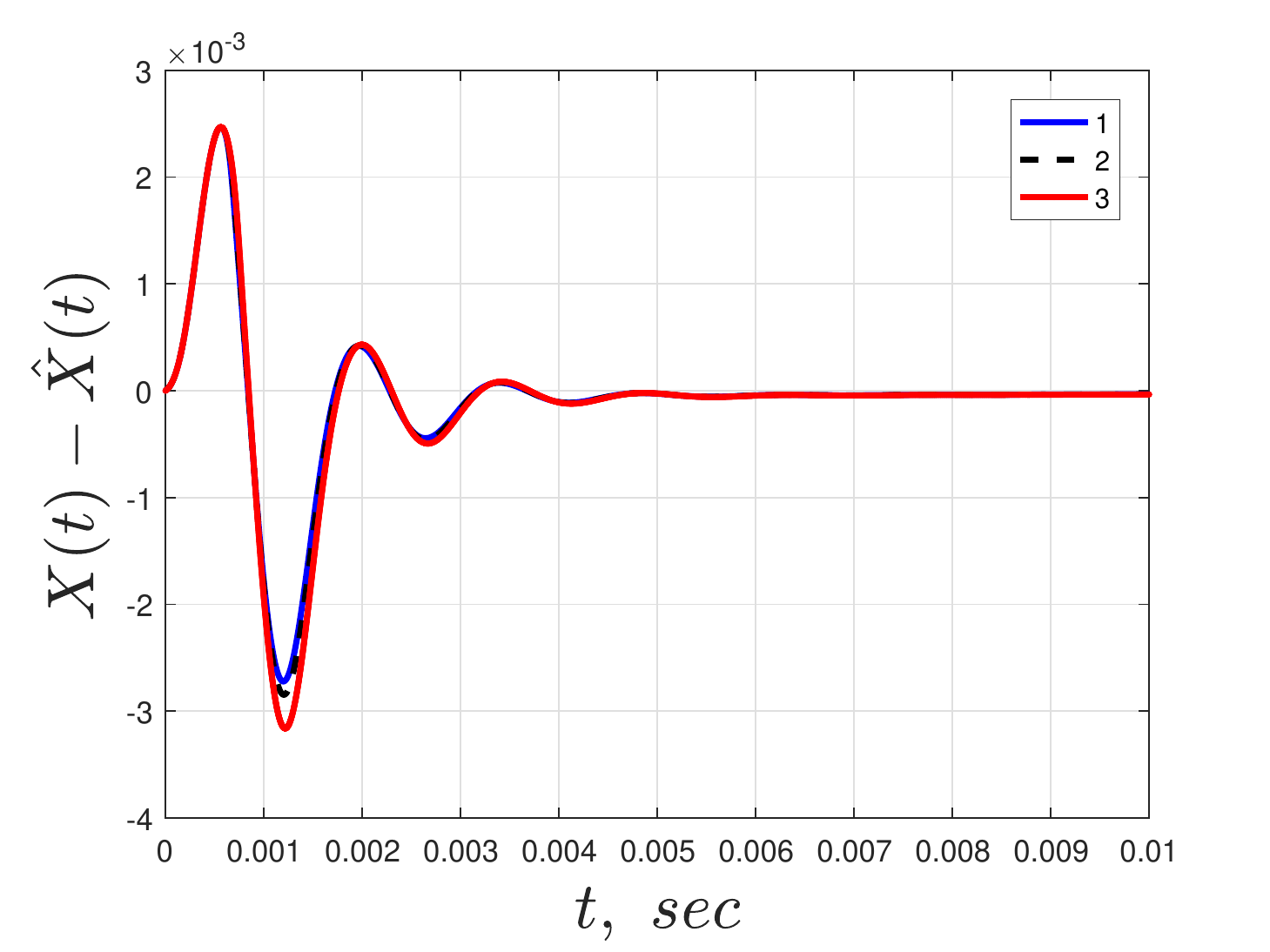}}
	\qquad
	\subcaptionbox{\label{fig58} Transients for $\dot X(t)-\hat v_X(t)$}{\includegraphics[width=0.37\textwidth]{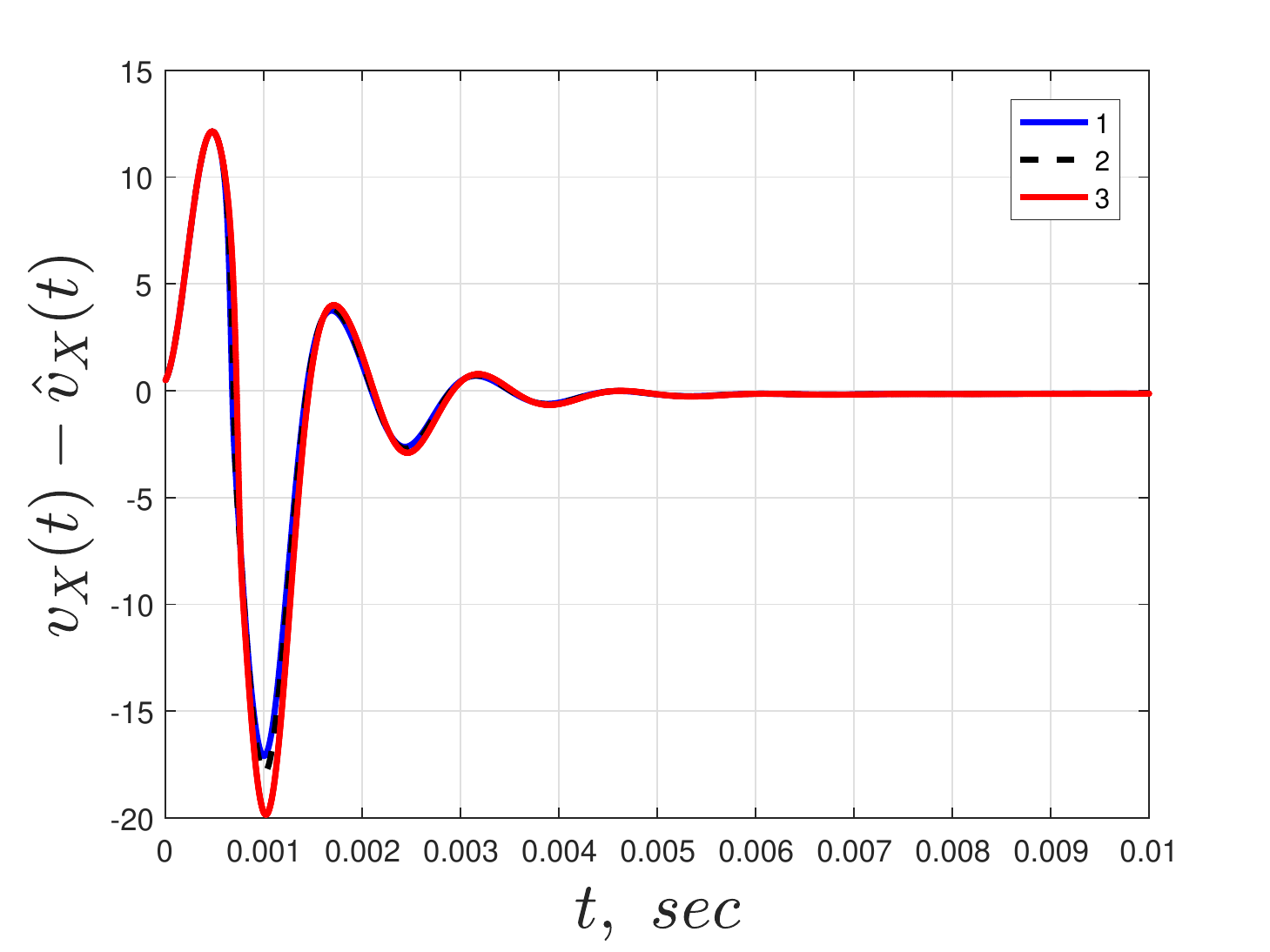}}
	
	\caption{\label{fig5} Behaviour of the observation errors of the system with the sensorless-based IDA-PBC: 1.~$\gamma_i=1000$, 2.~$\gamma_i=500$, 3.~$\gamma_i=100$, i=1,...,4}
\end{figure*}

\begin{figure*}[htp]
\centering
	\subcaptionbox{\label{fig61} Transients for $\lambda_1(t)-\hat\lambda_1(t)$}{\includegraphics[width=0.37\textwidth]{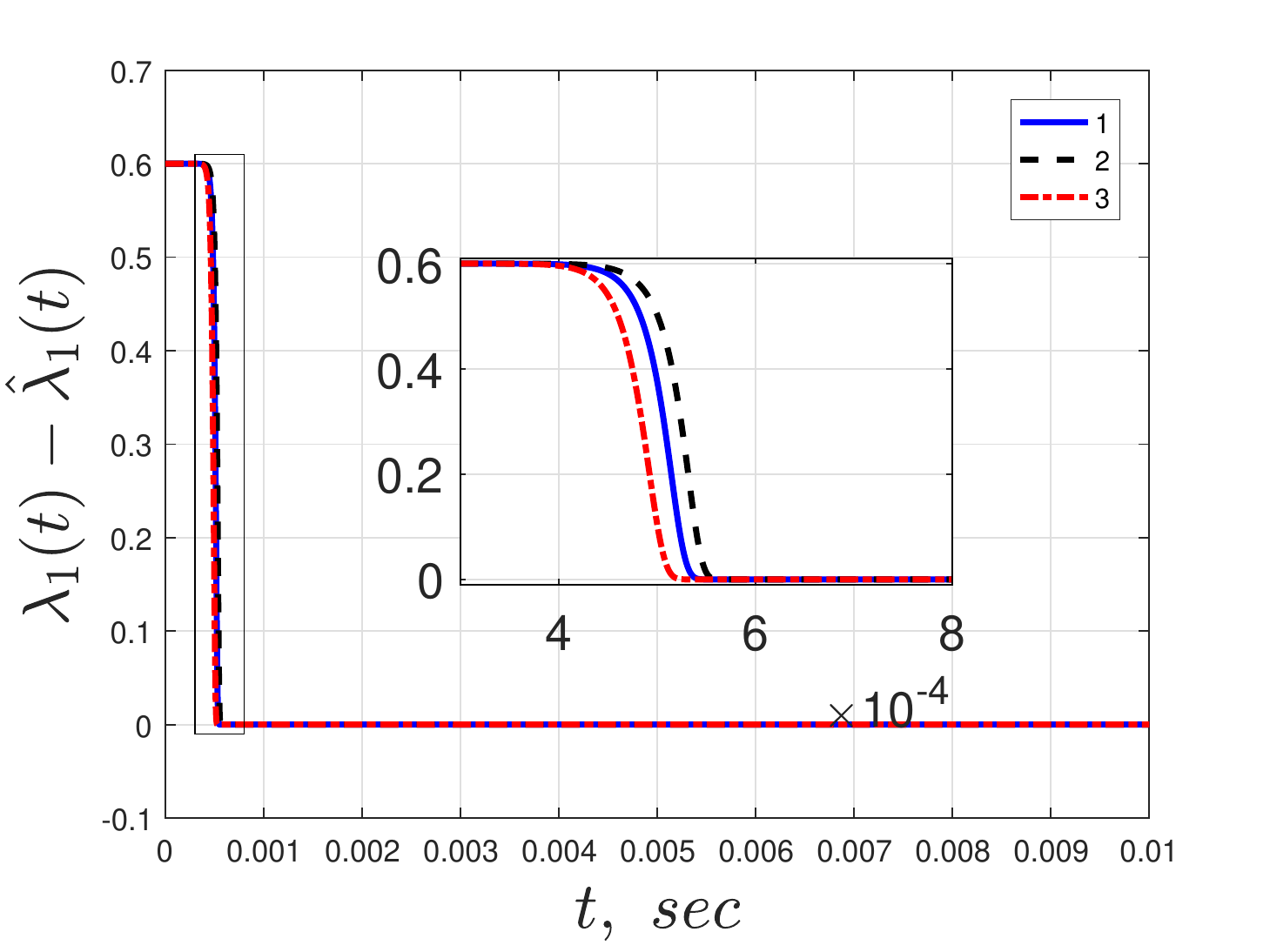}}
	\qquad
	\subcaptionbox{\label{fig62} Transients for $\lambda_2(t)-\hat\lambda_2(t)$}{\includegraphics[width=0.37\textwidth]{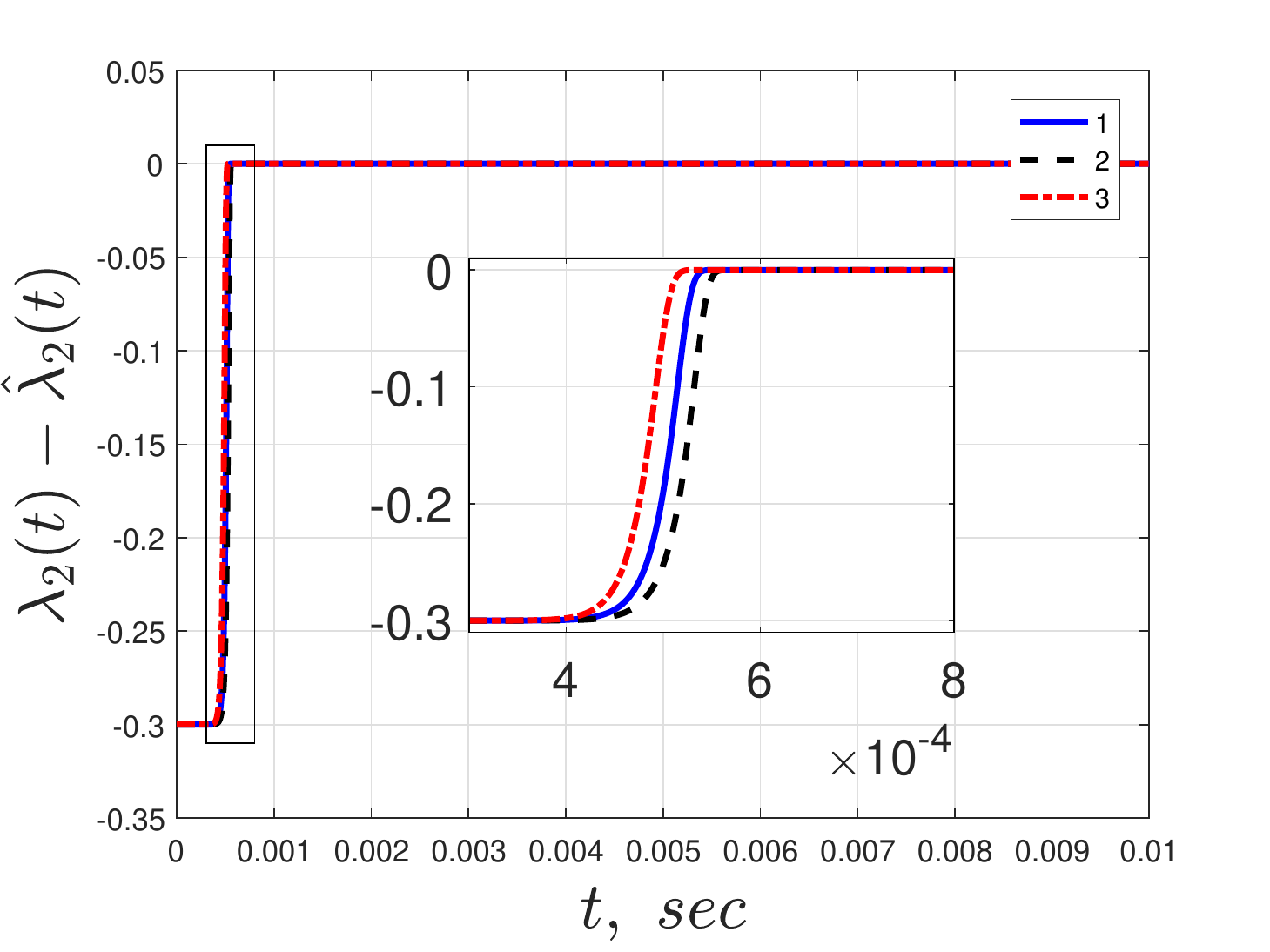}}
	\\
	\subcaptionbox{\label{fig63} Transients for $Y(t)-\hat Y(t)$}{\includegraphics[width=0.37\textwidth]{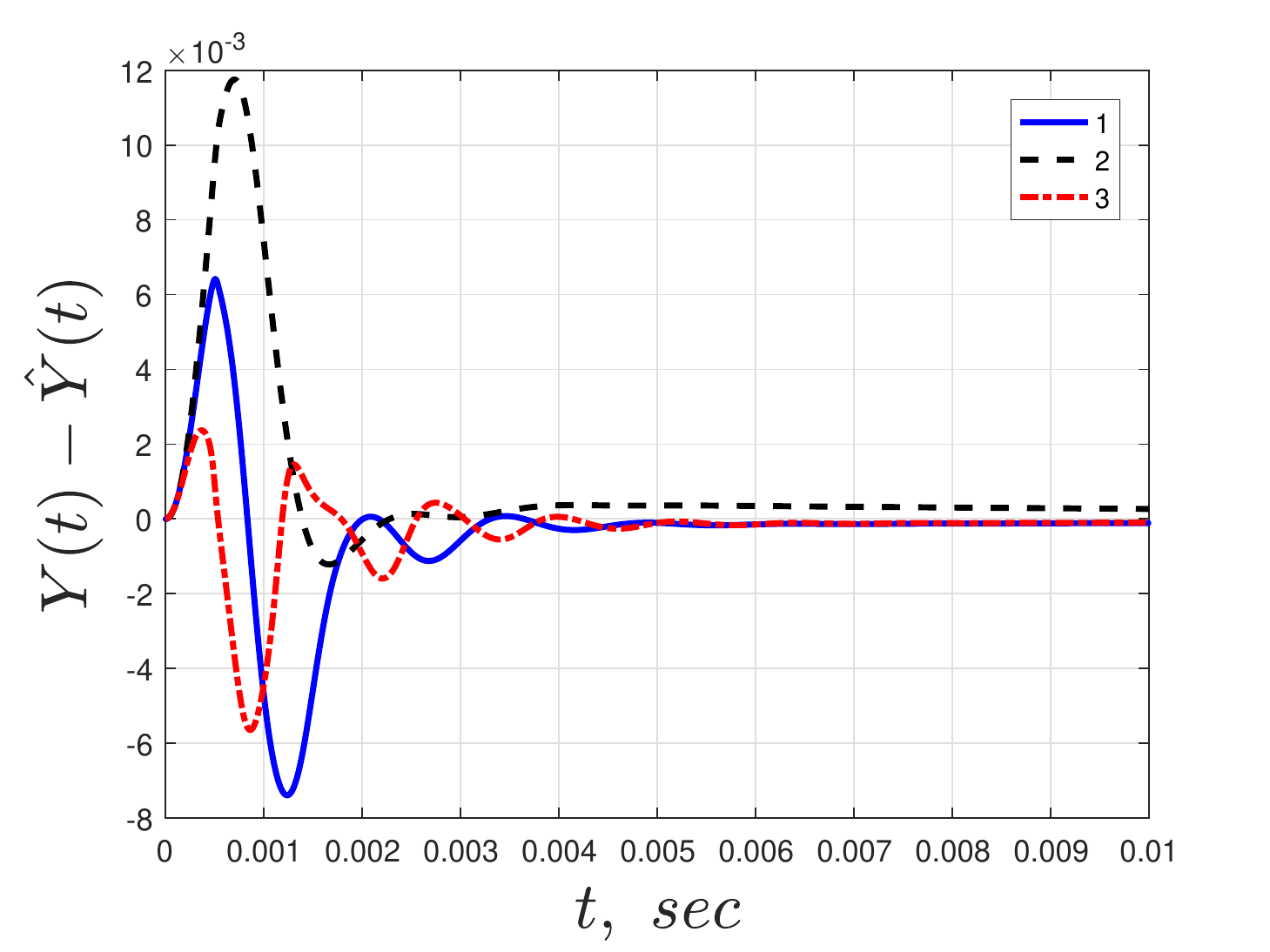}}
	\qquad
	\subcaptionbox{\label{fig64} Transients for $\dot Y(t)-\hat v_Y(t)$}{\includegraphics[width=0.37\textwidth]{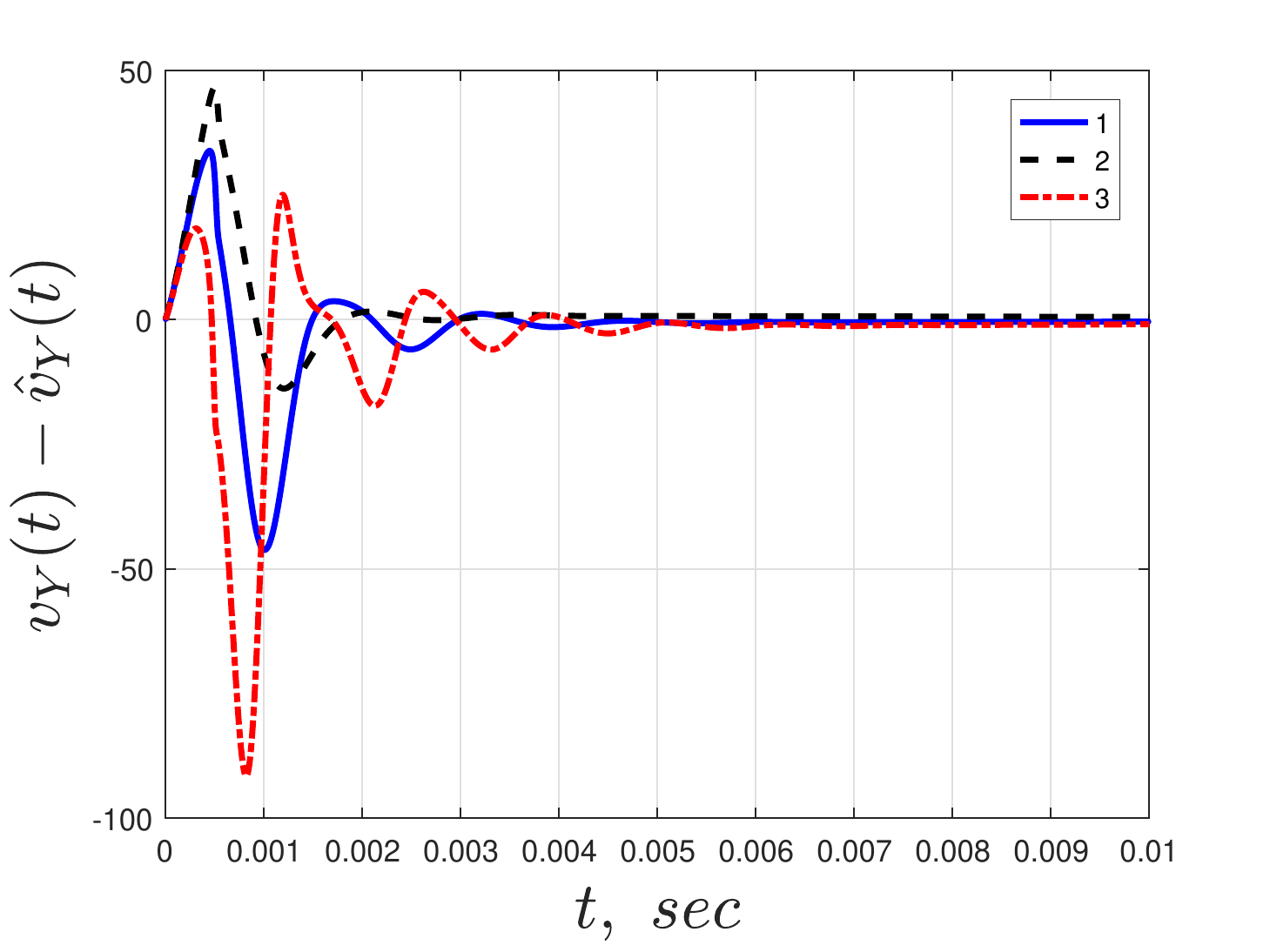}}
	\\
	\subcaptionbox{\label{fig65} Transients for $\lambda_3(t)-\hat\lambda_3(t)$}{\includegraphics[width=0.37\textwidth]{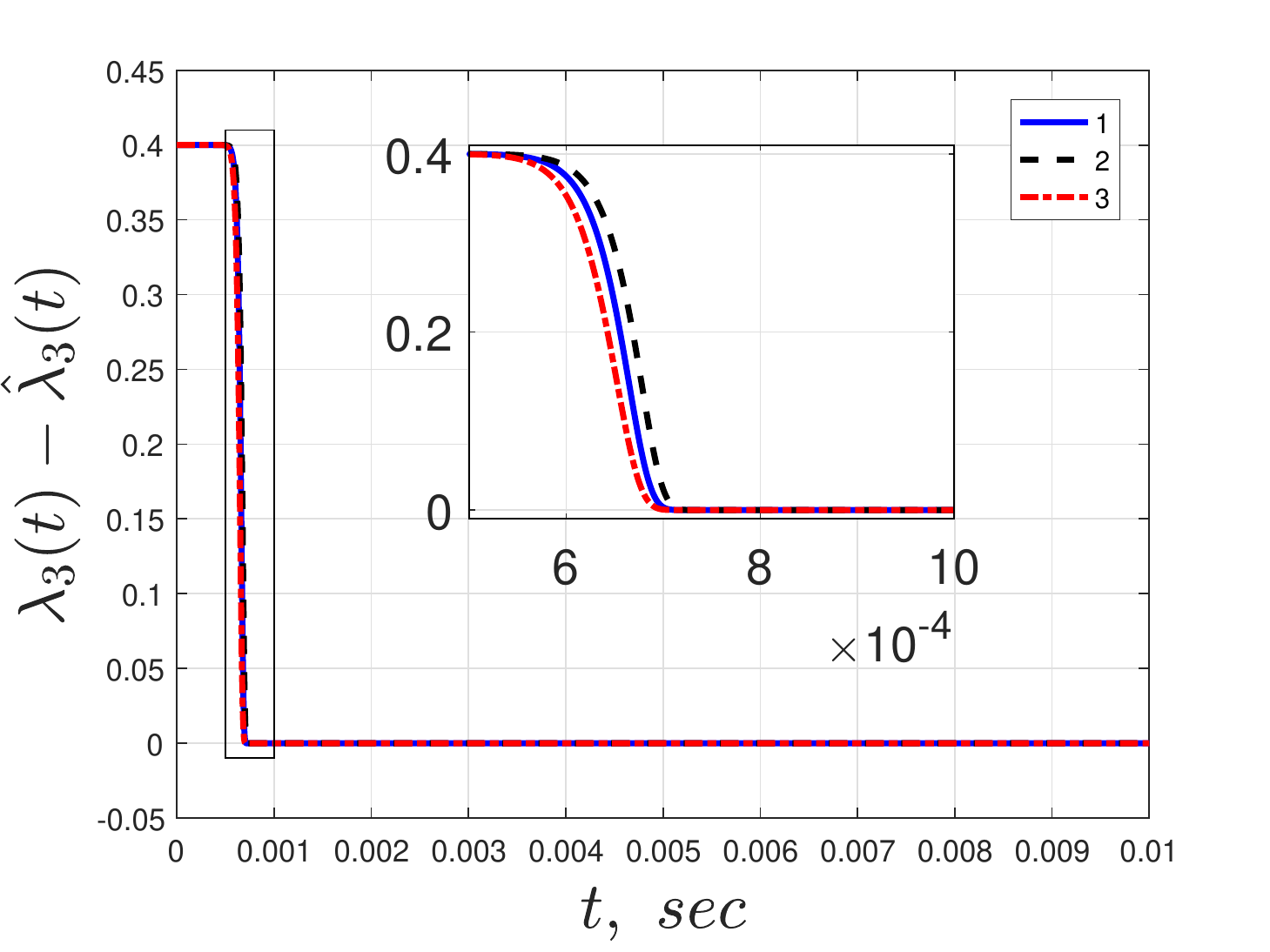}}
	\qquad
	\subcaptionbox{\label{fig66} Transients for $\lambda_4(t)-\hat\lambda_4(t)$}{\includegraphics[width=0.37\textwidth]{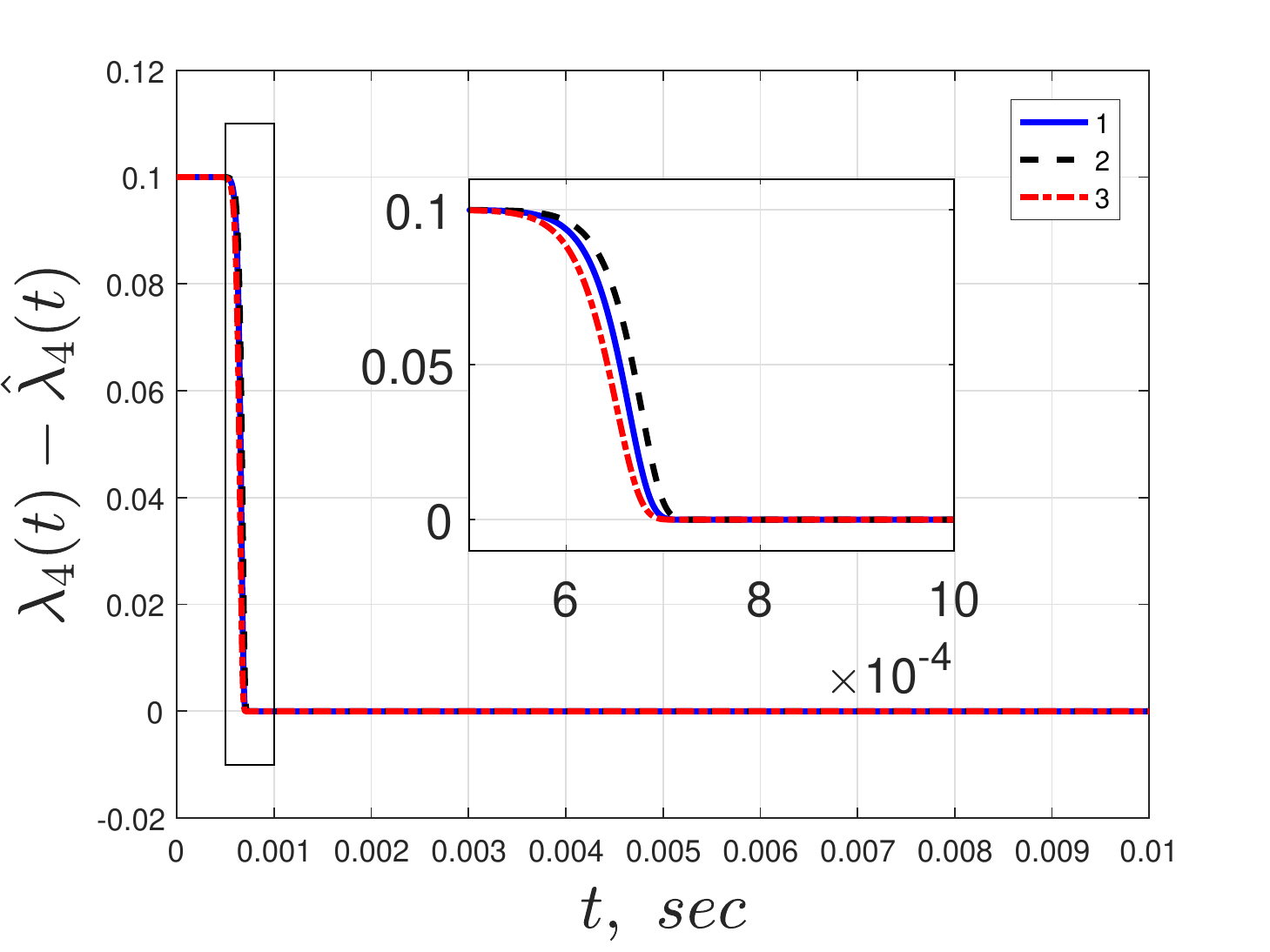}}
	\\
	\subcaptionbox{\label{fig67} Transients for $X(t)-\hat X(t)$}{\includegraphics[width=0.37\textwidth]{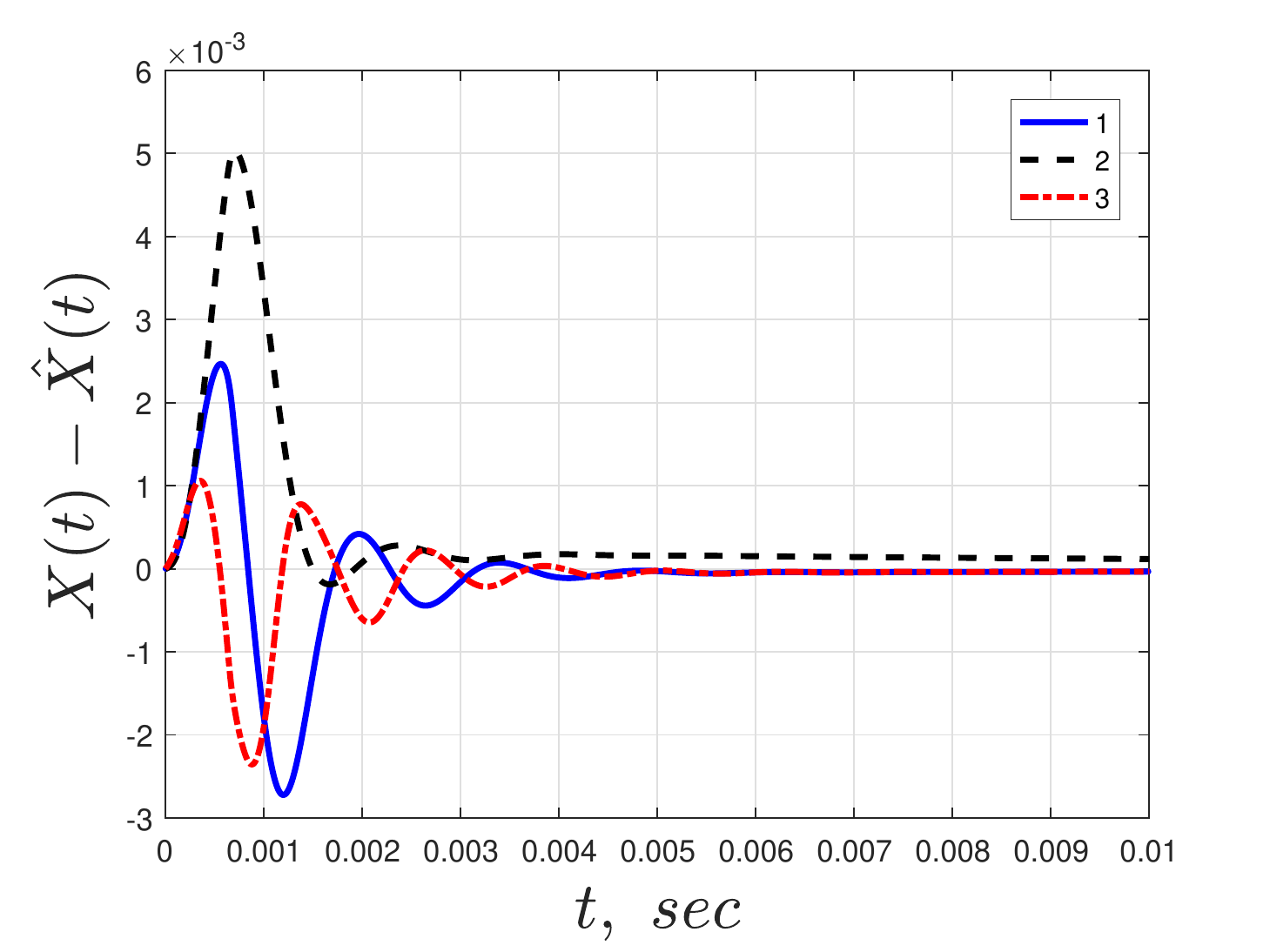}}
	\qquad
	\subcaptionbox{\label{fig68} Transients for $\dot X(t)-\hat v_ X(t)$}{\includegraphics[width=0.37\textwidth]{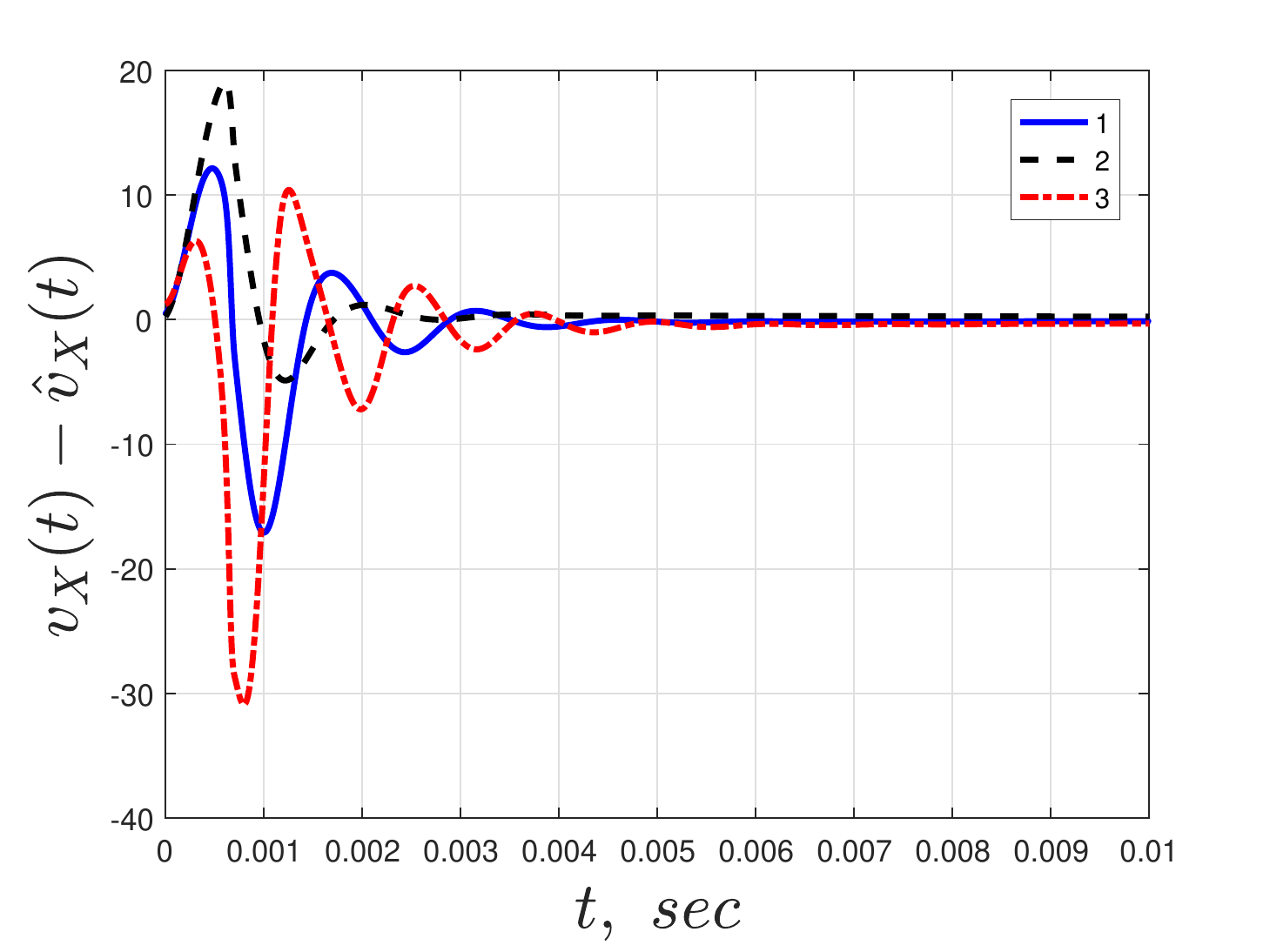}}
	
	\caption{\label{fig6} Behaviour of the observation errors of the system with the sensorless-based IDA-PBC:: 1. $\mu_{X}=\mu_Y=\gamma_{X}=\gamma_Y=2000$, 2. $\mu_{X}=\mu_Y=\gamma_{X}=\gamma_Y=1000$, 3. $\mu_{X}=\mu_Y=\gamma_{X}=\gamma_Y=5000$}
\end{figure*}

\begin{figure*}[htp]
\centering
	\subcaptionbox{\label{fig71} Transients for $\lambda_1(t)-\hat\lambda_1(t)$}{\includegraphics[width=0.37\textwidth]{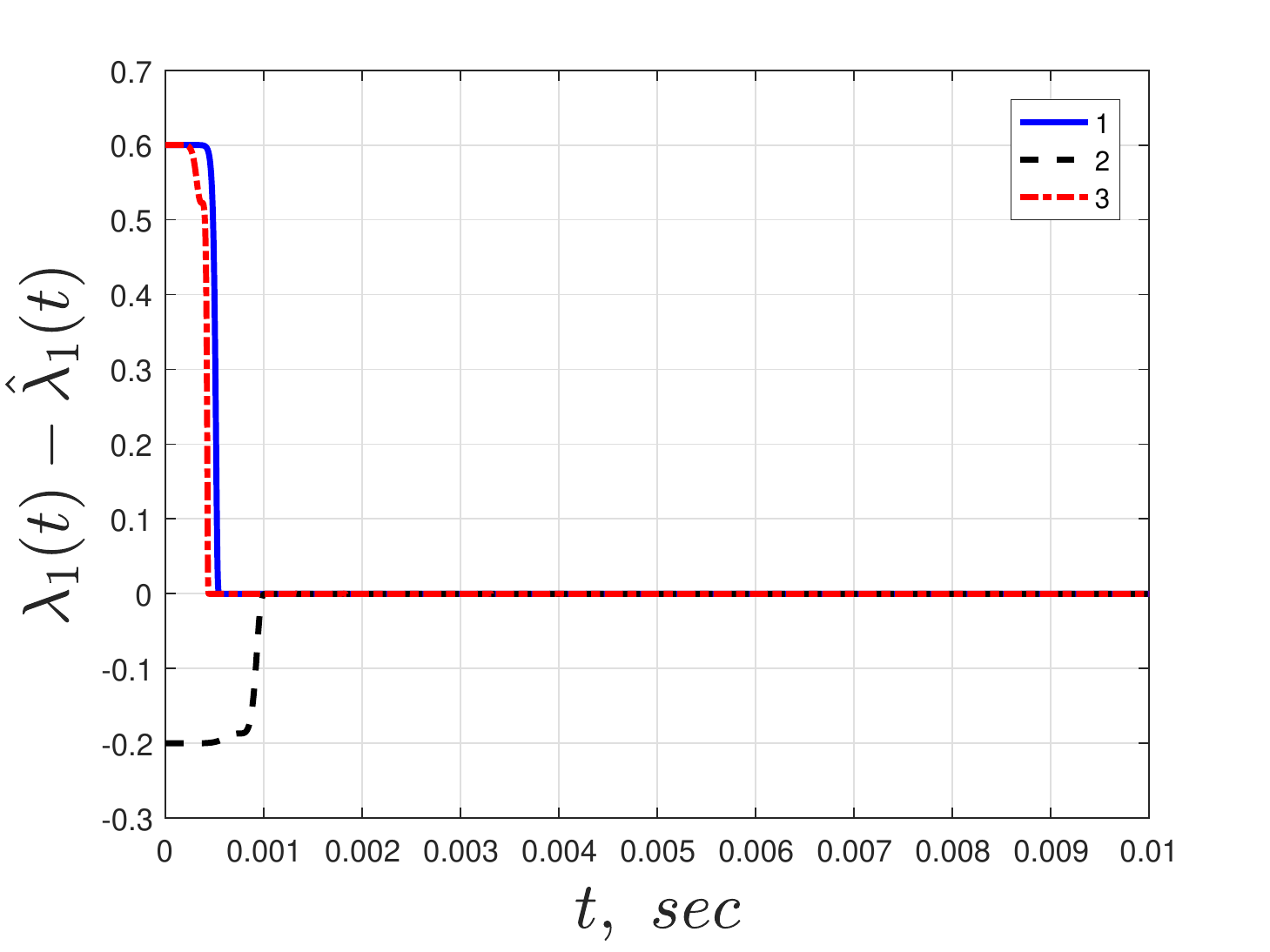}}
	\qquad
	\subcaptionbox{\label{fig72} Transients for $\lambda_2(t)-\hat\lambda_2(t)$}{\includegraphics[width=0.37\textwidth]{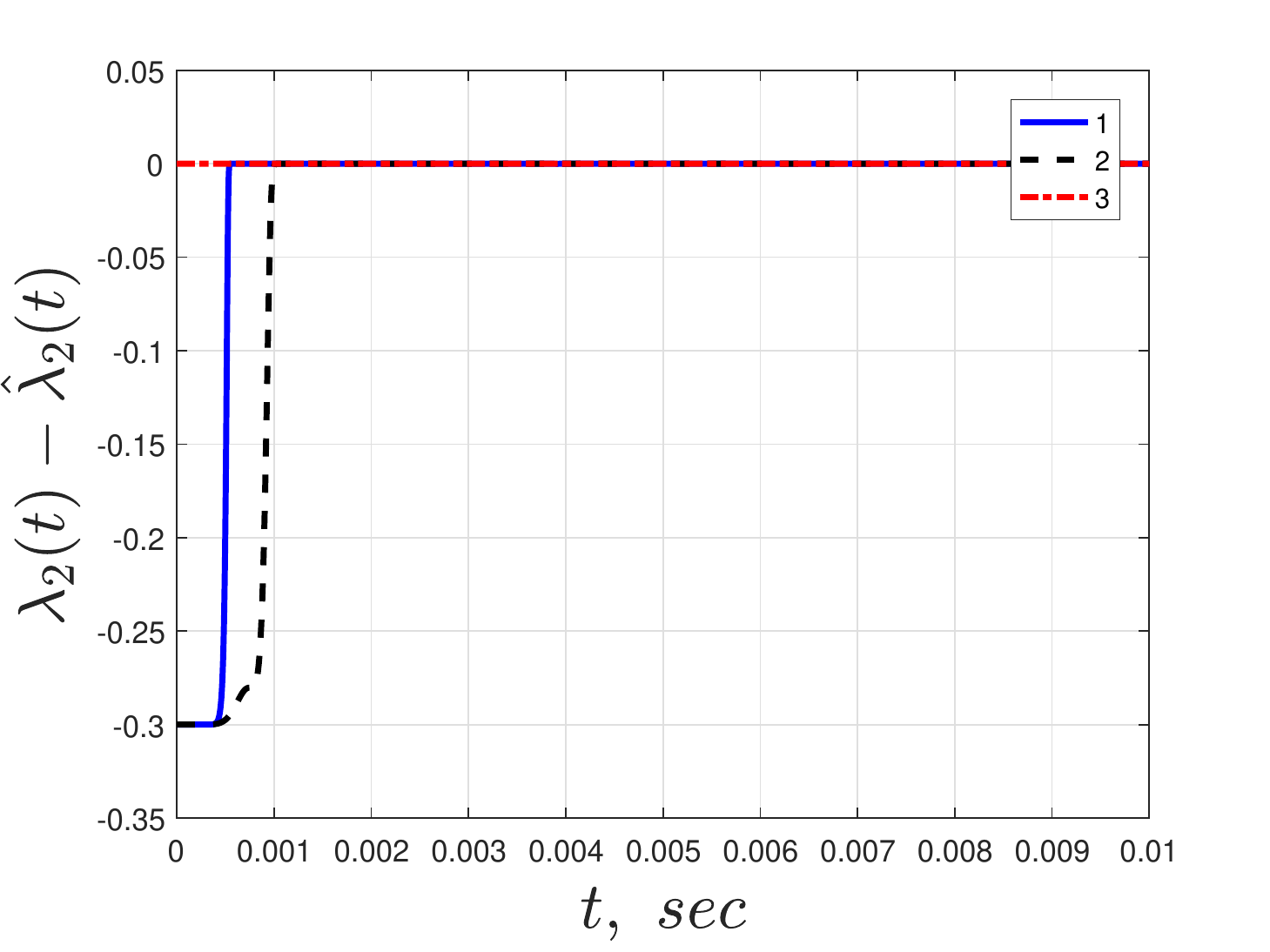}}
	\\
	\subcaptionbox{\label{fig73} Transients for $Y(t)-\hat Y(t)$}{\includegraphics[width=0.37\textwidth]{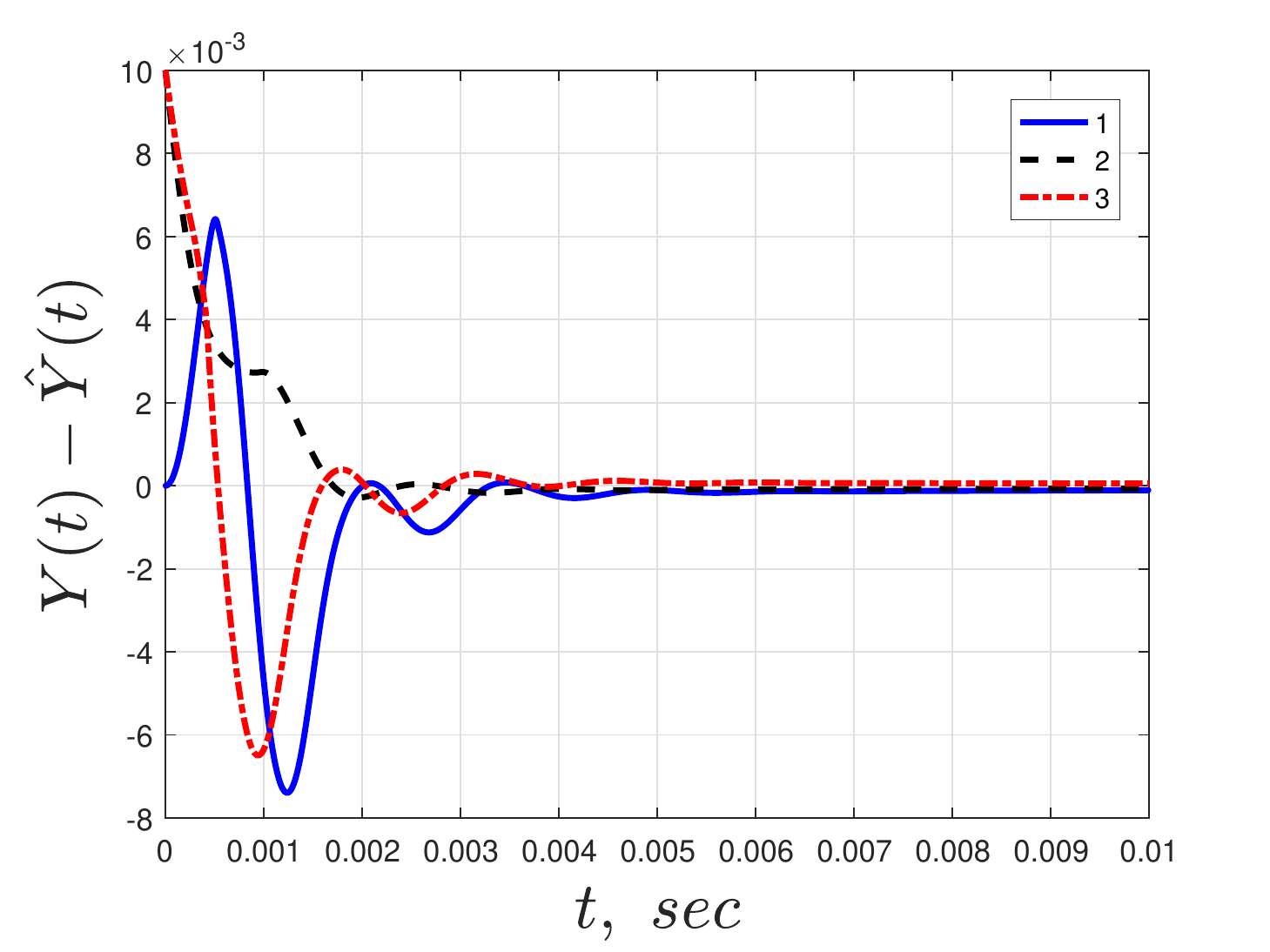}}
	\qquad
	\subcaptionbox{\label{fig74} Transients for $\dot Y(t)-\hat v_Y(t)$}{\includegraphics[width=0.37\textwidth]{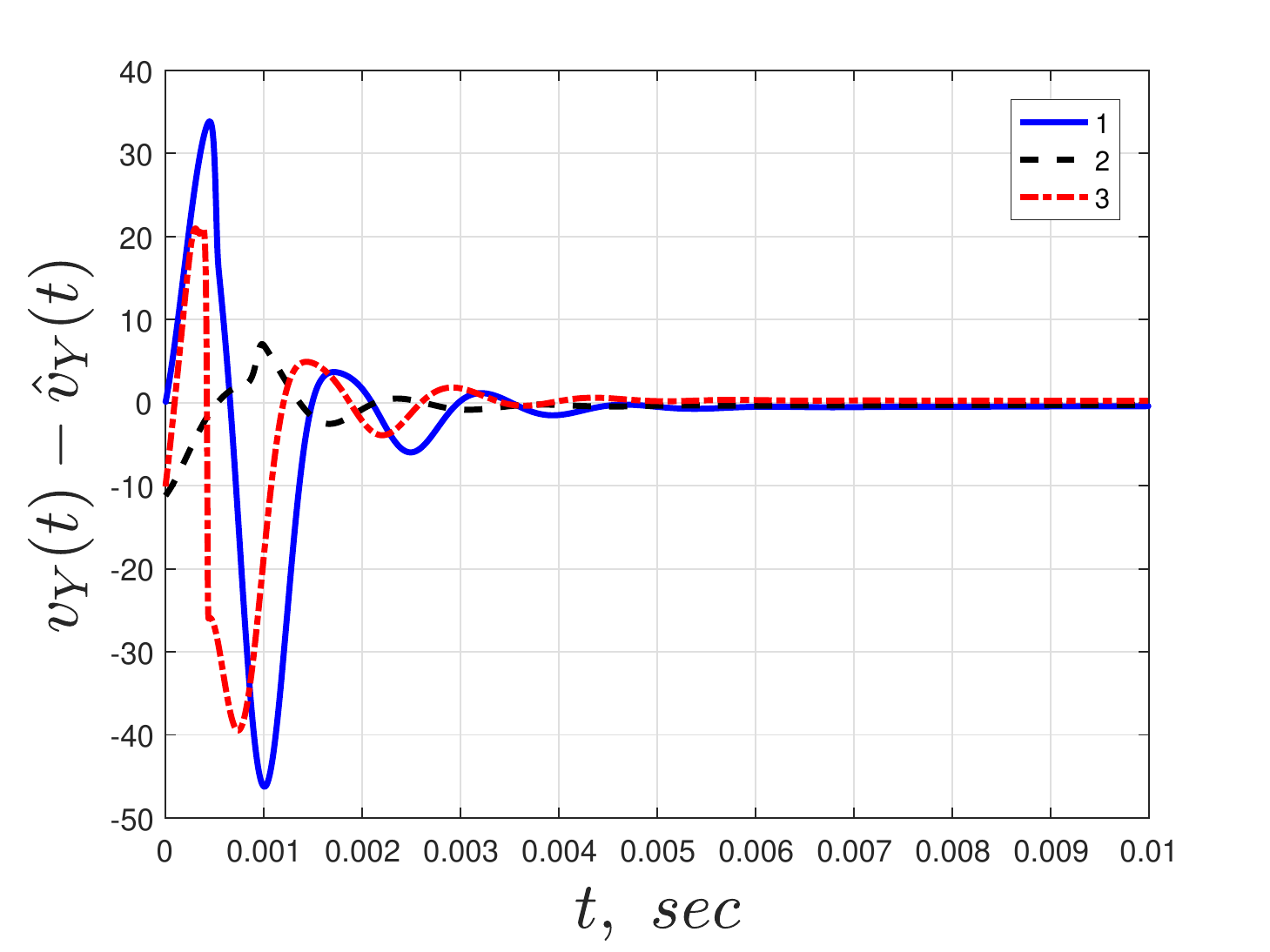}}
	\\
	\subcaptionbox{\label{fig75} Transients for $\lambda_3(t)-\hat\lambda_3(t)$}{\includegraphics[width=0.37\textwidth]{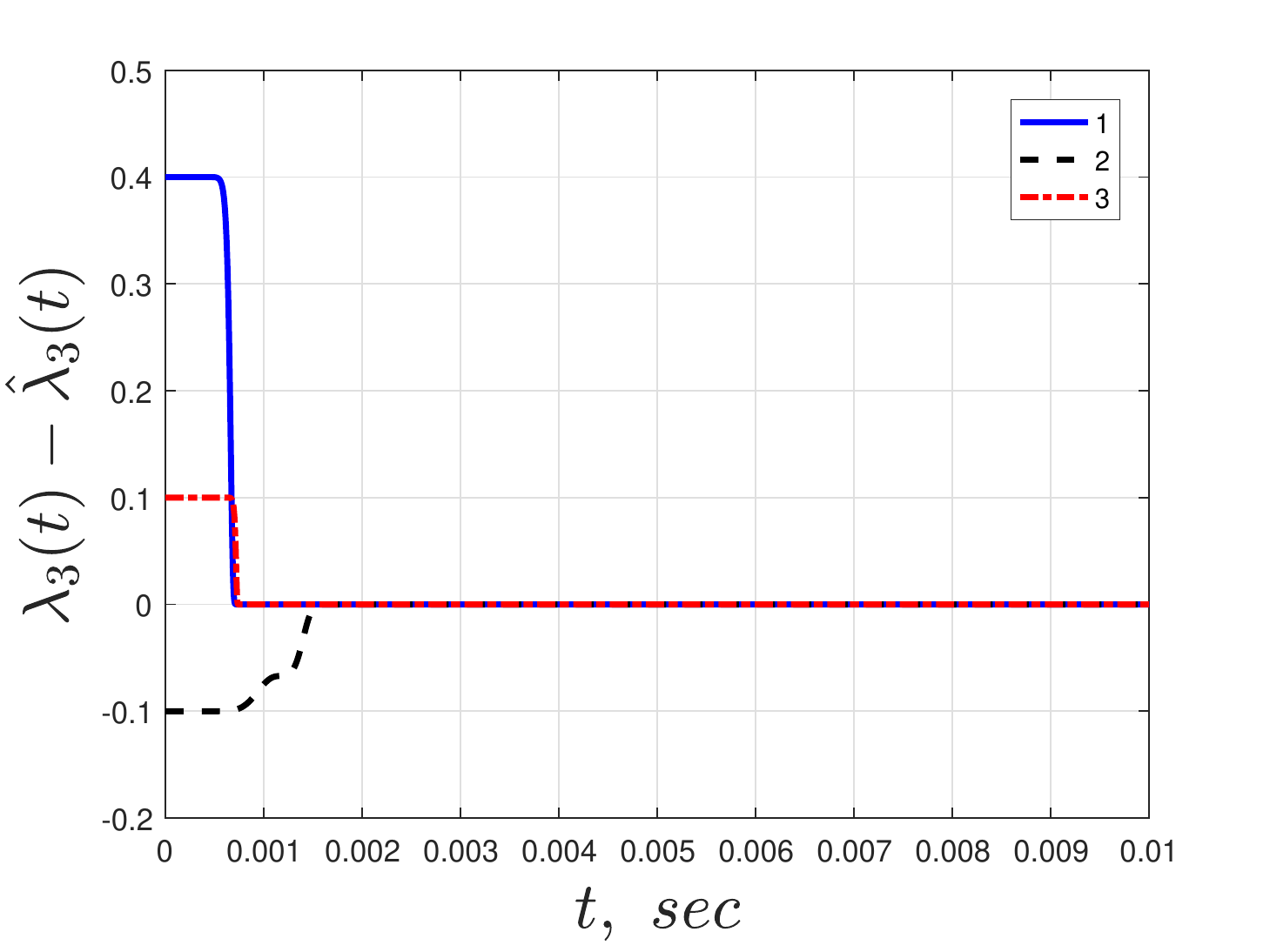}}
	\qquad
	\subcaptionbox{\label{fig76} Transients for $\lambda_4(t)-\hat\lambda_4(t)$}{\includegraphics[width=0.37\textwidth]{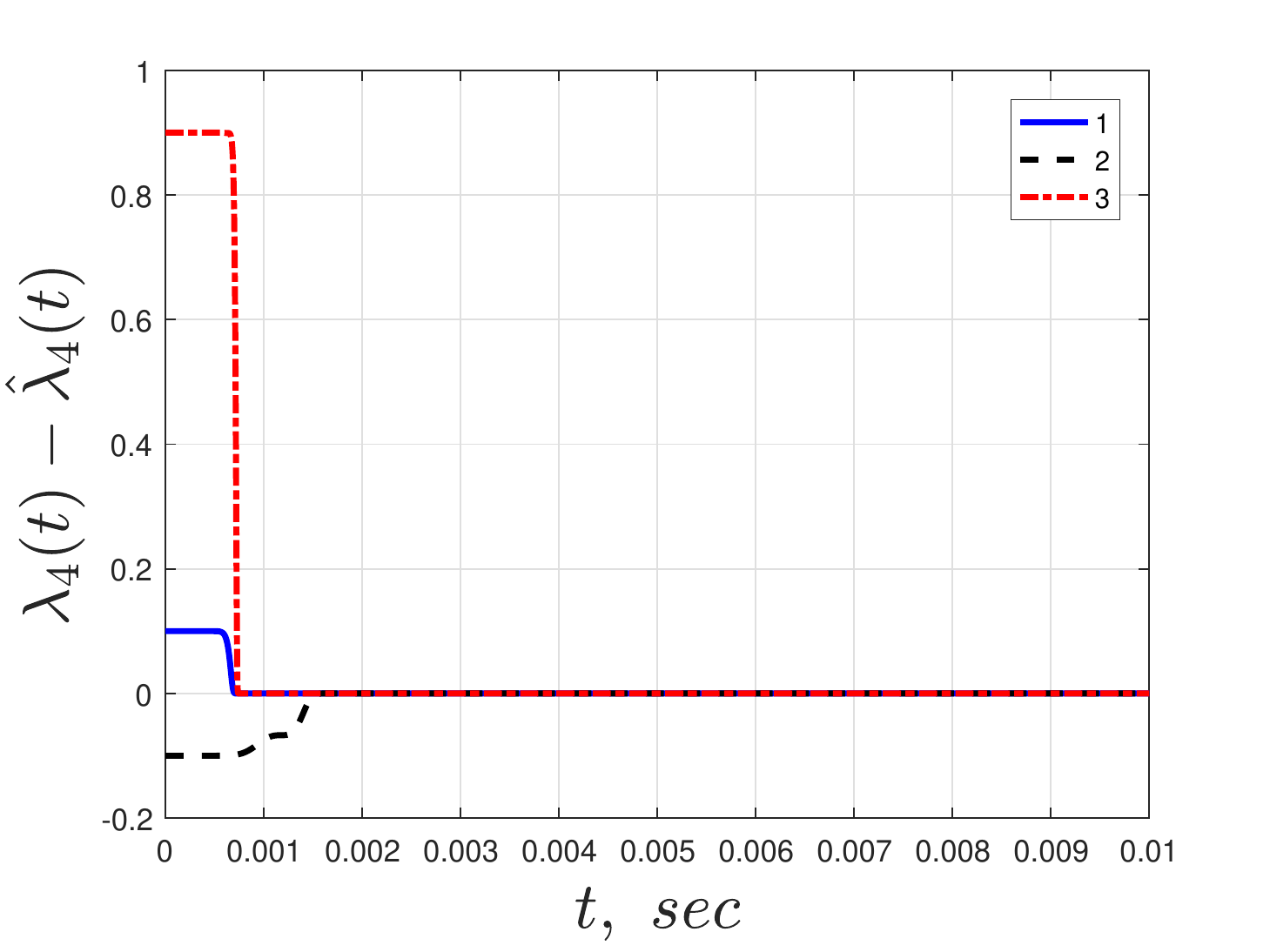}}
	\\
	\subcaptionbox{\label{fig77} Transients for $X(t)-\hat X(t)$}{\includegraphics[width=0.37\textwidth]{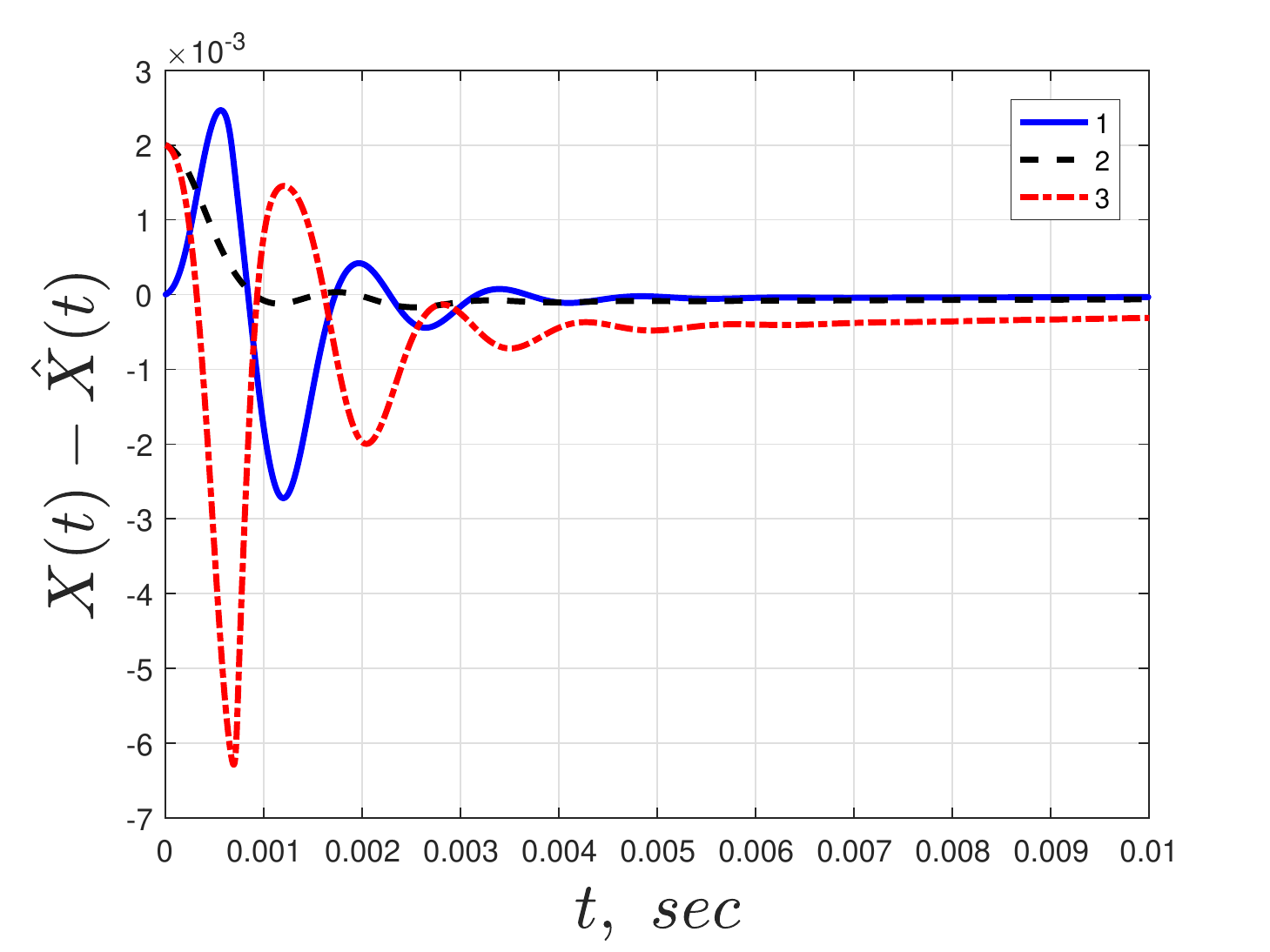}}
	\qquad
	\subcaptionbox{\label{fig78} Transients for $\dot X(t)-\hat v_X(t)$}{\includegraphics[width=0.37\textwidth]{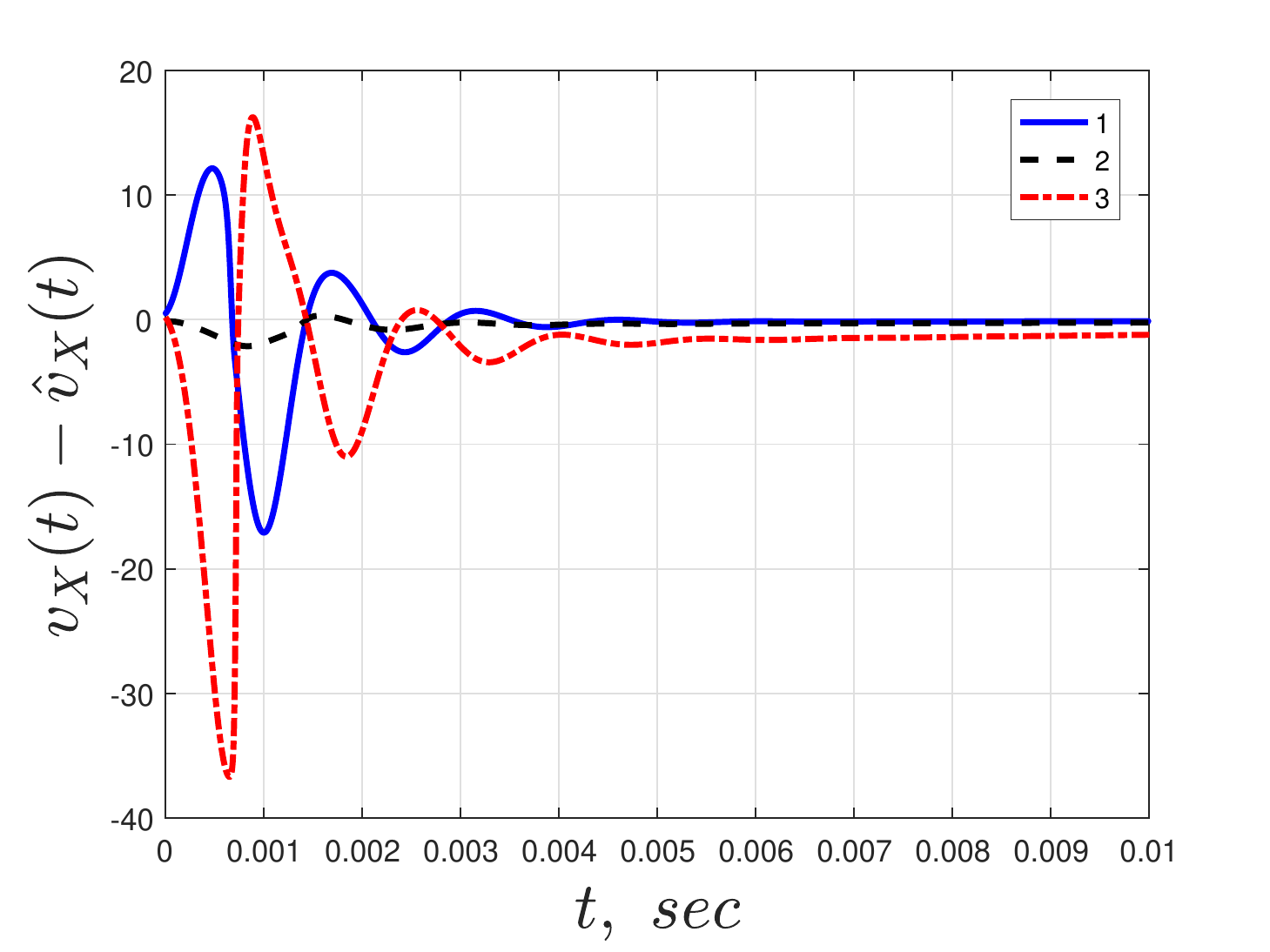}}
	
	\caption{\label{fig7} Behaviour of the observation errors of the system with the sensorless-based IDA-PBC for different initial conditions}
\end{figure*}

Fig. \ref{fig4} shows the results, for the same parameters, but for a circle trajectory defined by
\begequ
\label{crc}
X_*(t)=0.1\sin(0.1t), \quad Y_*(t)=0.1\cos(0.1t).
\endequ
In Figs. \ref{fig5}, \ref{fig6} we evaluated the effect of changing the adaptation gains $\gamma_i$ and the observer gains $(\mu_{X},\mu_{Y},\gamma_{X},\gamma_{Y})$, respectively. In Fig. \ref{fig7} different initial conditions, given in Table \ref{tab_2}, were taken.

\subsection{ $1$-dof Maglev system}
\label{subsec52}
%

The $1$-dof Maglev system \eqref{lamdot}--\eqref{f} in closed-loop with the sensorless version of the FLC \eqref{feelin} was simulated with the following plant parameters: $m = 0.0844$, $k = 1$, $R = 2.52$, $c = 0.005$. The filters used in DREM were implemented with the gains $\rho=0.01$, $\mu=10$, while the parameters of the FLC were fixed at $k_0=1000$, $k_1=300$, $k_2=30$, which corresponds to a pole location of the ideal closed-loop dynamics of $s_1 = s_2 = s_3 = -10$. For all experiments the default initial conditions are $\lambda(0) = \eta$, $\psi(0) = 0$, $\hat \lambda(0) = 0$, $Y(0)=-1$,  $\dot{Y}(0) = 0.5$,  $\hat{Y}(0) = 0$, $\hat v_Y(0)=0$, $\hat{\eta}(0)=0.0001$. 

Two reference signals for $Y$ were considered: filtered sum of sinusoids and filtered steps: $Y^*(t) = \frac{\nu^4}{(p+\nu)^4}Y^*_0(t) $ with
\begin{align}
	Y^*_0(t) &= \sin t + \sin 2t + 0.5\sin(3.7t+\pi/3),
\\
\textnormal{and}\qquad\quad&\nonumber
\\
\label{steps2}
Y^*_0(t)&=\left\{\begin{array}{l}
0, \text{ for } 0\le t\le 1 \text{ sec, }\\
2, \text{ for } 1\le t\le 3 \text{ sec, }\\
0, \text{ for } 3\le t\le 5 \text{ sec, }\\
3, \text{ for } t\ge 5 \text{ sec. }
\end{array}
\right.
\end{align} 
where $\nu=10$ for the sinusoids and $\nu=1$ for the steps. 

\begin{figure*}[htp]
\centering
	\subcaptionbox{\label{fig:1dof_errors_sin_gamma_lambda} Transients for $\lambda(t)-\hat\lambda(t)$}{\includegraphics[width=0.31\textwidth]{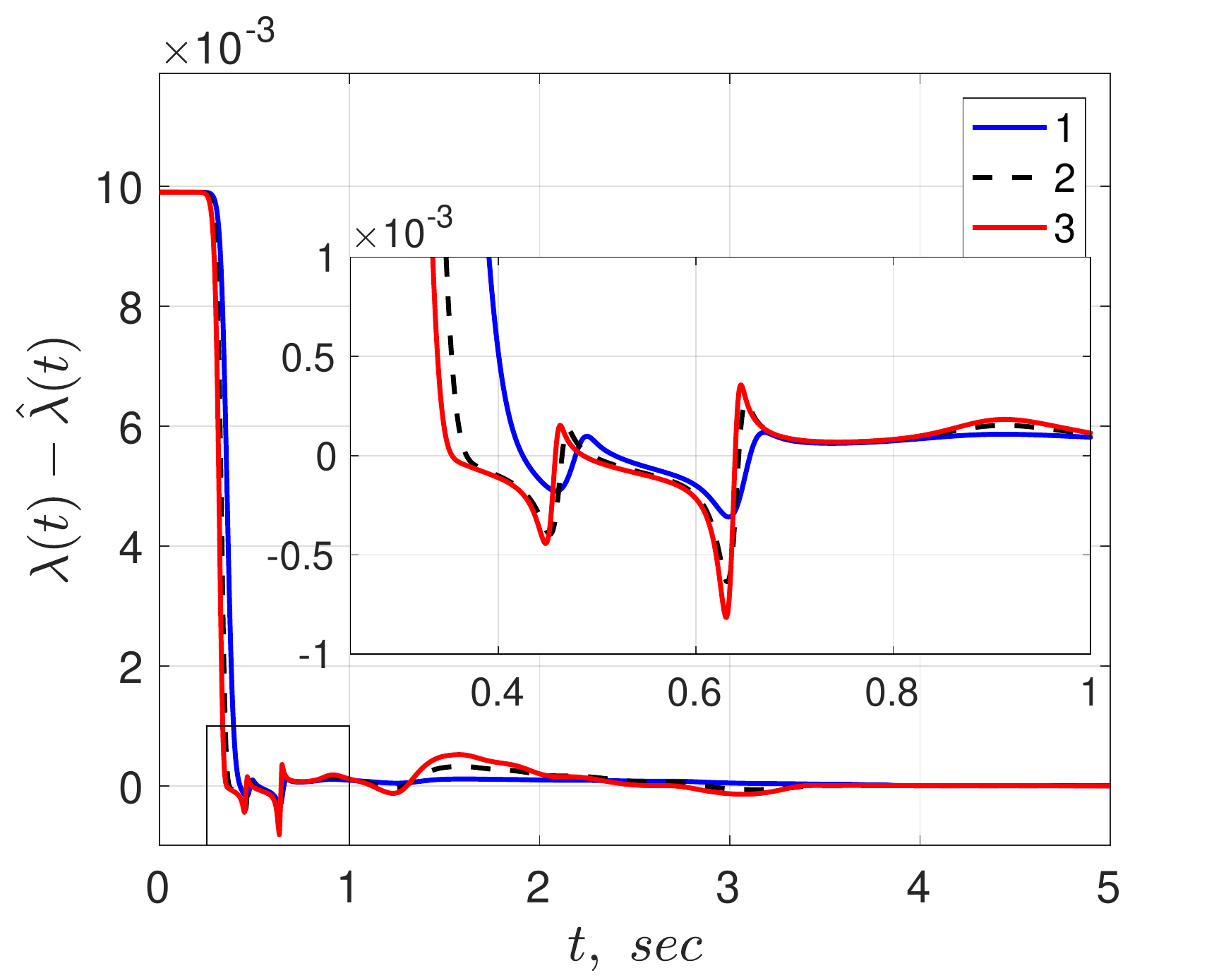}}
	\
	\subcaptionbox{\label{fig:1dof_errors_sin_gamma_Y} Transients for $Y(t)-\hat Y(t)$}{\includegraphics[width=0.31\textwidth]{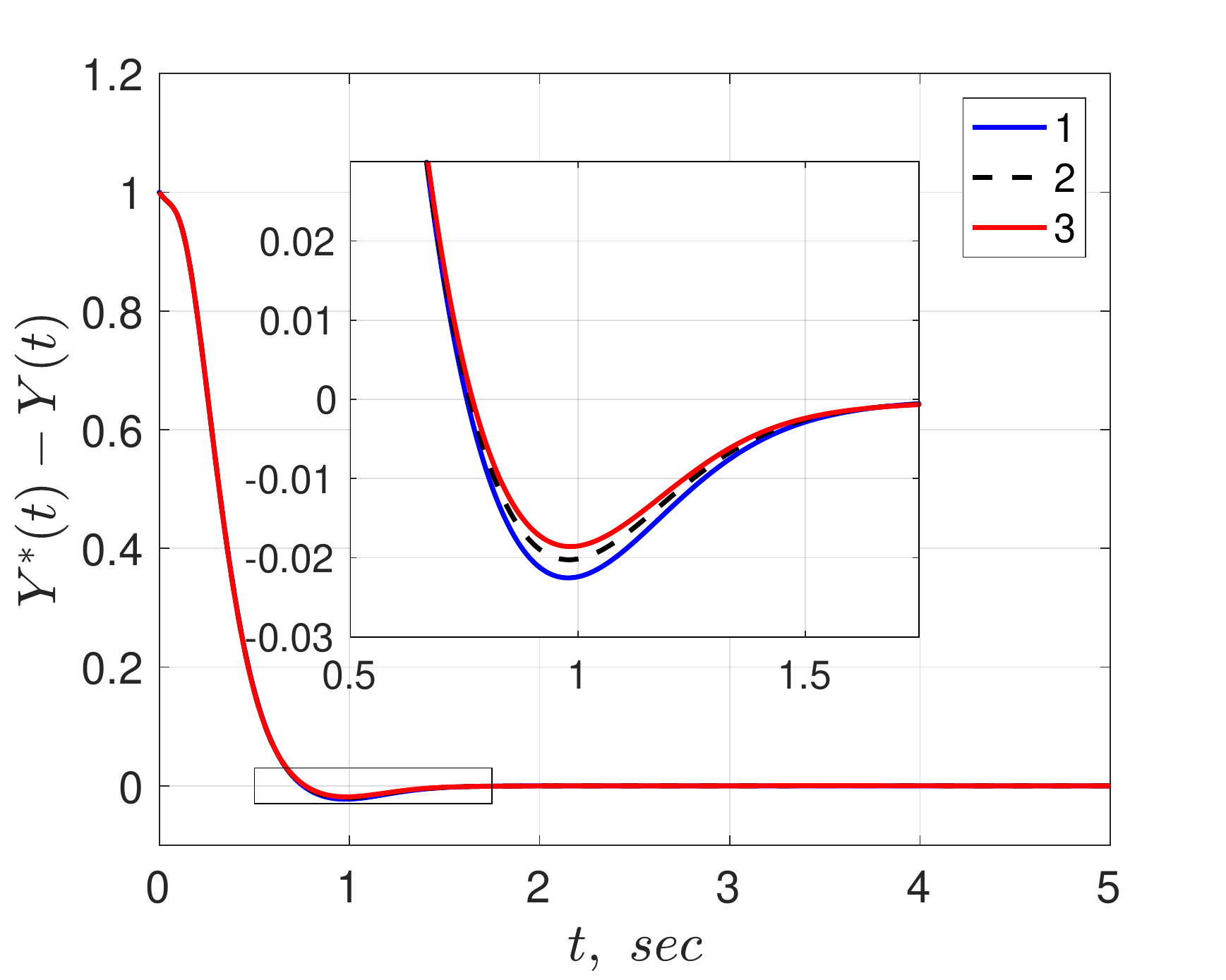}}
	\
	\subcaptionbox{\label{fig:1dof_errors_sin_gamma_dY} Transients for $\dot Y(t)-\hat v_{Y}(t)$}{\includegraphics[width=0.31\textwidth]{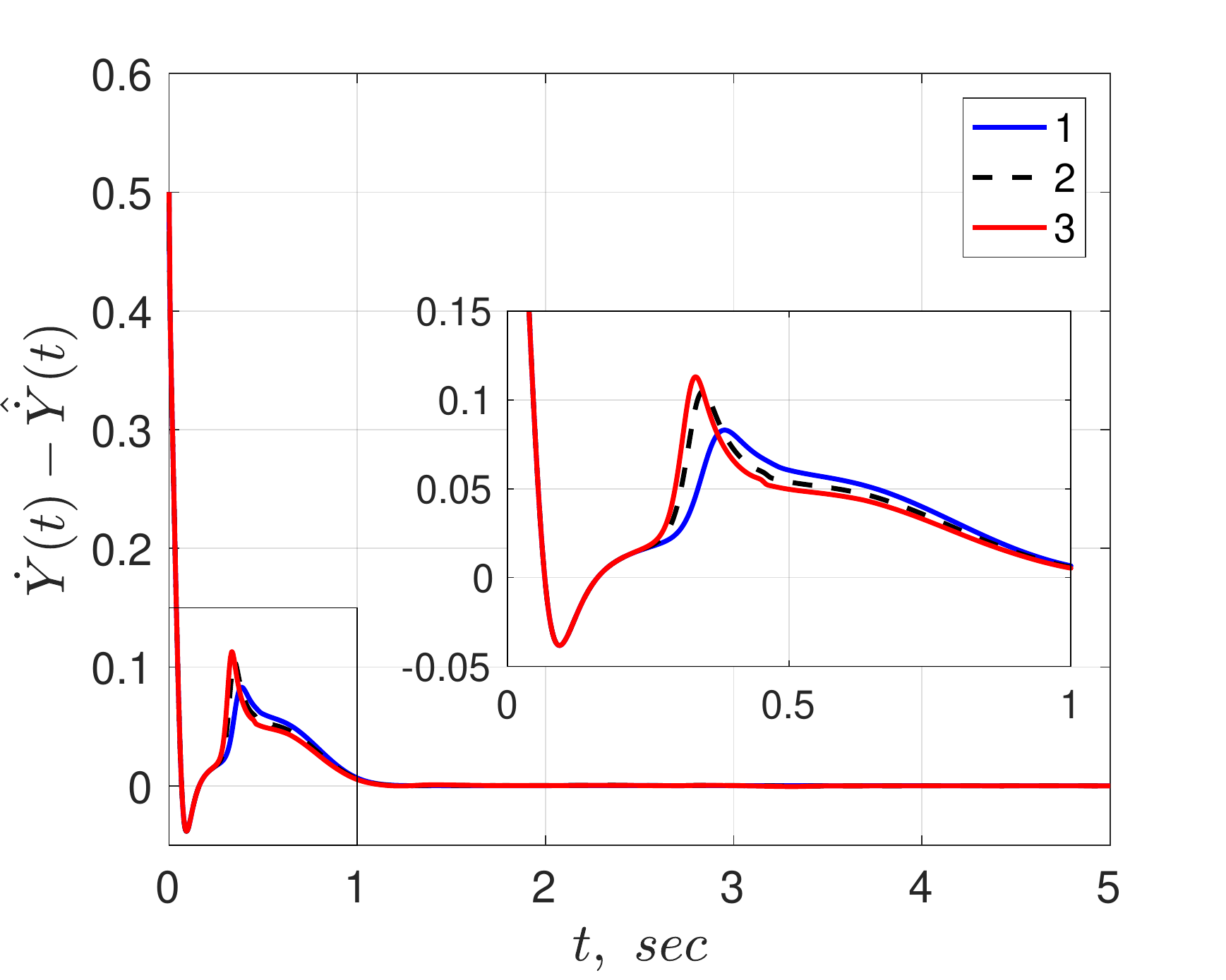}}
	\vspace{-2mm}
	\caption{\label{fig:1dof_errors_sin_gamma} Errors with the sensorless-based FLC for the sinusoidal position reference: 1. $\gamma=1$, 2. $\gamma=5$, 3. $\gamma=10$}
\end{figure*}

\begin{figure*}[htp]
\centering
	\subcaptionbox{\label{fig:1dof_errors_step_gamma_lambda} Transients for $\lambda(t)-\hat\lambda(t)$}{\includegraphics[width=0.31\textwidth]{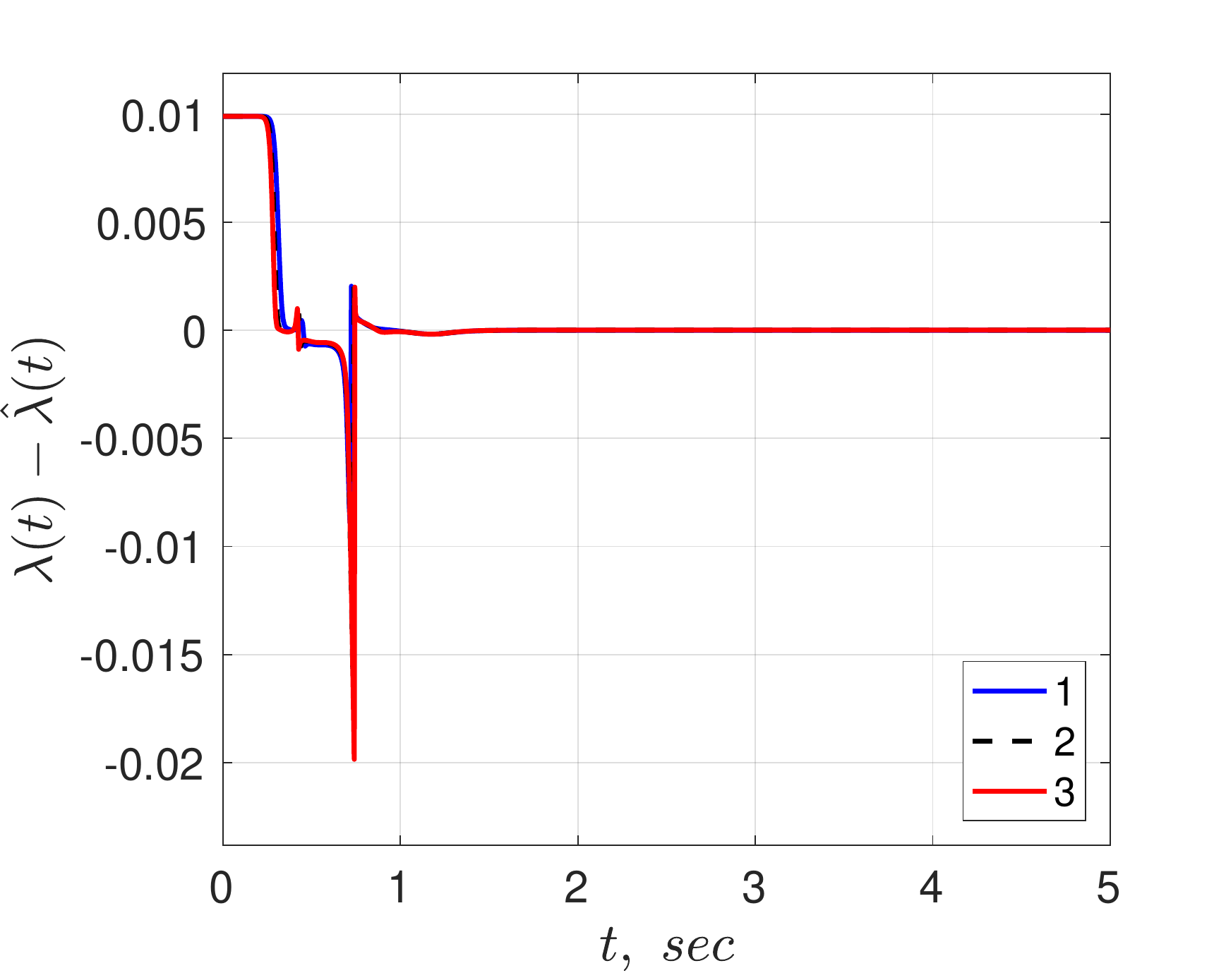}}
	\
	\subcaptionbox{\label{fig:1dof_errors_step_gamma_Y} Transients for $Y(t)-\hat Y(t)$}{\includegraphics[width=0.31\textwidth]{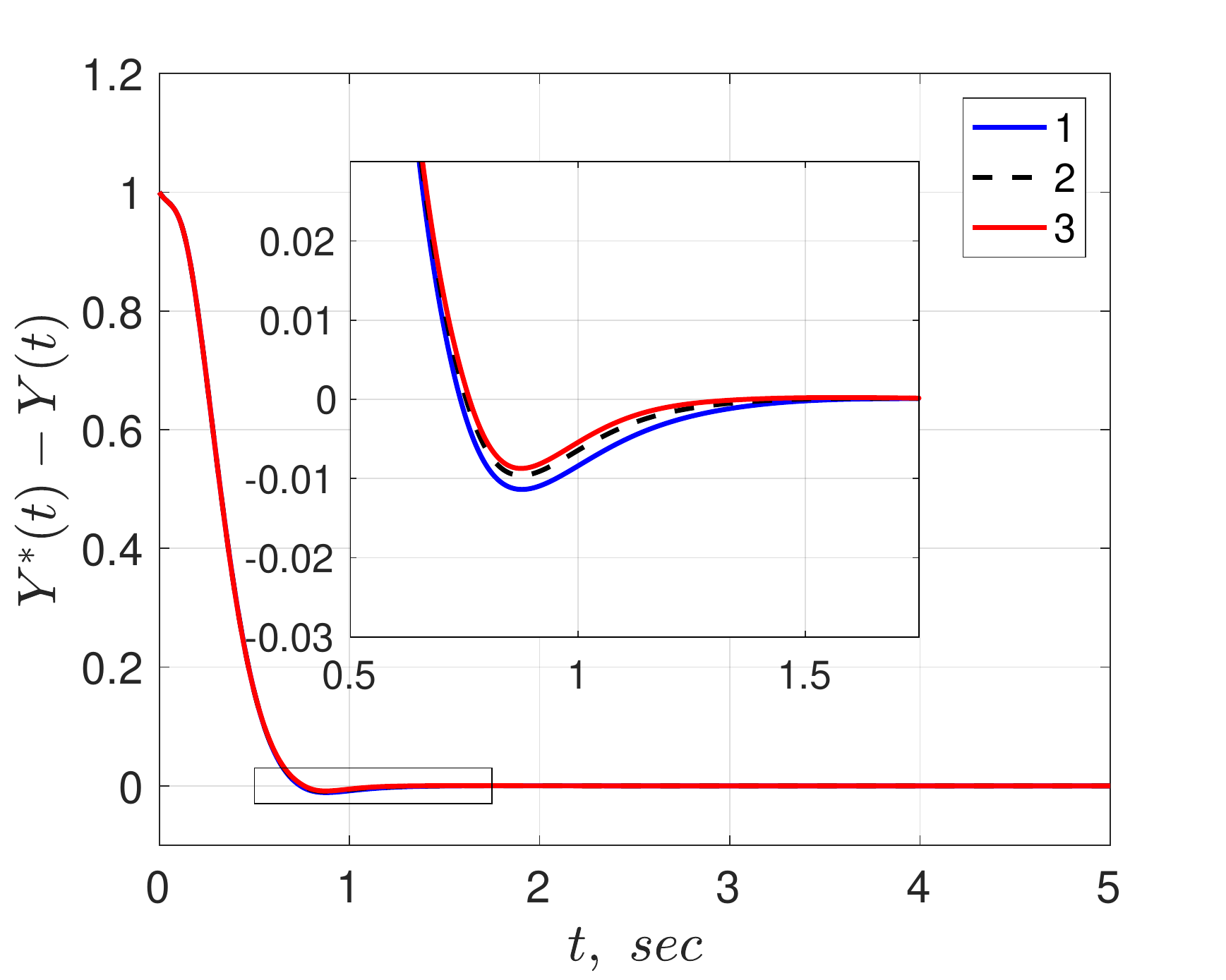}}
	\
\subcaptionbox{\label{fig:1dof_errors_step_gamma_dY} Transients for $\dot Y(t)-\hat v_{Y}(t)$}{\includegraphics[width=0.31\textwidth]{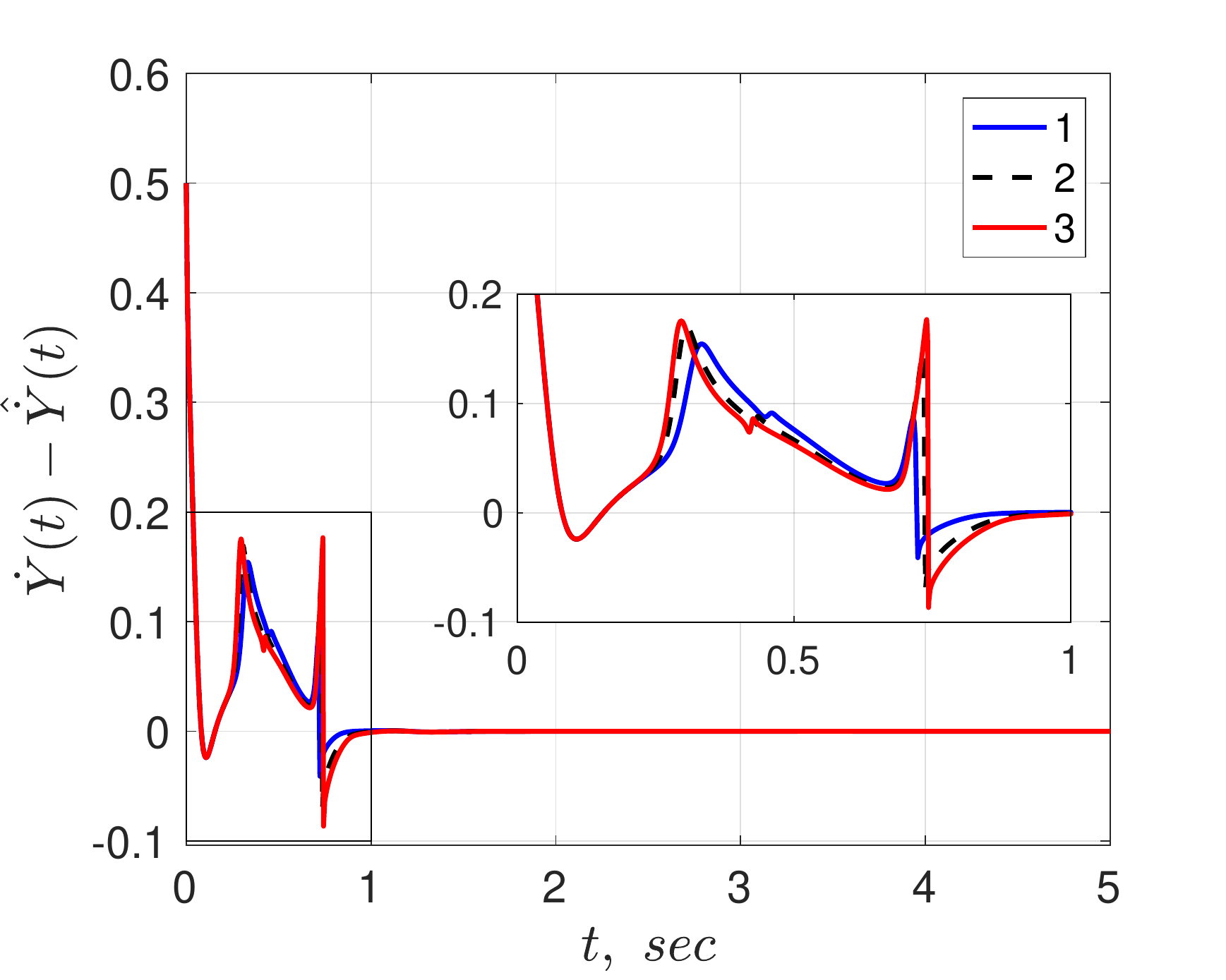}}
	\vspace{-2mm}
	\caption{\label{fig:1dof_errors_step_gamma} Errors with the sensorless-based FLC for the steps position reference: 1. $\gamma=1000$, 2. $\gamma=5000$, 3. $\gamma=10000$}
\end{figure*}

\begin{figure*}[htp]
\centering
	\subcaptionbox{\label{fig:1dof_errors_sin_l0_lambda} Transients for $\lambda(t)-\hat\lambda(t)$}{\includegraphics[width=0.31\textwidth]{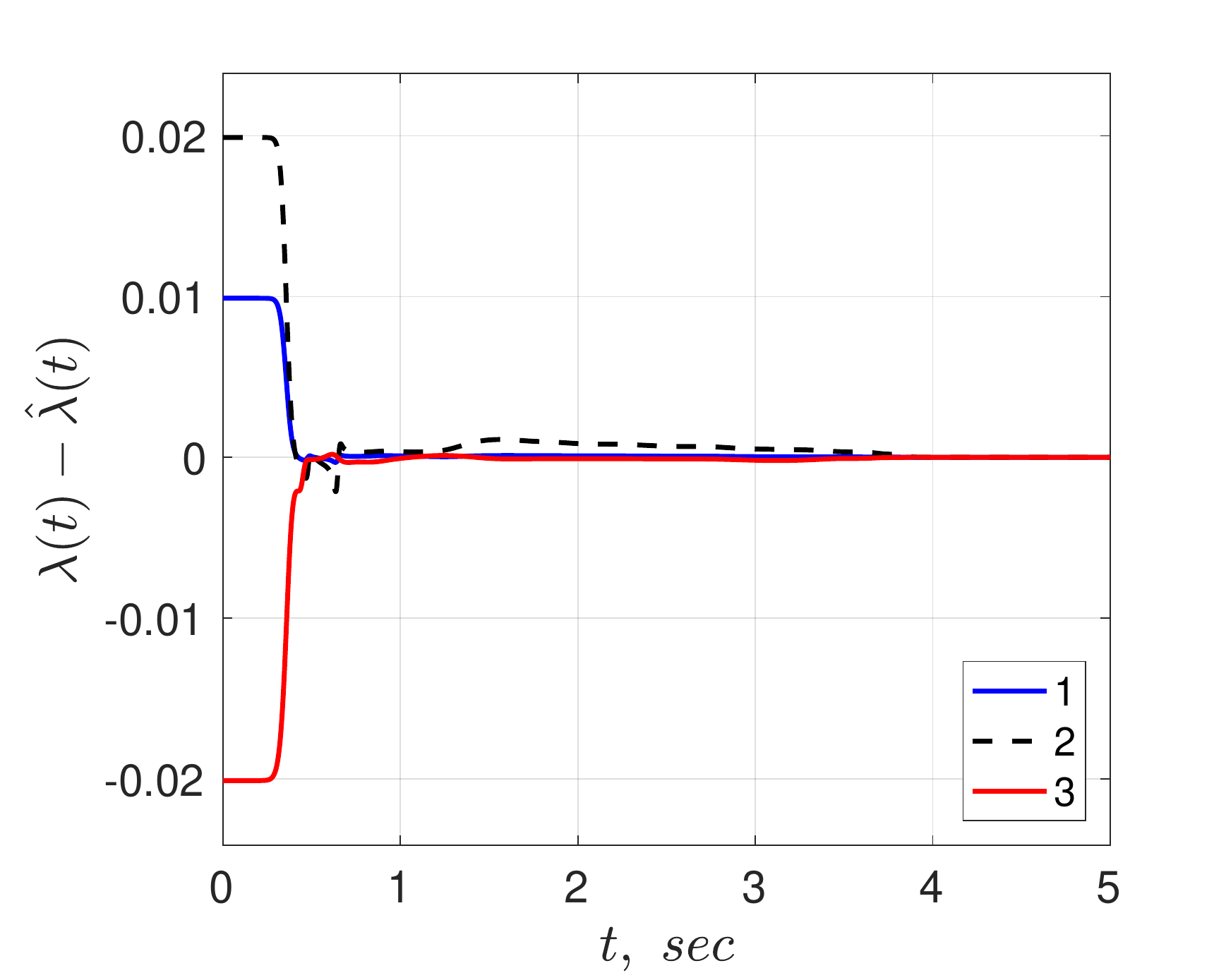}}
	\
	\subcaptionbox{\label{fig:1dof_errors_sin_l0_Y} Transients for $Y(t)-\hat Y(t)$}{\includegraphics[width=0.31\textwidth]{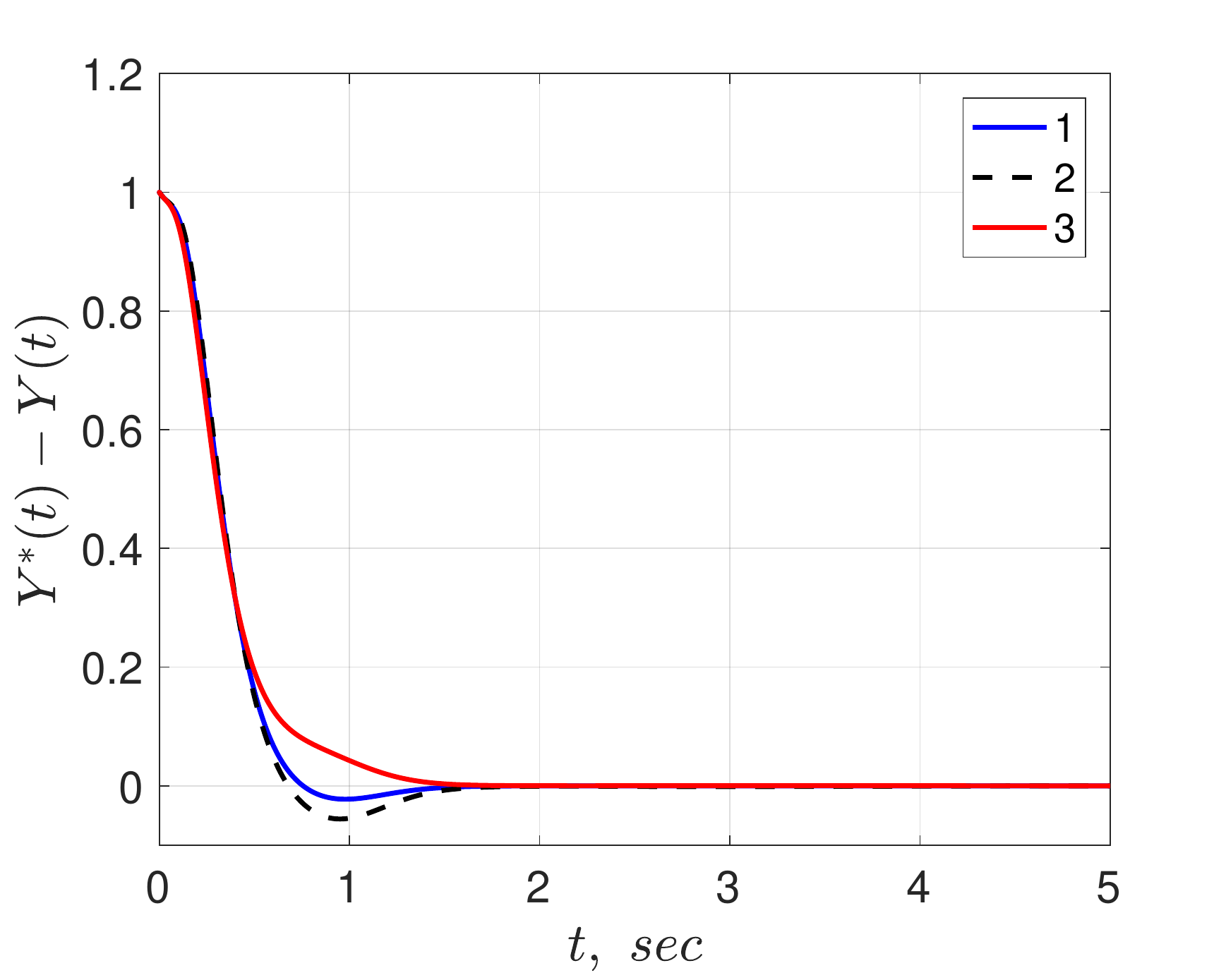}}
	\
	\subcaptionbox{\label{fig:1dof_errors_sin_l0_dY} Transients for $\dot Y(t)-\hat v_{Y}(t)$}{\includegraphics[width=0.31\textwidth]{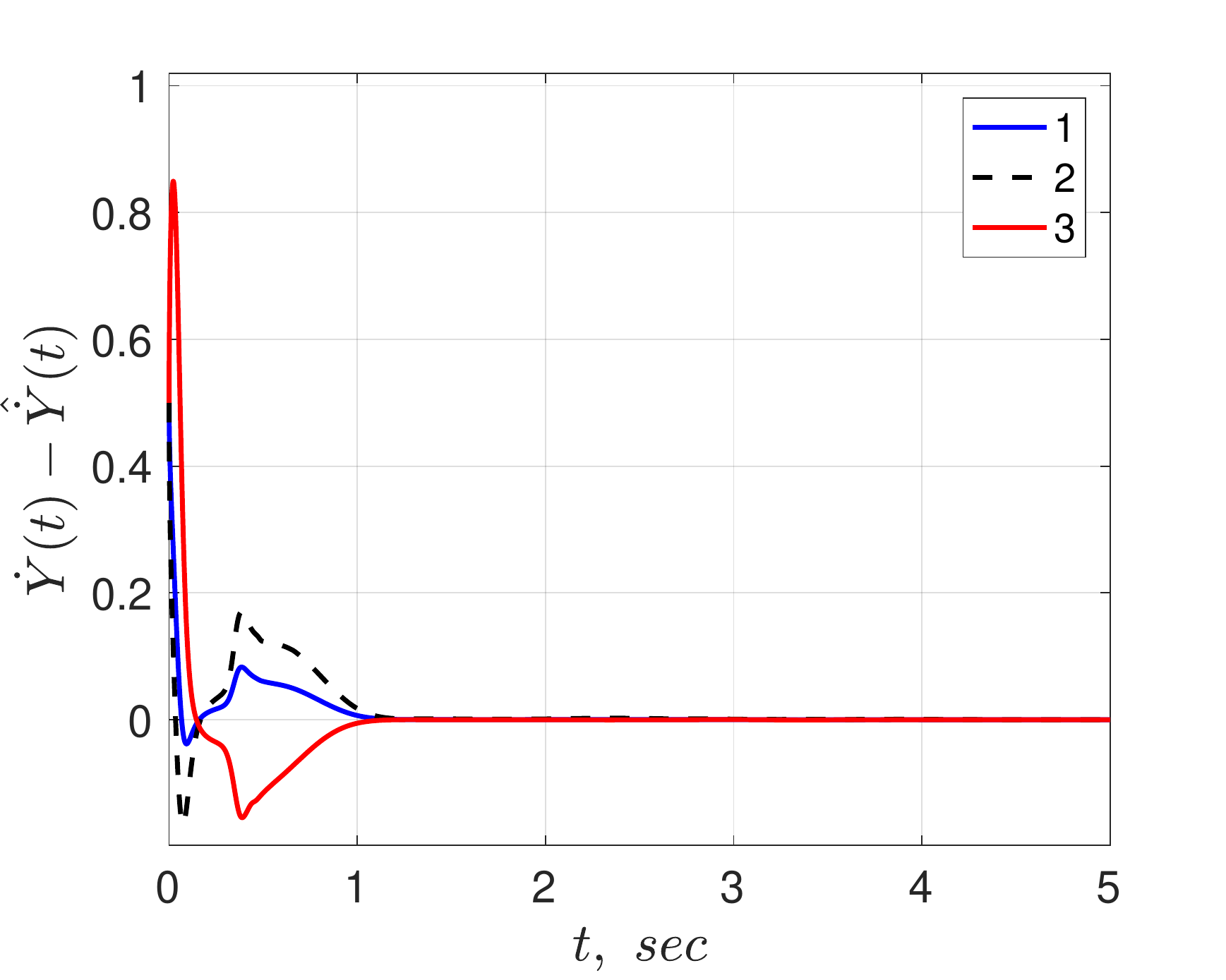}}
	\vspace{-2mm}
	\caption{\label{fig:1dof_errors_sin_l0} Errors with the sensorless-based FLC for the sinusoidal position reference: 1. $\eta=0.01$, 2. $\eta=0.02$, 3. $\eta=-0.02$}
\end{figure*}

\begin{figure*}[htp]
\centering
	\subcaptionbox{\label{fig:1dof_errors_step_l0_lambda} Transients for $\lambda(t)-\hat\lambda(t)$}{\includegraphics[width=0.31\textwidth]{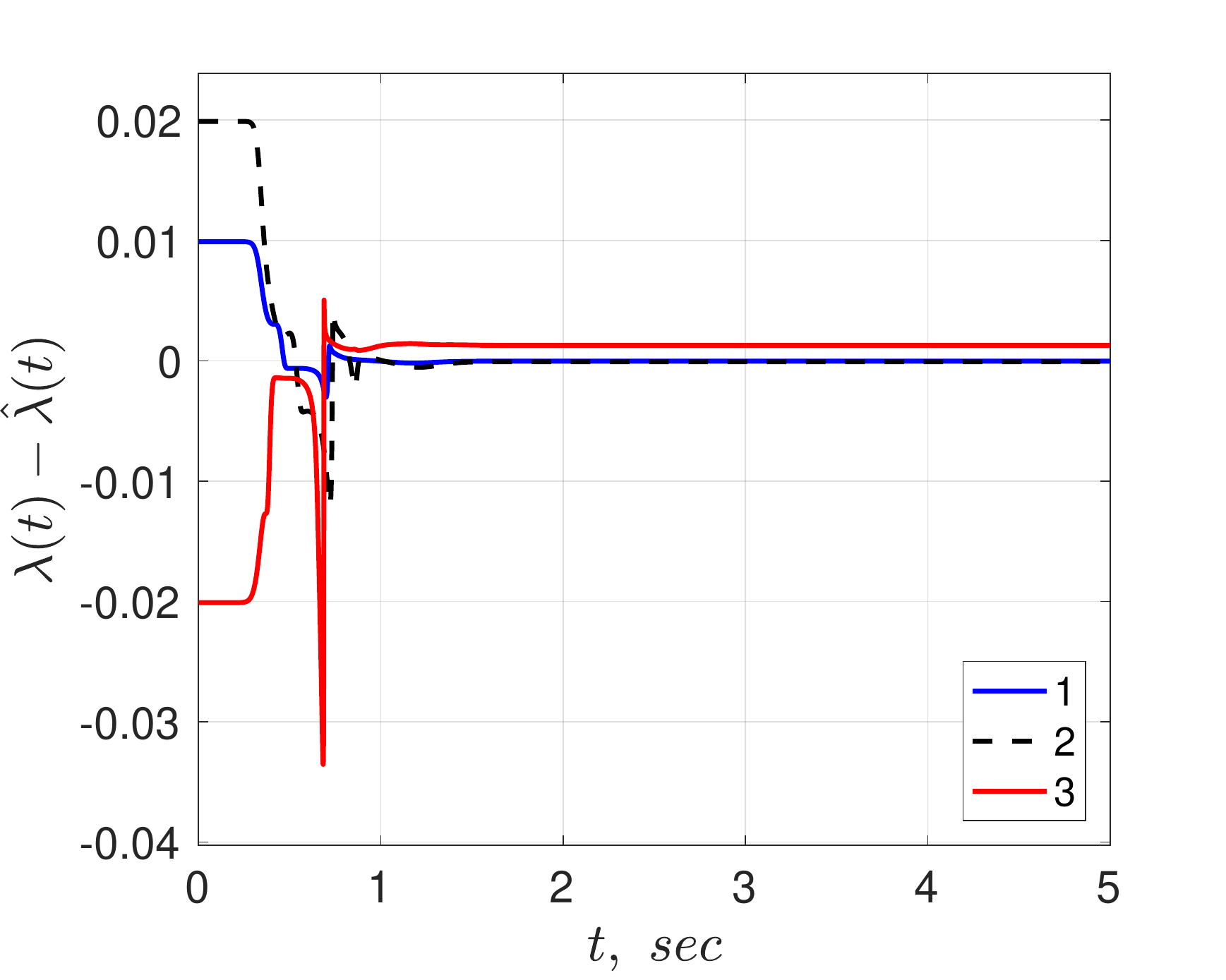}}
	\
	\subcaptionbox{\label{fig:1dof_errors_step_l0_Y} Transients for $Y(t)-\hat Y(t)$}{\includegraphics[width=0.31\textwidth]{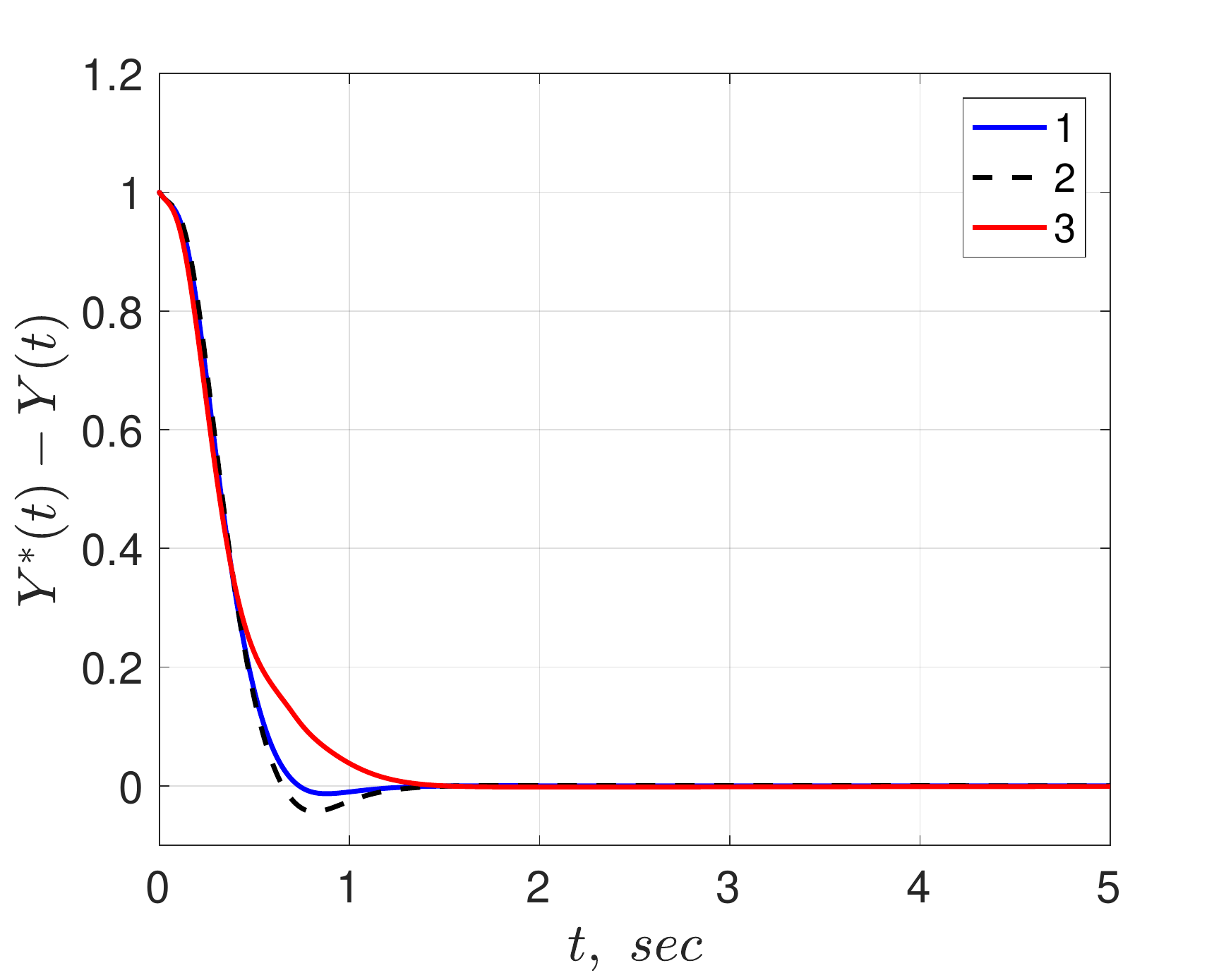}}
	\
	\subcaptionbox{\label{fig:1dof_errors_step_l0_dY} Transients for $\dot Y(t)-\hat v_{Y}(t)$}{\includegraphics[width=0.31\textwidth]{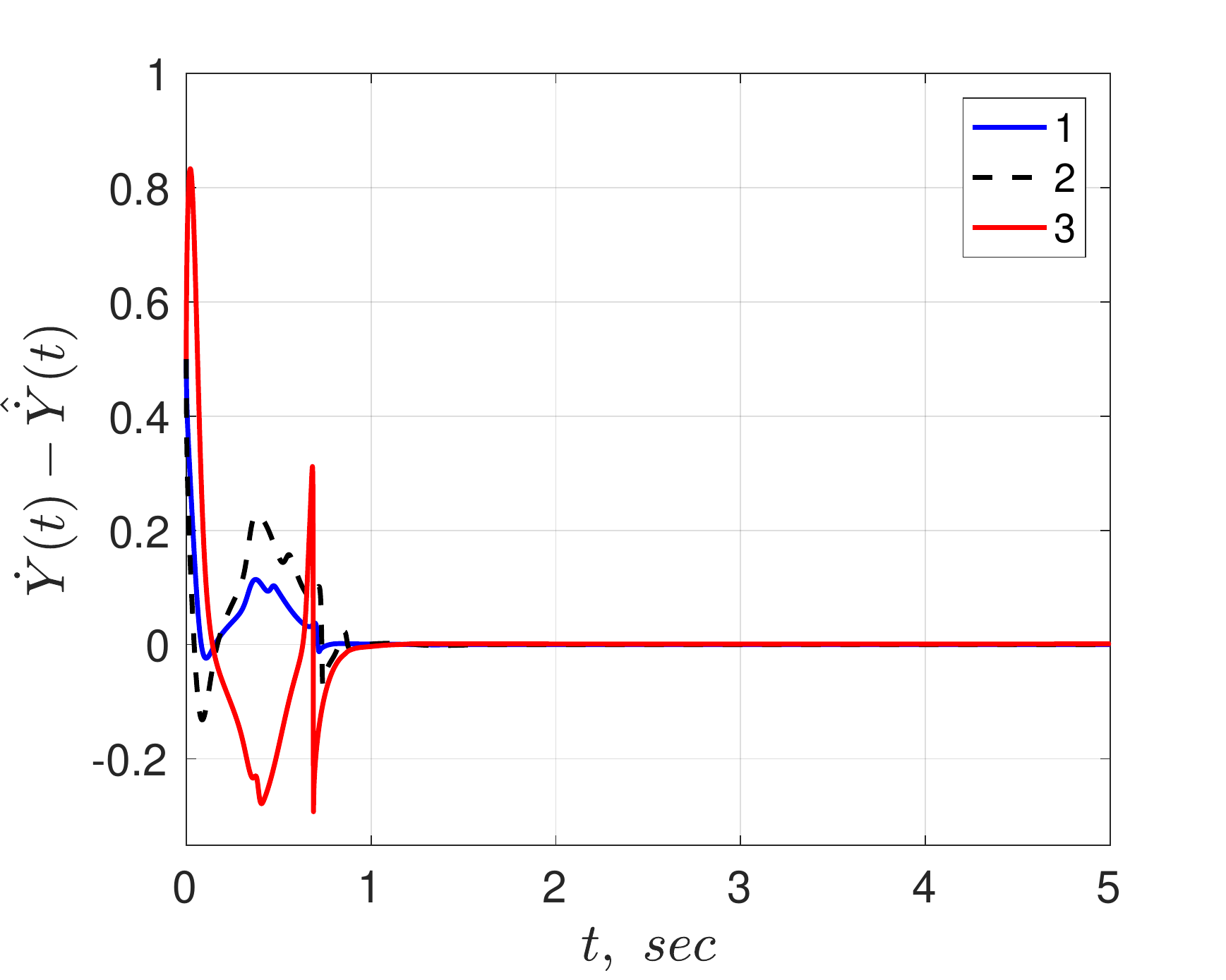}}
	\vspace{-2mm}
	\caption{\label{fig:1dof_errors_step_l0} Errors with the sensorless-based FLC for the smooth steps position reference: 1. $\eta=0.01$, 2. $\eta=0.02$, 3. $\eta=-0.02$}
\end{figure*}

In Figures \ref{fig:1dof_Y} we compare the behaviour of the position for the two desired trajectories with the difference in the initial conditions of $\lambda$ and $\psi$ such that $\eta = 0.01$: $\lambda(0) = 0.01$ and $\psi(0) = 0$.
In Figs.~\ref{fig:1dof_errors_sin_gamma} and  \ref{fig:1dof_errors_step_gamma} we evaluated the effect on the observation errors of changing the adaptation gain $\gamma$. In Figs. \ref{fig:1dof_errors_sin_l0} and \ref{fig:1dof_errors_step_l0} the behaviour of the observer for different values of $\eta$ is showed. In last figure we observe that there is a steady state error, which increases for bigger adaptation gains. This reveals that the condition $\Delta \notin \call_2$ is not satisfied, but the overall performance is still satisfactory.

\begin{table}[!hb]
\begin{center}
\caption{Initial conditions}

\begin{tabular}{|c|c|c|c|}
\hline
$\lambda_1(0)=0.5$&
$\hat\lambda_1(0)=0.1$ &
$Y(0)=-0.001$&
$\hat Y(0)=0$
\\
$\lambda_2(0)=0.6$ &
$\hat\lambda_2(0)=0.5$ &
$\dot Y(0)=0$&
$\hat v_Y(0)=0$
\\
$\lambda_3(0)=0.7$ &
$\hat\lambda_3(0)=0.1$ &
$X(0)=0$&
$\hat X(0)=0$
\\
$\lambda_4(0)=0.2$&
$\hat\lambda_4(0)=0.5$&
$\dot X(0)=0$&
$\hat v_X(0)=0$
\\
\hline
\end{tabular}
\label{tab_1}
\end{center}
\end{table}
\begin{table}[!hb]
\caption{Initial conditions}
\resizebox{\columnwidth}{!}{%
\begin{tabular}{|c|c|c|c|}
\hline
\multicolumn{4}{|c|}{First case} \\
\hline
$\lambda_1(0)=0.5$&
$\hat\lambda_1(0)=0.1$ &
$Y(0)=-0.001$&
$\hat Y(0)=0$
\\
$\lambda_2(0)=0.6$ &
$\hat\lambda_2(0)=0.5$ &
$\dot Y(0)=0$&
$\hat v_Y(0)=0$
\\
$\lambda_3(0)=0.7$ &
$\hat\lambda_3(0)=0.1$ &
$X(0)=0$&
$\hat X(0)=0$
\\
$\lambda_4(0)=0.2$&
$\hat\lambda_4(0)=0.5$&
$\dot X(0)=0$&
$\hat v_X(0)=0$
\\
\hline
\hline
\multicolumn{4}{|c|}{Second case} \\
\hline
$\lambda_1(0)=0.1$&
$\hat\lambda_1(0)=0.2$ &
$Y(0)=0.001$&
$\hat Y(0)=-0.001$
\\
$\lambda_2(0)=0.2$ &
$\hat\lambda_2(0)=0.3$ &
$\dot Y(0)=0.01$&
$\hat v_Y(0)=0$
\\
$\lambda_3(0)=0.3$ &
$\hat\lambda_3(0)=0.5$ &
$X(0)=0.01$&
$\hat X(0)=0$
\\
$\lambda_4(0)=0.5$&
$\hat\lambda_4(0)=0.8$&
$\dot X(0)=0.01$&
$\hat v_X(0)=0$
\\
\hline
\hline
\multicolumn{4}{|c|}{Third case} \\
\hline
$\lambda_1(0)=0.6$&
$\hat\lambda_1(0)=0.5$ &
$Y(0)=0.001$&
$\hat Y(0)=-0.001$
\\
$\lambda_2(0)=0.6$ &
$\hat\lambda_2(0)=-0.3$ &
$\dot Y(0)=0.01$&
$\hat v_Y(0)=-0.01$
\\
$\lambda_3(0)=0.8$ &
$\hat\lambda_3(0)=0.2$ &
$X(0)=0.03$&
$\hat X(0)=0.02$
\\
$\lambda_4(0)=0.1$&
$\hat\lambda_4(0)=0.1$&
$\dot X(0)=0.02$&
$\hat v_X(0)=0.04$
\\
\hline
\end{tabular}
}
\label{tab_2}
\end{table}

%
\section{Concluding Remarks and Future Research}
\label{sec6}
\vspace{-2mm}
%
We have presented in this paper a potential solution to the challenging problem of designing a sensorless controller for MagLev systems. Instrumental for the development of the theory was the use of PEBO and DREM parameter estimators---which were recently reported in the control literature---to estimate the flux and the mechanical coordinates of the system. The sensorless controller is then obtained replacing the estimated state in a full-state feedback IDA-PBC for the 2-dof system and in a FLC for the 1-dof case. It should be underscored that these controllers can be replaced with any other full-state feedback stabilizing controller. Simulation results show the excellent behaviour of the proposed observer. Consequently, the regulation performance of the sensorless controller is very similar to the one obtained with the full-state feedback scheme. 

The convergence proof of the proposed observers relies on the excitation conditions \eqref{exccon} and  \eqref{exccon0}, which has a clear energy interpretation. Assuming that the energy of the flux is unbounded is reasonable, since it holds true for a moving levitated object. However, the assumption on the functions $\Delta_i,\;i=1,2,$ is hard to verify {\em a-priori} and is critically dependent on the choice of the filters that generate the extended regressors~---~see \cite{ARAetaltac,GERetal,ORTetalaut} for some discussion on this important issue. Notice that if the signals $\xi$, defined in \eqref{zxi}, are persistently exciting this condition will hold true for any choice of the filters. Since $\psi$ is generated via \eqref{dotpsi}, the level of excitation of $\xi$ is essentially determined by the excitation of $U$ and $I$. This fact suggests the addition of a high-frequency probing signal to the voltage to enforce the excitation condition, which is common practice in technique-oriented sensorless schemes.  On the other hand, as indicated in \cite{SCHMAS}, an important feature of most practical MagLev systems is that the amplifiers
driving the coils are switching amplifiers, which induce high frequency perturbations to the coil currents and tend to produce periodically perturbed bias flux. Hence, one might expect that this switching ripple could produce the required excitation conditions. 

Several open questions are currently being investigated.  
\begite
\item The computational complexity of the proposed observer is relatively high for this kind of application---particularly for the $1$-dof system. Controller approximation techniques should be tried to obtain a practical design. 
\item Experimental validation is currently under way, but is being hampered by the computational complexity mentioned above.
\item It would be interested to compare our proposal with existing technique-oriented methods as well as the signal injection-based PEBO reported in \cite{YIetal}.
\item Saturation effects, which may degrade the systems performance, have been neglected in our analysis. It seems possible to incorporate this consideration in the controller design.
\item As mentioned in Remark \ref{rem5} a potential difficulty of DREM is the use of open-loop integration. This problem is particularly important in for noisy signals. It should be mentioned that, in spite of this potential drawback, several successful experimental validations of the effectiveness of PEBO, which incorporate some {\em ad-hoc} ``safety-nets" to PEBO, have been reported, see {\em e.g.}, \cite{BOBetalaut,BOBetalijc,CHOetal}. Finding the right safety nets for the MagLev application will be needed in the experimental test. See also \cite{PYRcst} where a variation of PEBO that avoids these robustness problems is presented. 
\endite
\vspace{-3mm}
\begin{ack}
\vspace{-3mm}
The third author expresses his gratitude to Dr Eric Maslen for some clarifications regarding his work. This research was partially supported by  Government of Russian Federation (074U01), the Ministry of Education and Science of Russian Federation (14.Z50.31.0031).
\end{ack}


\newpage

%

%
\appendix
%
\section{Proof of Proposition \ref{pro5}}
\label{propro5}
%
To simplify the expressions we write the model \eqref{I1dof} in state-space form with the state variables $x=\col(x_1, x_2, x_3):=\col(Y, m\dot Y, \lambda)$ and denote the measurable signal $y:=i$. This yields,
\begin{align}
\label{x1}
\dot x_1&={1\over m}x_2,\\
\label{x2}
\dot x_2 &={1\over 2k}x_3^2-mg,\\
\label{x3} 
\dot x_3 & = -Ry+u,\\
\label{1DOF_y}
y & = {1\over k}(c-x_1)x_3.
\end{align}

From \eqref{dotpsi0} and \eqref{x3} we get
\begin{align}
\label{x3_par}
x_3(t)=\eta+\psi(t),
\end{align}
where $\eta=x_3(0)-\psi(0)$. The essence of the proof is to,  using \eqref{x3_par}, manipulate the systems equations \eqref{x1}--\eqref{1DOF_y} to establish an algebraic relation that depends only on the signals $y$ and $u$ ---and filtered combinations of them---and a function of the unknown parameter $\eta$. 

Instrumental to carry out this task is the Swapping Lemma, see {\em e.g.}, Lemma 3.6.5 of \cite{SASBOD}, that is used in this proof in the following way
$$
{\mu\over p+\mu}[a(t)b(t)]=b(t) {\mu\over p+\mu}[a(t)]+ {\mu\over p+\mu}[\dot b(t){1 \over p+\mu}[a(t)]],
$$
where $a$ and $b$ are some scalar functions of time and $\mu>0$. 

First, compute $\dot y$
\begin{align}
\label{ydot}
\dot y & = -{1\over k}\dot x_1x_3+{1\over k}(c-x_1)\dot x_3
\end{align}
and consider $y\dot x_3-\dot y x_3$ together with \eqref{x1}:
\begin{align}
\label{reg1}
y\dot x_3-\dot y x_3 & = {1\over k}\dot x_1x_3^2=
{1\over km} x_2x_3^2.
\end{align}
Substituting \eqref{x3_par} into \eqref{reg1} we get
\begin{align}
\label{reg3}
y\dot x_3-\dot y \psi &=\dot y \eta+
{1\over km} x_2x_3^2.
\end{align}
Applying the operator\footnote{To simplify the notation, In the sequel we omit the argument $p$ from the operator $W(p)$.}
$$
W(p):={\mu\over p+\mu}.
$$
to \eqref{reg3} we get
\begin{align}
\label{reg4}
W [y\dot x_3-\dot y \psi] &=W \dot y \eta+
W {1\over km} x_2x_3^2.
\end{align}
Define the signal
\begin{align}
q_1 :& =  W \,y\dot x_3-W \,\dot y \psi \nonumber\\
	& = W\, y(u-Ry)-\psi\,W \,\dot y+ {1\over p+\mu}\,\left[\dot\psi W \,\dot y\right]\nonumber\\
	& = W\, y(u-Ry)-\psi\,{\mu\, p\over p+\mu} \, y \nonumber \\
	& \quad + {1\over p+\mu}\,\left[(u-Ry) {\mu\, p\over p+\mu} \, y\right]\nonumber\\
	& = W\, y(u\!-\!Ry)\!-\!\psi\,\omega_1 \! +\! {1\over p\!+\!\mu}\,\left[(u\!-\!Ry) \omega_1\right],
\label{q1}
\end{align}
where the Swapping Lemma  was applied to the term $W \,\dot y \psi$ to get the second identity and we defined the (measurable) signal 
\begin{align}
\label{omega1}
\omega_1:=W \, y.
\end{align}
Note that $q_1$ may be computed based on $y$ and $u$ only.
Replacing \eqref{q1} in \eqref{reg4} we get
\begin{align}
\label{reg5}
km\,q_1 &= \eta km \omega_1
+
W x_2x_3^2
\end{align}
and after applying the Swapping Lemma again to the term $W x_2x_3^2$ we get
\begin{align}
km\,q_1 & = \eta \,km\,\omega_1 + x_2W x_3^2-{1\over p+\mu}\left[\dot x_2W x_3^2\right] \nonumber \\
	& = \eta \,km\,\omega_1 + x_2W x_3^2 \nonumber \\
	& \quad - {1\over p+\mu}\left[\left({1\over 2k}x_3^2-mg\right)W x_3^2\right] \nonumber \\
	& = \eta \,km\,\omega_1 + x_2\phi_1 \nonumber \\
	& \quad - {1\over p+\mu}\left[\left({1\over 2k}x_3^2-mg\right)\phi_1\right]
\label{reg6}
\end{align}
where we defined the signal
\begin{align}
\label{phi1}
\phi_1:=W x_3^2.
\end{align}
Define a second auxiliary signal
\begin{align}
\label{q2}
q_2 :& = W \,km\,q_1 \nonumber \\
	& = \eta \,km\,W\,\omega_1 + {W\,x_2\phi_1} \nonumber \\
	& \quad - {\mu\over (p+\mu)^2}\left[\left({1\over 2k}x_3^2-mg\right)\phi_1\right] \nonumber \\
	& = \eta \,km\,W\,\omega_1 + {x_2W\,\phi_1-{1\over p+\mu}\left[\dot x_2W\,\phi_1\right]} \nonumber \\
	& \quad - {\mu\over (p+\mu)^2}\left[\left({1\over 2k}x_3^2-mg\right)\phi_1\right] \nonumber \\
	& = \eta \,km\,\omega_2 + x_2\phi_2 - {1\over p+\mu}\left[\left({1\over 2k}x_3^2-mg\right)\phi_2\right] \nonumber \\
	& \quad - {\mu\over (p+\mu)^2}\left[\left({1\over 2k}x_3^2-mg \right)\phi_1\right]
\end{align}
where we used \eqref{reg6} in the second equation, applied the Swapping Lemma to the term $W x_2\phi_1$ to get the third identity, used \eqref{x2} in the fourth one and  
\begin{align}
	\omega_2:=W\,\omega_1,
	\\
	\phi_2:=W\,\phi_1.
	\label{phi2}
\end{align}
Consider the following identity
\begin{align}
	& (km\,q_1\phi_2 - q_2\phi_1)2k\mu = \eta\,(\omega_1\phi_2-\omega_2\phi_1)2k\mu \nonumber \\
	& \quad - \phi_2\,W \left[x_3^2\phi_1-2mgk\phi_1\right] + \phi_1\,W\left[x_3^2\phi_2-2mgk\phi_2\right] \nonumber \\
	& \quad	+ \phi_1\,{\mu\over (p+\mu)^2}\left[x_3^2\phi_1-2mgk\phi_1\right],
\label{eq:q3}
\end{align}
where we replaced $q_1$ and $q_2$ with \eqref{reg6} and~\eqref{q2} respectively to obtain right-hand side. 
Signals $x_3^2$, $\phi_1$, and $\phi_2$ cannot be computed based on the measurable signals $y$ and $u$, but can be replaced by combination of the measurable signal $\psi$ and unknown parameter $\eta$ using ~\eqref{x3_par}, \eqref{phi1}, and~\eqref{phi2}
\begin{align}
\label{x32}
x_3^2 &= \eta^2 + 2\eta\psi + \psi^2, \\
\label{phi1_par}
\phi_1 &= \eta^2 + 2\eta(W\psi) + (W\psi^2) + \varepsilon(t),\\
\label{phi2_par}
\phi_2 &= \eta^2 + 2\eta(W^2\psi) + (W^2\psi^2) + \varepsilon(t).
\end{align}

Neglecting exponential decaying terms $\varepsilon(t)$ in~\eqref{phi1_par}--\eqref{phi2_par} and substituting with~\eqref{x32} into~\eqref{eq:q3} after lenghty, but straightfoward, calculations we get a linear regression model:
\begin{align}
\label{eq:1dof_reg}
z^0 = \eta\varphi^0_1 + \eta^2\varphi^0_2
  + \eta^3\varphi^0_3 + \eta^4\varphi^0_4
  + \eta^5\varphi^0_5 + \eta^6,
\end{align}
where
\break
\vspace{-3mm}
\begin{align}
z^0 & := 2 k \mu ( k m q_1 \phi_2  -  q_2 \phi_1 ) \nonumber \\
  & \quad - W[\psi^2] \Big( W[(\psi^2 - 2 m g k) W^2[\psi^2]] \nonumber \\
  & \quad \quad + W^2[(\psi^2 - 2 m g k) W[\psi^2]] \Big) \nonumber \\
  & \quad + W^2[\psi^2] W[(\psi^2 - 2 m g k) W[\psi^2]],
\nonumber \\
\varphi^0_1  & := 2 k^2 m \mu ( \omega_1 W^2[\psi^2] - \omega_2 W[\psi^2] ) \nonumber \\
  & \quad + 2 W[\psi]\Big(W[(\psi^2 - 2 m g k) W^2[\psi^2]]  \nonumber \\
  & \quad \quad + W^2[(\psi^2 - 2 m g k) W[\psi^2]]\Big) \nonumber \\
  & \quad + 2 W[\psi^2]\Big( W[\psi W^2[\psi^2]] + W[(\psi^2 - 2 m g k) W^2[\psi]] \nonumber \\
  & \quad \quad + W^2[\psi W[\psi^2]] + W^2[(\psi^2 - 2 m g k)W[\psi]] \Big) \nonumber \\
  & \quad - 2 W^2[\psi] W[(\psi^2 - 2 m g k) W[\psi^2]] \nonumber \\
  & \quad - 2 W^2[\psi^2] ( W[\psi W[\psi^2]] + W[(\psi^2 - 2 m g k) W[\psi]] ),
\nonumber 
\\
\varphi^0_2 & := 4 k^2 m \mu ( \omega_1 W^2[\psi] - \omega_2 W[\psi] ) + (W[\psi_2])^2 \nonumber \\
  & \quad + 2 m g k ( W^2[\psi_2] - 2 W[\psi_2])  \nonumber \\
  & \quad + 4 W[\psi] \Big( W[\psi W^2[\psi_2]] +  W[(\psi^2 - 2 m g k) W^2[\psi]] \nonumber \\
  & \quad \quad	+  W^2[\psi W[\psi_2]] +  W^2[(\psi^2 - 2 m g k) W[\psi]]\Big) \nonumber \\
  & \quad + 2 W[\psi_2] W\Big[W^2[\psi_2] + 2 \psi W^2[\psi] + 2 W [\psi W[\psi]]\Big] \nonumber \\
  & \quad - 4 W^2 \psi ( W[\psi W[\psi_2]] +  W[(\psi^2 - 2 m g k) W[\psi]]) \nonumber \\
  & \quad - (W^2[\psi_2])^2 +  W[(\psi^2 - 2 m g k) W^2[\psi_2]] \nonumber \\
  & \quad + W^2[(\psi^2 - 2 m g k) W[\psi_2]] - 4 W^2[\psi_2] W[\psi W[\psi]] \nonumber \\
  & \quad - W[(\psi^2 - 2 m g k) W[\psi_2]],
\nonumber 
\\
\varphi^0_3 & := 2 k^2 m \mu ( \omega_1 - \omega_2 ) + 4 m g k (W^2[\psi] - 2 W[\psi]) \nonumber \\
  & \quad + 4 W[\psi] \Big( W[\psi^2] + W^3[\psi^2] + 2 W[\psi W^2[\psi]] \nonumber \\
  & \quad \quad + 2 W^2[\psi W[\psi]] \Big) + 4 W[\psi^2] W^3[\psi] \nonumber \\
  & \quad - 4 W^2[\psi] ( W^2[\psi^2] + 2 W[\psi W[\psi]] ) \nonumber \\
  & \quad - 2 W[\psi W[\psi^2]] + 2 W[\psi W^2[\psi^2]] + \nonumber
\\
  & \quad + 2 W[(\psi^2 - 2 m g k)(W^2[\psi] - W[\psi]])] \nonumber \\
  & \quad + 2 W^2[(\psi^2 - 2 m g k) W[\psi] + \psi W[\psi^2]],
\nonumber \\
\varphi^0_4 & := - 2 m g k + 2 W [\psi^2] - W^2[\psi^2] + 2 W^3[\psi^2] \nonumber \\
  & \quad - 4 W[\psi(W[\psi]] + W^2 [\psi]])] + 4 W[\psi] W^2[\psi]  \nonumber \\
  & \quad - 4 (W^2[\psi])^2 + 4 W[\psi] \Big( W[\psi] + 8 W^3[\psi] \Big),
\nonumber \\
\varphi^0_5 & := 4 W[\psi] - 2 W^2[\psi] + 4 W^3 [\psi].
\nonumber
\end{align}

The proof is completed applying to the regression model \eqref{eq:1dof_reg} the filter ${\rho p\over p+\rho}$ to get the new regression model \eqref{reg9}, where we defined ${(\cdot)}={\rho p\over p+\rho}(\cdot)^0$. Notice that, due to the derivative action of the filter, the constant term $\eta^6$ in \eqref{eq:1dof_reg} has been removed in \eqref{reg9}. This eliminates a constant term (a one) from the regressor, whose excitation conditions for parameter convergence are strictly weaker.

\end{document}